\documentclass[18pt]{article}
\usepackage{arxiv}
\usepackage[utf8]{inputenc} % allow utf-8 input
\usepackage[T1]{fontenc}    % use 8-bit T1 fonts
\usepackage{hyperref}       % hyperlinks
\usepackage{url}            % simple URL typesetting
\usepackage{booktabs}       % professional-quality tables
\usepackage{amsfonts}       % blackboard math symbols
\usepackage{nicefrac}       % compact symbols for 1/2, etc.
\usepackage{microtype}      % microtypography
\usepackage{lipsum}
\usepackage{bm}
\usepackage{graphicx}
\usepackage{enumerate}
\usepackage{siunitx}
\usepackage[]{algorithm2e}
\usepackage{amsmath}
\usepackage{amsbsy}
\usepackage{bm}
\usepackage{amssymb}
\usepackage{color}
\usepackage{subfigure}
\usepackage{wrapfig}
\usepackage{appendix}
\title{Linear stability analysis of particle laden planar jet in the dilute suspension limit }

\author{
  Srikumar Warrier\\
  Interdisciplinary Center for Energy Research\\
  Indian Institute of Science\\
  Bangalore, India \\
   \And
  Gaurav Tomar\thanks{Corresponding author.} \\
  Department of Mechanical Engineering\\
  Indian Institute of Science\\
  Bangalore, India \\
  \texttt{gtom@iisc.ac.in} \\
}

\begin{document}
\maketitle

\begin{abstract}
 Particle laden flows are commonly seen in many industrial applications such as fluidized beds in process industry, air laden with abrasive particles in abrasive machining and particle laden plumes in chemical industries. In the present work, we perform local analysis of a particle laden planar jet in the dilute suspension regime. Unladen parallel planar jets have been extensive studied using normal modes and is shown to have two unstable modes namely sinuous and varicose modes. Sinuous modes are found to be more unstable compared to the varicose modes. In the present study, we investigate the effect of particles on the stability of planar jets.  Addition of particles at low Stokes numbers ($St$) (fine particles) results in higher growth rates than that of the unladen jet. In the intermediate Stokes number regime, addition of particles have a stabilizing effect on both the sinuous and the varicose modes. Interestingly for $St\sim10$, the unstable varicose mode is completely damped. Increasing the Stokes number by increasing the particle size, both sinuous and varicose modes show increasing growth rates, while increasing density ratio has a stabilizing effect on the flow. For non uniform particle loading, additional modes apart from the sinuous and varicose modes are observed. These modes suggests occurrence of compositional instability with an increased particle accumulation in the shear layer that is an order of magnitude higher compared to that of the sinuous and varicose modes.
\end{abstract}


\section{Introduction}
\label{sec:Intro}
Particle laden flows find a variety of industrial applications for different types of dispersed phases carried by the flow. Generally by particle laden flows, we mean gas-solid suspension flows. Classic examples of such systems include fluidized beds, particle plumes and particle laden gas jets for abrasive jet machining, and in pollution control devices like cyclone separators. Particle laden flows involving liquid - gas phases are employed in food processing industries, for instance, milk injected into a stream of high temperature gas to obtain dry powdered milk.
\newline
 Mixing layers are characterized by the presence of organised vortical structures \cite{brown_roshko_1974}. Such flow structures make it difficult to control transport of particles in turbulent flows.
In the context of design of propulsion devices for aerospace applications, the stability of open shear flows like planar jets of gas/air carrying atomized fuel droplets plays a considerable role in determining the efficiency of combustion.
Particle laden turbulent mixing layers and planar jets, exhibit preferential concentration of particles which is shown to have significant effect on droplet combustion and also on the turbulence of the carrier fluid  \cite{EATON1994169}.  Tatsumi and Kakutani \cite{tatsumi_kakutani_1958} performed local stability analysis of an unladen planar jet and predicted the flapping mode (sinuous mode) to be dominant compared to the bulging mode (varicose mode) of instability. Sato and Sakao \cite{sato_sakao_1964} performed experimental investigation of planar jet and confirmed that the  most dominant mode of instability is indeed  the flapping mode. Matsubara et al. \cite{matsubara_alfredsson_segalini_2020} performed local and weakly non parallel stability analysis on a turbulent planar jet and captured both qualitatively and quantitatively the features of coherent structures. However the effect of particles in the flow on the stability of the jet has not been studied in the past.
Experimentally it has been observed that preferential concentration of particles is mainly caused by coherent structures in turbulent flows where dense particles are centrifuged away from the vortex cores and accumulates in the vicinity of the shear layers. For instance in dilute particle laden mixing layers, particles accumulate in the braid region, outside the vortex cores, as observed by Meiberg et al. \cite{meiburg2000vorticity}. Fan et al.\cite{Fan2004}, performed direct numerical simulation of a compressible planar particle laden turbulent jet and observed that when Stokes number is O$(1)$, particles accumulate near the outer edges of large-scale vortex structures. For small Stokes numbers $St= 0.01$, particles tend to follow the fluid closely and are seen to have considerable dispersion in the lateral direction.   
At large Stokes numbers of $St= 10$ and $50$, particles tend to  pass though the vortex structures as the particle response times are larger compared to the convective time scale of the flow.

   Saffman \cite{saffman_1962} studied the stability of flow of a dusty gas in a channel. By performing normal mode analysis, he showed that fine particles (low Stokes number) destabilize the flow while coarse particles (large Stokes number) stabilize the flow. Klinkenberg et al.\cite{Klinkenberg2014LinearSO} performed modal and non-modal stability analyses of particle laden channel flows to study the effect of added mass and Basset force in addition to the Stokes drag considered by Saffman \cite{saffman_1962}. They found that particles lighter than the fluid resulted in decrease of critical Reynolds number for modal stability, while heavier particles  increased the critical Reynolds number.  Non-modal analysis revealed that the generation of stream-wise streaks was the most dominant disturbance-growth mechanism in flows laden with particles. Introducing Basset force did not result in any qualitative change in the instability. In the context of shear layer flows with particles, Michael \cite{michael_1965} used Saffman's\cite{saffman_1962} formulation using which he analysed a planar vortex sheet with dust particles. He considered the steady state motion of dust with different mass concentration and relaxation times in the upper and lower streams along with the gas on each side of the vortex sheet. He found that for small disturbances, the growth rate reduces compared to the unladen case but the vortex sheet remains unstable regardless of the concentration of dust. Dimas and Kriger \cite{Dimas_Kiger} performed inviscid spatial analysis on a $tanh$ model of flow with the assumption that the mean particle velocity is equal to the mean local fluid velocity (which was also assumed by  Saffman\cite{saffman_1962}). This work was soon followed by Narayanan et al. \cite{Narayananetal2002}, who performed temporal analysis by considering viscous terms and derived an equivalent Orr-Sommerfeld equation to evaluate the temporal growth rates. Both Dimas and Kriger \cite{Dimas_Kiger} and Narayanan et al. \cite{Narayananetal2002} followed the formulation of  Zhang and Prosperreti \cite{ZHANG1997425} %\citeauthor{\citeauthor{ZHANG1997425}(\citeyear{ZHANG1997425}) 
  which was simplified for the case of dilute suspensions. 
  Xie et al. \cite{xie_09} performed linear analysis of Bickley jet profile for low and intermediate Stokes numbers with uniform particle loading. However the study was limited to finding the neutral curves for low ($St < 0.1$) and intermediate Stokes numbers ($1 < St < 10$). They used the modified Orr-sommerfeld equations derived by Saffman \cite{saffman_1962}. Xie et al. \cite{xie_09} saw that adding particles with $St > 10$ has a stabilizing effect on the planar jet.
  
 We note that, although with the advent of large supercomputers, direct numerical simulations of such flows has become possible, but in order to delineate the various mechanisms of the instability leading to the spatiotemporal growth of disturbances developing on a laminar base state, it is instructive to perform linear stability analysis. In this paper we perform linear stability analysis of particle laden planar jet and explore the linear stability characteristics at different particle Stokes numbers (small, intermediate and large), for constant and variable particle loading numerically using Chebyshev polynomials. We perform a detailed analysis of the unstable eigenmodes using momentum budget  for better understanding. As mentioned above the aspect of particle migration is also looked into in the linear regime. We show that non uniform particle loading gives rise to additional modes in the large Stokes limit. The approach taken in the present work is that of local volume averaging.  

 This paper is divided into 6 sections. Section (\ref{sec:Intro}) is the introduction. Governing equations are introduced along with the non dimensionalization of flow variables in  section (\ref{sec:formulation}). Perturbed equations are presented in section (\ref{sec: LVAE}). Numerical method and validation cases are reported in section (\ref{sec:Numerics}). Results are discussed in section (\ref{sec: results}). Paper concludes with section  (\ref{sec:conclusion}). Brief sketch of the derivation is given in the appendix (\ref{appendix_a}).

\section{Governing equations}
\label{sec:formulation}
\begin{figure}[!ht]
	\centering
	\subfigure[]{\includegraphics[width=0.3\textwidth]{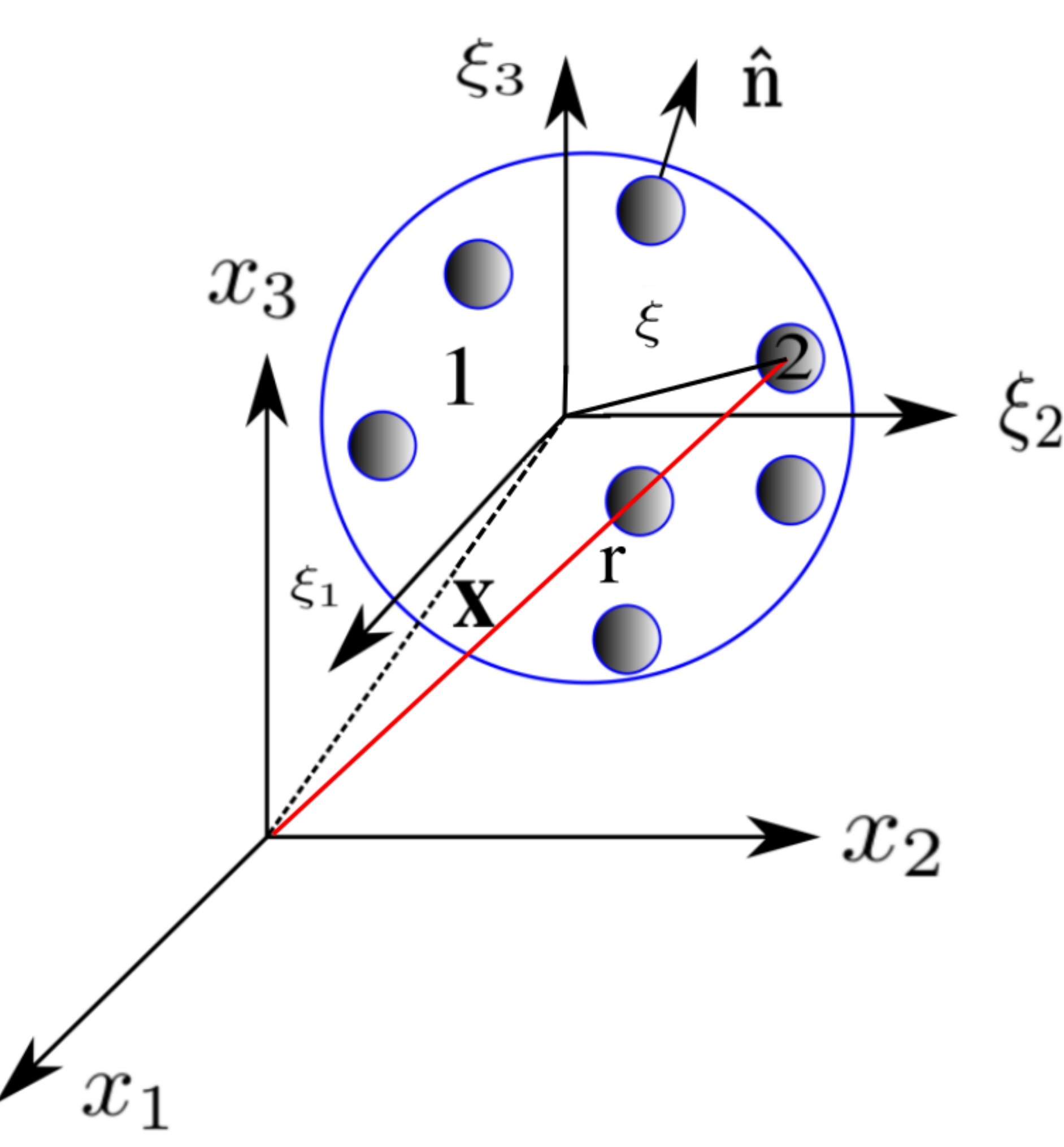}} 
	\subfigure[]{\includegraphics[width=0.55\textwidth]{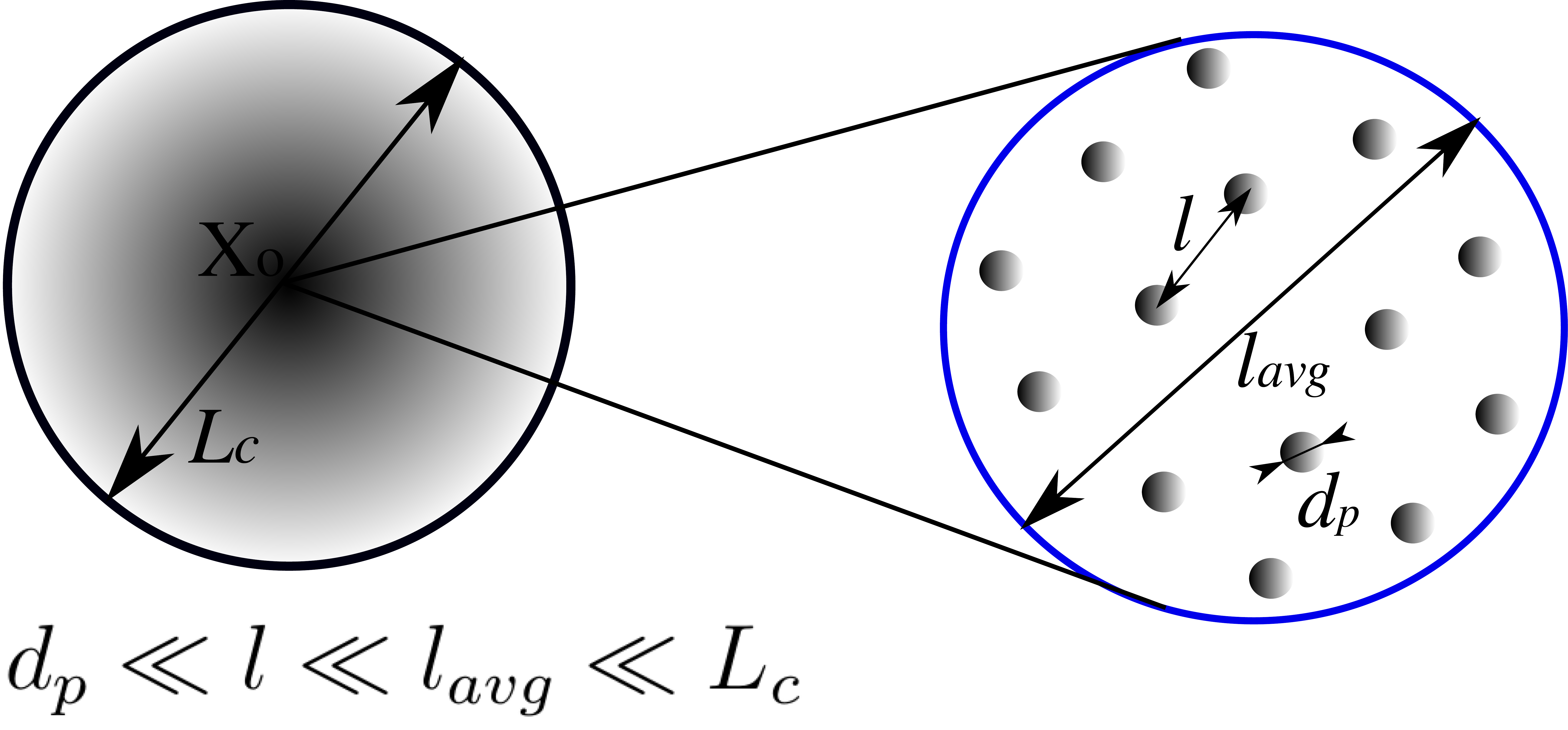}}
	\caption{(a) shows a two phase system with phase 2 (particles) suspended in phase 1 (gas) in a representative sampling volume. The center of the averaging volume is at a distance $\mathbf{x}$ from the origin. The local coordinate $\boldsymbol{\xi}$ is used to specify the coordinates within the averaging volume. $\boldsymbol{n}^{j}$ is the unit outward normal from $j^{th}$ phase. $\boldsymbol{r}$ is the position vector of phase 2 with respect to the origin of the global coordinate axes. (b) shows the sampling volume for averaging. For the point $\mathbf{x}_{0}$ in the flow, we associate an averaging volume to it. $d_{p}$ is the diameter of the particle, $l$ is the distance between any two particles. The dimension of the averaging volume ($l_{avg}$) is such that it is much greater than the inter particle distance but far smaller compared to the length scale ($L_{c}$) over which macroscopic changes occur (the characteristic length scale of the flow). The condition $d_{p}\ll l$ implies dilute suspension, i.e., the size of the particle is far smaller compared to the inter particle distance. }
	\label{coordinate system}  
\end{figure}

In order to model particle laden flows, we employ volume averaged equations that convert the Lagrangian description for particles into an Eulerian formulation resulting in a volume fraction field. Volume averaging method was initially developed by  Whitaker \cite{Whitaker_1967} and Slattery \cite{Slattery1967} for porous media flows. The basic idea behind volume averaging is that the governing equations defined only for the continuous phase(phase 1 in figure \ref{coordinate system}) can be spatially smoothed, resulting in a set of equations that are valid for the entire domain. As an example, Whitaker \cite{whittaker} used the idea of volume averaging to study the transport of water through pores to the external surface where it is removed by warm and dry air. Instead of solving the problem in terms of equations and boundary conditions that are defined only in the pores, Whitaker \cite{whittaker} used the pore scale information to obtain local volume averaged equations, accounting for the boundary conditions at the solid boundaries. The approach was rooted in the fact that there are more than one length scale involved as water sweeps through the pores and the disparity in scales allowed Whitaker to define a representative elemental volume (REV) over which the "smoothing procedure" could be applied. Gray and Lee \cite{gray1977} and Gray et al. \cite{Gray_1993} provided an alternative derivation of the local averaging theorems using distribution functions (Heaviside functions) which is unity in the dispersed phase and zero in the continuous phase. Hassanizadeh and Gray  \cite{Hassanizadeh_1979a}, \cite{Hassanizadeh_1979b}  applied volume averaging theorem to the balance laws and derived the microscopic interfacial conditions in the absence of surface properties. Cushman \cite{CUSHMAN1982248} provided a formal rigorous proof of the averaging theorems derived earlier by Gray and Lee \cite{gray1977}. A systematic derivation of averaging for particle laden flows in the framework of ensemble averaging can be found in Zhang and Prosperetti \cite{ZHANG1997425}. They derived the averaged momentum and energy equations for disperse two phase flows resulting in a two-fluid model. In the context of particle laden flows, volume averaging has been applied to the governing equations to obtain an Eulerian-Eulerian formulation for the suspension (also see Balachandar and Eaton \cite{Balachandar_eaton} and  Luca Brandt et al. \cite{Luca_arfm_2022}).
 
 Before we perform linear stability analysis of a particle laden jet, we derive the equations governing the motion of a particle-gas suspension following Chu and Prosperetti \cite{CHU2016176}. Consider a system with two phases $1$ and $2$,  with phase $2$ suspended in phase $1$ as shown in figure \ref{coordinate system}a. In order to find the average of a quantity (say velocity, particle concentration etc which are governed by conservation equations) defined for say, phase 1, for each point in space, we associate an averaging volume and integrate the governing equation over phase 1.
This is shown in figure \ref{coordinate system}b  where the point $\mathbf{x}_{0}$ is associated with an averaging volume that contains both the phases. Volume average of any quantity $\mathbf{q}^{(j)}$, over the averaging volume $V$ at time $t$, where $j$ indicates the phase 1,2 is defined as
\begin{equation}
\label{vol_avg}
\left\langle \mathbf{q}^{(j)}\left(\mathbf{x},t\right)\right\rangle =\frac{1}{V}\int_{V}\mathbf{q}\left(\mathbf{x}+\boldsymbol{\xi},t\right)\gamma^{(j)}\left(\mathbf{x}+\boldsymbol{\xi},t\right)dV_{\xi},
\end{equation}

where $V=V_{1}+V_{2}$ is the total volume and $V_{1},V_{2}$ are the volume occupied by the individual phases. Figure \ref{coordinate system}b  shows the coordinate system used. The coordinates $\mathbf{x} = (x_{1},x_{2},x_{3})$ denotes the center of volume over which averaging is performed. The local coordinate $\boldsymbol{\xi}$ corresponds to a point inside the averaging volume. 

The conservation equation for the quantity $\mathbf{q}^{(j)}$ defined at the microscopic scales is given by,
\begin{equation}
  \frac{\partial \boldsymbol{q}^{j}}{\partial t}=\nabla\cdot\left(-\boldsymbol{q}^{j}\mathbf{u}^{j}\right)+\mathbf{\nabla}\cdot \pmb{\phi}^{j},
  \label{eq:superficialAvg}
\end{equation}
where $\mathbf{u}^{j}$ is the velocity of the phase $j$, and $\pmb{\phi}^{j}$ is the non-convective flux of $\mathbf{q}^{j}$ (diffusive flux or source at the surface of the dispersed phase).  Volume averaging the entire equation and using the averaging theorems (see Appendix \ref{appendix_a}), we get,

\begin{equation}
\frac{\partial}{\partial t}\left(\left\langle \boldsymbol{q}^{j}\left(\mathbf{x},t\right)\right\rangle \right)+\nabla\cdot \left\langle \boldsymbol{q}^{j}\left(\mathbf{x},t\right)\mathbf{u}^{j}\right\rangle ={\nabla\cdot \left\langle \pmb{\phi}^{j}\right\rangle }+\frac{1}{V}\intop_{s^{i}}\left(-\boldsymbol{q}^{j}\left(\mathbf{u}^{j}-\mathbf{w}\right)\right)\cdot\mathbf{n}^{j}ds^{i}+{\frac{1}{V}\intop_{s^{i}}\pmb{\phi}^{j}\cdot\mathbf{n}^{j}ds^{i}},
\end{equation}
where, $S^i$ is the interfacial area between the dispersed and the continuous phase and $\mathbf{n}$ is the normal at the interface pointing away from the $j$th phase. Here $\mathbf{w}$ is the interface velocity.

It has to be noted that the definition of volume average of a quantity  $\boldsymbol{q}^{j}$ used in Chu and Prosperreti \cite{CHU2016176} is that of {\it intrinsic averaging}, that is,volume averaging performed over the volume occupied by the $j^{th}$ phase. The notation used in the present work  for intrinsic average of a quantity $\boldsymbol{q}$ is that followed by Gray and Lee \cite{gray1977}, Whittaker \cite{whittaker} and is denoted as $\left \langle \boldsymbol{q}^{j}\right \rangle^{(j)}$. Following Gray and Lee \cite{gray1977}, we define intrinsic averaging as the volume averaging performed over the volume occupied by the $j^{th}$ phase,

\begin{equation}
\left\langle \boldsymbol{q}^{(j)}\right\rangle ^{(j)}=\frac{1}{V^{(j)}}\intop_{V}q\left(\boldsymbol{x}+\boldsymbol{\xi},t\right)\gamma^{(j)}\left(\boldsymbol{x}+\boldsymbol{\xi},t\right)dV_{\xi}.
\end{equation}

Volume fraction of the $j^{th}$ phase is then given by,

\begin{equation}
\alpha^{(j)}=\frac{\left\langle \boldsymbol{q}^{(j)}\right\rangle }{\left\langle \boldsymbol{q}^{(j)}\right\rangle ^{(j)}}=\frac{V^{(j)}}{V},
\end{equation}

In terms of intrinsic average quantities,

\begin{equation}
  \frac{\partial}{\partial t}\left(\alpha^{(j)}\left\langle \mathbf{q}^{(j)}\right\rangle ^{(j)}\right)+\nabla\cdot\left(\alpha^{(j)}\left\langle \mathbf{q}^{(j)}\mathbf{u}^{(j)}\right\rangle ^{(j)}\right)=\nabla\cdot\left(\alpha^{j}\left\langle \mathbf{\pmb{\phi}}^{(j)}\right\rangle ^{(j)}\right)+\frac{1}{V}\int_{s^{(i)}}\mathbf{q}^{(j)}\left(\mathbf{u}^{(j)}-\mathbf{w}^{(j)}\right)\cdot\mathbf{n}^{(j)}ds^{i}+\frac{1}{V}\int_{s^{(i)}}\pmb{\phi}^{(j)}\cdot\mathbf{n}^{(j)}ds^{i}.
  \label{eq:IntrinsicAvg}
\end{equation}
At the interface we have,

\begin{equation}
\left(\pmb{\phi}^{(1)}-\pmb{\phi}^{(2)}\right)\cdot\boldsymbol{n}^{(1)}=\gamma_{s},
\end{equation} 

where $\gamma_{s}$ is the surface source term and $\boldsymbol{n}^{(2)}=-\boldsymbol{n}^{(1)}$. In case of momentum equation $\boldsymbol{\phi}$ is the stress and $\gamma_{s}$ is the net force due to surface forces such as surface tension, electric field, etc. The superficial averaging equation Eq. (\ref{eq:superficialAvg}) in terms of the corresponding intrinsic averages (Eq.\ref{eq:IntrinsicAvg}) is similar to that in Chu and Prosperetti \cite{CHU2016176} (see Eq. (10) therein). For the next part of the derivation we follow the procedure given in Chu and Prosperetti \cite{CHU2016176} and briefly mention the steps here. Let,

\begin{equation}
\boldsymbol\Phi^{j}={\nabla\cdot V^{(j)}\left\langle \pmb{\phi}^{(j)}\right\rangle }^{(j)}+{\intop_{s^{i}}\pmb{\phi}^{(j)}\cdot\mathbf{n}^{(j)}ds^{i}}.
\label{PHI_eq}
\end{equation}

\begin{equation}
\frac{\partial}{\partial t}\left(V^{(j)}\left\langle \boldsymbol{q}^{(j)}\right\rangle^{(j)} \right)+\nabla\cdot V^{(j)}\left\langle \boldsymbol{q}^{(j)}\mathbf{u}^{(j)}\right\rangle^{(j)}  = \boldsymbol{\Phi}^{(j)}+\intop_{s^{i}}\left(-\boldsymbol{q}^{(j)}\left(\mathbf{u}^{(j)}-\mathbf{w}^{(j)}\right)\right)\cdot\mathbf{n}^{(j)}ds^{i}.
\label{chu_present}
\end{equation}

Defining a new quantity $\overline{\pmb{\phi}}$ such that it represents the large scale effects due to $\left\langle \pmb{\phi}^{\left(j\right)}\right\rangle$ varying slowly over averaging volume as,

\begin{equation}
\overline{\pmb{\phi}}=\alpha^{\left(1\right)}\left\langle \pmb{\phi}^{\left(1\right)}\right\rangle^{(1)} +\alpha^{\left(2\right)}\left\langle \pmb{\phi}^{\left(2\right)}\right\rangle^{(2)},
\end{equation}

and  $\left(\pmb{\phi}^{\left(j\right)}\right)^{'}$ be the contribution to $\pmb{\phi}^{\left(j\right)}$ due to local small scale fluctuations in the averaging volume,
\begin{equation}
\pmb{\phi}^{\left(j\right)}=\overline{\pmb{\phi}}+\left(\pmb{\phi}^{\left(j\right)}\right)^{'}.
\end{equation}

We note that $\alpha^{\left(1\right)} + \alpha^{\left(2\right)} = 1$ and $\left\langle \pmb{\phi}^{\left(j\right)^{'}}\right\rangle^{(j)} = 0$.
Rewriting Eq. (\ref{chu_present}), for phase-1 ($j = 1$), with the above considerations, we have,

\begin{equation*}
\frac{\partial}{\partial t}\left(\alpha^{\left(1\right)}\left\langle \boldsymbol{q}^{\left(1\right)}\right\rangle^{(1)} \right)+\nabla\cdot\alpha^{\left(1\right)}\left\langle \boldsymbol{q}^{\left(1\right)}\mathbf{u}^{\left(1\right)}\right\rangle^{(1)} =\alpha^{\left(1\right)}\left(\nabla\cdot\overline{\pmb{\phi}}\right)-\frac{1}{V}\oint_{s^{j}+s^{i}}\left(\pmb{\phi}^{\left(2\right)}\right)^{'}\cdot\mathbf{n}^{\left(2\right)}ds
\end{equation*}

\begin{equation}
+\frac{1}{V}\intop_{s^{i}}\left(-\boldsymbol{q}^{\left(1\right)}\left(\mathbf{u}^{\left(1\right)}-\mathbf{w}^{(i)}\right)+\gamma_{s}\right)\cdot\mathbf{n}^{\left(1\right)}ds^{i}.
\label{eq:Chu_prosperetti_form}
\end{equation}

where the term $\frac{1}{V}\oint_{s^{j}+s^{i}}\left(\pmb\phi^{\left(2\right)}\right)^{'}\cdot\mathbf{n}^{\left(2\right)}ds$ represents the local interaction between the two phases over the averaged volume. We call this the two way coupling term.  In case of the momentum conservation equation this term represents the average force acting on the particle due to the continuous phase, $\mathbf{F}_{p}$, and is equal and opposite to the average force acting on the continuous phase due to the particle:  
\begin{equation}
\frac{1}{V}\oint_{s^{j}+s^{i}}\left(\pmb\phi^{\left(2\right)}\right)^{'}\cdot\mathbf{n}^{\left(2\right)}ds=\frac{1}{V}\sum\mathbf{f}^{\left(d\right)}=\mathbf{F}_{p},
\end{equation}
where $\mathbf{f}_p$ is the force due to individual particles. The surface tension term is ignored (i.e. $\gamma_s = 0$) in the present work as we are dealing with gas-solid suspensions.

Equation governing the mass conservation of both the phases have $\pmb q^{j}=\rho^{j}$ and in the case of non-reacting phases $\pmb\phi^{j}=0$. In the case of constant density of individual phases and in the absence 
of mass transfer across the interface we have,

\begin{equation}
\frac{\partial\alpha^{j}}{\partial t}+\nabla\cdot\alpha^{j}\left\langle \mathbf{u}^{j}\right\rangle^{(j)} =0.
\end{equation}

Putting $j=1,2$ and adding the continuous and dispersed equations we get,
\begin{equation}
\nabla\cdot\left(\alpha^{\left(1\right)}\left\langle \mathbf{u}^{\left(1\right)}\right\rangle^{(1)} +\alpha^{\left(2\right)}\left\langle \mathbf{u}^{\left(2\right)}\right\rangle^{(2)} \right)=0.
\end{equation}

Defining $\overline{\mathbf{u}}=\alpha^{\left(1\right)}\left\langle \mathbf{u}^{\left(1\right)}\right\rangle^{(1)} +\alpha^{\left(2\right)}\left\langle \mathbf{u}^{\left(2\right)}\right\rangle^{(2)} $, we can write,
\begin{equation}
\nabla\cdot\overline{\mathbf{u}}=0.
\label{eq:superficial_velocity}
\end{equation}
The velocity field, $\overline{\mathbf{u}}$, is often referred to as superficial velocity. We note that the superficial velocity field is divergence free whereas the individual averaged velocity field is not divergence free.
\newline
For momentum equation we have $\pmb q^{j}=\rho^{j}\mathbf{u}^{j}$ and $\pmb\phi^{j}=\pmb\sigma^{j}$, where $\pmb\sigma^{j}$ is the total fluid stress. The momentum conservation equation for the continuous phase ($j=1$) is thus given by,
\begin{equation}
\frac{\partial}{\partial t}\left(\alpha^{\left(1\right)}\rho^{\left(1\right)}\left\langle \mathbf{u}^{1}\right\rangle^{(1)} \right)+\nabla\cdot\alpha^{\left(1\right)}\rho^{\left(1\right)}\left\langle \mathbf{u}^{\left(1\right)}\mathbf{u}^{\left(1\right)}\right\rangle^{(1)} =\alpha^{\left(1\right)}\left(\nabla\cdot\overline{\pmb\sigma}\right)-\mathbf{F}_{p}.
\label{fluid_mom_vec}
\end{equation}

The expression for $\mathbf{F}_{p}$ is derived by summing the Stokes drag over $N$ number of particles averaged over the averaging volume. Let $\mathbf{v}^{i}$ be the velocity of the $i^{th}$ particle of radius $a$, $\mu$ and $\mathbf{u}$ are the viscosity and velocity of the continuous phase, respectively,
\begin{equation}
 \sum_{i}^{N}\mathbf{f}^{d}	=\sum_{i}^{N}m^{i}\frac{d\mathbf{v}^{i}}{dt}=6\pi\mu a\sum_{i}^{N}\left(\mathbf{u}-\mathbf{v}^{i}\right).
\end{equation}

%\[
%\mathbf{F}_{p}=\frac{1}{V}\sum_{i}^{N}\mathbf{f}^{\left(d\right)}=\frac{1}{V}\sum_{i}^{N}\frac{\rho_{s}V_{i}N}{N}\frac{d\mathbf{v}^{i}}{dt}=\frac{6\pi\mu aN}{VN}\sum_{i}^{N}\left(\mathbf{u}-\mathbf{v}^{i}\right),
%\mbox{		Multiply and dividing by N on both sides, we have,}\]
Thus, the net force $\mathbf{F}_{p}$ can be written as,

%\[
%\frac{d}{dt}\left(\rho_{s}\alpha\frac{\sum_{i}^{N}\mathbf{v}_{i}}{N}\right)=\frac{d}{dt}\left(\rho_{s}\alpha\mathbf{v}_{avg}\right)=\frac{6\pi\mu a}{N}\frac{\alpha}{V_{i}}\sum_{i}^{N}\left(\mathbf{u}-\mathbf{v}^{i}\right)
%,\]
%
%\[
%\mathbf{F}_{p}=\zeta\alpha\left(\mathbf{u}-\frac{1}{N}\sum_{i}^{N}\mathbf{v}^{i}\right),
%\]

\begin{equation}
\mathbf{F}_{p}=\zeta\alpha\left(\mathbf{u}-\mathbf{u}_{p}\right).
\end{equation}

where $\zeta=\frac{6\pi\mu a}{V_{i}}$ and $\left\langle \boldsymbol{u}^{\left(2\right)}\right\rangle =\boldsymbol{u}_{p}=\frac{1}{N}\sum_{i}^{N}\mathbf{v}^{i}$. Here $V_{i}$ is volume of the i$^{th}$-particle.

The first term in the R.H.S of Eq. (\ref{fluid_mom_vec}) is $\nabla\cdot\overline{\pmb\sigma}$ which
is the divergence of the mean stress:

\begin{equation}
\overline{\sigma}_{ik} =-\overline{p}\delta_{ik}+2 \mu \overline{e}_{ik},
\end{equation}
where $\overline{e}_{ik} = (\frac{\partial\overline{u}_{i}}{\partial x_{k}} +  \frac{\partial\overline{u}_{k}}{\partial x_{i}})/2$.

%The divergence of the mean stress is given by,
%\begin{equation}
%\nabla\cdot\overline{\sigma}=\frac{\partial\overline{\sigma}_{ik}}{\partial x_{i}}=-\frac{\partial\overline{p}}{\partial x_{i}}+2\mu\frac{\partial\overline{e}_{ik}}{\partial x_{k}}=-\frac{\partial\overline{p}}{\partial x_{i}}+2\mu\frac{\partial}{\partial x_{k}}\left(\frac{\partial\overline{u}_{k}}{\partial x_{i}}+\frac{\partial\overline{u}_{i}}{\partial x_{k}}\right)=-\frac{\partial\overline{p}}{\partial x_{i}}+2\mu\frac{\partial}{\partial x_{k}}\left(\frac{\partial\overline{u}_{i}}{\partial x_{k}}\right).
%\end{equation}

Chu and Prosperetti \cite{CHU2016176} recommend that $\mu$ should be the effective viscosity, but we argue that $\mu$ should be the fluid viscosity and not the effective viscosity of the suspension. Considering effective viscosity
of the suspension would mean to double count the effect of particles
that has already taken into account through the force term $\mathbf{F}_{p}$
in the volume averaged fluid momentum equation. We note that the force term was not added in an \textit{ad-hoc} manner, it naturally comes out as part of the volume averaging procedure. We assume same average pressure to be acting on both the phases. This results in average pressure  $\overline{p}=p$.  Now the term $\left\langle \mathbf{u}^{j}\mathbf{u}^{j}\right\rangle $ in the L.H.S of Eq.  (\ref{fluid_mom_vec}) can be written as,
%\end{itemize}
\begin{equation}
\left\langle \mathbf{u}^{j}\mathbf{u}^{j}\right\rangle =\left\langle \mathbf{u}^{\left(j\right)}\right\rangle \left\langle \mathbf{u}^{\left(j\right)}\right\rangle +\underset{\mbox{closure term}}{\underbrace{\left\langle \left(\mathbf{u}^{j}-\left\langle \mathbf{u}^{\left(j\right)}\right\rangle \right)\left(\mathbf{u}^{j}-\left\langle \mathbf{u}^{\left(j\right)}\right\rangle \right)\right\rangle}.}
\label{eq:closure_term}
\end{equation}

For turbulent flows,  mixing length models have been proposed but Dimas and Kiger \cite{Dimas_Kiger} used ensemble averaged equation derived by Zhang and Prosperetti \cite{CHU2016176}. They argued that the closure term  can be ignored for linear stability of mixing layer flows of dilute suspensions, when the base state is steady and laminar,  and all the particles initially have the same velocity. However in the context of volume averaging for dilute suspensions with low particle Reynolds number (based on relative velocity), the closure term can be ignored by arguing that there is fore-aft flow symmetry and assuming that the distribution of particles is random, the cross correlation terms can be assumed to be small.

%\newline
Thus, the momentum equation for the continuous phase in the dimensional form is given by, 
\begin{equation}
\frac{\partial}{\partial t^{*}}\left(\alpha^{\left(1\right)}\rho^{\left(1\right)}\left\langle \mathbf{u}^{\left(1\right)*}\right\rangle \right)+\nabla^{*}\cdot\left(\alpha^{\left(1\right)}\rho^{\left(1\right)}\left\langle \mathbf{u}^{\left(1\right)*}\right\rangle \left\langle \mathbf{u}^{\left(1\right)*}\right\rangle\right) =\alpha^{\left(1\right)}\left(\nabla\cdot\overline{\pmb\sigma}^{*}\right)-\mathbf{F}_{p}^{*}.
\label{eq_averaged_fluid_momentum}
\end{equation}

Particle momentum equation is given by,
\begin{equation}
\frac{\partial}{\partial t^{*}}\left(\alpha^{\left(2\right)}\rho^{\left(2\right)}\left\langle \mathbf{u}^{\left(2\right)*}\right\rangle \right)+\nabla^{*}\cdot\left(\alpha^{\left(2\right)}\rho^{\left(2\right)}\left\langle \mathbf{u}^{\left(2\right)*}\right\rangle \left\langle \mathbf{u}^{\left(2\right)*}\right\rangle\right) =\alpha^{\left(2\right)}\left(\nabla\cdot\overline{\pmb\sigma}^{*}\right)+\mathbf{F}_{p}^{*}.
\label{eq:averaged_particle_momentum}
\end{equation}

We non dimensionalize the above equation as,

\begin{equation}
\begin{aligned}
\mathbf{u}=\frac{\mathbf{u}^{*}}{U_{ref}}, \\
\mathbf{x}=\frac{\mathbf{x}^{*}}{L}, \\
p=\frac{p^{*}}{\rho^{\left(1\right)}U_{ref}^{2}}, \\
\nabla=\frac{\nabla^{*}}{L}, \\
\end{aligned}
\label{non_dimension}
\end{equation}
where $L$ is the reference length scale and $U_{ref}$ is the reference velocity. For planar jets, we choose the centreline jet velocity as the reference velocity and shear layer thickness as the reference length scale. For particle mixing layers, the reference velocity $U_{ref}$ is the mean velocity of the two streams and reference length $L$ is the shear layer thickness.
%As part of validation given in section \ref{sec:Numerics}, for particle laden mixing layers, the reference velocity $U_{ref}$ is the mean velocity of the two streams, reference length $L$ is the half vorticity thickness. 
%
We shall denote $\left\langle \boldsymbol{u}^{\left(1\right)}\right\rangle =\boldsymbol{u}$ and $\left\langle \boldsymbol{u}^{\left(2\right)}\right\rangle =\boldsymbol{u}_{p}$ henceforth.

% and   substituting Eq. (\ref{non_dimension}) into Eq. (\ref{continuous_phase_mom}) gives, 
%\[
%\frac{\partial}{\partial t}\left(\left(1-\alpha\right)u_{i}\right)+\frac{\partial}{\partial x_{k}}\left(\left(1-\alpha\right)u_{i}u_{k}\right)=\left(\frac{-\rho_{f}U_{ref}^{2}}{L\rho_{f}}\right)\frac{L}{U_{ref}^{2}}\left(1-\alpha\right)\frac{\partial p}{\partial x_{i}}+
%\]
%
%\begin{equation}
%\left(1-\alpha\right)\left(\frac{\mu U_{ref}}{L\rho_{f}}\right)\left(\frac{L}{U_{ref}^{2}}\right)\frac{\partial^{2}}{\partial x_{k}^{2}}\left(\left(1-\alpha\right)u_{i}\right)+\frac{\left(1-\alpha\right)\mu}{\rho_{f}L}\left(\frac{L}{U_{ref}^{2}}\right)\frac{\partial^{2}}{\partial x_{k}^{2}}\left(\alpha u_{pi}\right)-\left(\frac{\zeta U_{ref}}{\rho_{f}}\frac{L}{U_{ref}^{2}}\right)\alpha\left(u_{i}-u_{pi}\right).
%\end{equation}

Non dimensionalising  Eq.(\ref{eq_averaged_fluid_momentum}) with the above reference length, velocity and pressure scales, we obtain the fluid momentum equation as,
%\begin{itemize}
%The non dimensional term $\frac{\zeta U_{ref}}{\rho_{f}}\frac{L}{U_{ref}^{2}}$
%can be written as
%%\end{itemize}
%\[
%\left(\frac{\zeta U_{ref}}{\rho_{f}}\frac{L}{U_{ref}^{2}}\right)=\frac{\frac{\mu}{\frac{2}{9}}\frac{1}{a^{2}}L}{\rho_{f}U_{ref}}=\left(\frac{\mu}{\frac{2}{9}}\frac{1}{a^{2}\rho_{s}}\right)\left(\frac{L}{U_{ref}}\right)\gamma=\frac{t_{c}}{\tau_{p}}\gamma=\frac{\gamma}{St} .
%\]
%where Stokes number ($St$) defined as the ratio of relaxation time $\left(\tau=\frac{2}{9}\frac{a^2\rho_{s}}{\mu}\right)$ to the flow convective time scale $\left(t_{c}=\frac{L}{U_{0}}\right)$.  Stokes number $St = \frac{\tau}{t_{c}} = \frac{2a^{2}\rho_{s}}{9\mu}$.
\begin{equation}
\frac{\partial}{\partial t}\left(\left(1-\alpha\right)u_{i}\right)+\frac{\partial}{\partial x_{k}}\left(\left(1-\alpha\right)u_{i}u_{k}\right)=-\left(1-\alpha\right)\frac{\partial p}{\partial x_{i}}+\frac{\left(1-\alpha\right)}{Re}\frac{\partial^{2}}{\partial x_{k}^{2}}\left(\left(1-\alpha\right)u_{i}\right)-\frac{\gamma}{St}\alpha\left(u_{i}-u_{pi}\right),
\label{fluid_mom}
\end{equation}
where the index $i$ denotes the $i^{th}$ component of the velocity vector.

%
%Non dimensionlising and expressing in indicial notation, 
%
%\[
%\frac{\partial}{\partial t}\left(\alpha u_{pi}\right)+\frac{\partial}{\partial x_{k}}\left(\alpha u_{pi}u_{pk}\right)=\underset{\ll1}{\underbrace{\left(\frac{-\rho_{f}U_{ref}^{2}}{\rho_{s}V_{ref}^{2}}\right)}}\alpha\frac{\partial p}{\partial x_{i}}+
%\frac{\zeta}{\rho_{s}}\alpha\left(u_{i}-u_{pi}\right).
%\]
%
%%\begin{itemize}
%The term $\frac{\rho_{f}U_{ref}^{2}}{\rho_{s}V_{ref}^{2}}$ is negligible
%as typically in gas - solid particle laden flows the density ratio
%$\gamma=\frac{\rho_{s}}{\rho_{f}}\sim1000$.

Non dimensionalised particle momentum equation, Eq. (\ref{eq:averaged_particle_momentum}), is given by

\begin{equation}
\frac{\partial}{\partial t}\left(\alpha u_{pi}\right)+\frac{\partial}{\partial x_{k}}\left(\alpha u_{pi}u_{pk}\right)=\frac{1}{St}\alpha\left(u_{i}-u_{pi}\right).
\label{particle_mom}
\end{equation}

where Stokes number $St = \frac{\tau}{t_{c}} = \frac{2a^{2}\rho_{s}}{9\mu}$, $\gamma$ is the ratio of density of particle to the gas $\left(\frac{\rho_{s}}{\rho_{f}}\right)$ and $Re$ is the fluid Reynolds number based on the convective length scale of the flow.
We note that the placement of the volume fraction term in the pressure and viscous terms in the volume averaged equation has been debated in previous studies (see \cite{Narayananetal2002,JANET2015106,Harlow1975,prosperetti2009,PATIL2015388,CHU2016176}). In the pressure term of the momentum equation, the volume fraction term must be outside the gradient operator in order to avoid unphysical results. This was first noted by Harlow and Amsden \cite{Harlow1975} and later by Prosperetti and Tryggvason \cite{prosperetti2009}. Harlow and Amsden \cite{Harlow1975} considered the case of a two phase system at static equilibrium in the absence of gravity. If the volume fraction is inside the gradient operator then for the case of static equilibrium we have,
\begin{equation}
\nabla\cdot\left(\alpha\left\langle \pmb{\sigma}\right\rangle \right)=\alpha\underset{=0}{\underbrace{\left(\nabla\cdot\left\langle \pmb{\sigma}\right\rangle \right)}}+\left\langle \pmb{\sigma}\right\rangle \cdot\nabla\alpha.
\label{eq:Harlow_amsden}
\end{equation}
This term would result in unphysical motion of particles due to spatial gradients in volume concentration which should not occur when the system is in static equilibrium. Further Harlow and Amsden \cite{Harlow1975} identified that for the case of motion of a wave front into a region of mixed vapor and droplets, the growth rates of the disturbance had considerable effect due to the pressure
gradient term being $\alpha\nabla p$ instead of $\nabla\left(\alpha p\right)$
%\item \citeauthor{doi:10.1063/1.869769} (\citeyear{doi:10.1063/1.869769}) performed inviscid local analysis of dilute particle laden mixing layers. Their formulation considered,
%\end{itemize}
Later Dimas and Kriger \cite{Dimas_Kiger} performed inviscid local analysis of dilute particle laden mixing layers. Their formulation considered,
\begin{equation}
\left(1-\alpha\right)\rho_{f}\left\{ \frac{\partial u_{i}}{\partial t}+\frac{\partial}{\partial x_{k}}\left(u_{i}u_{k}\right)\right\} =-\left(1-\alpha\right)\frac{\partial p}{\partial x_{i}}-F_{pi},
\label{eq:Dimas_kriger1}
\end{equation}

which was then simplified as, 
\begin{equation}
\rho_{f}\left\{ \frac{\partial u_{i}}{\partial t}+\frac{\partial}{\partial x_{k}}\left(u_{i}u_{k}\right)\right\} =-\frac{\partial p}{\partial x_{i}}-\frac{\gamma}{St}\alpha\left(u_{i}-u_{pi}\right),
\label{eq:Dimas_kriger2}
\end{equation}

for small volume fractions. They have taken $\alpha$ as a constant in the Eq.\ref{eq:Dimas_kriger1} which can lead to issues as pointed by Jackson \cite{jackson2000dynamics}. Jackson \cite{jackson2000dynamics} considered the case of bubbles rising in a fluidized bed as air is passed from underneath a fluidized bed of particles and pointed out certain irregularities in the pressure field as a result of assuming $\alpha$ to be constant.
However in the case of linear stability analysis of dilute suspensions, this
might not make a considerable difference as far as the pressure gradient
term is concerned. However, the viscous term should contain the volume fraction term $\alpha$, inside and outside the divergence as discussed in Chu and Prosperetti \cite{CHU2016176}. The complete equation is given by,
\begin{equation}
\frac{\partial}{\partial t}\left(\left(1-\alpha\right)u_{i}\right)+\frac{\partial}{\partial x_{k}}\left(\left(1-\alpha\right)u_{i}u_{k}\right)=-\left(1-\alpha\right)\frac{\partial p}{\partial x_{i}}+\frac{\left(1-\alpha\right)}{Re}\frac{\partial^{2}}{\partial x_{k}^{2}}\left(\left(1-\alpha\right)u_{i}\right)-\frac{\gamma}{St}\alpha\left(u_{i}-u_{pi}\right),
\end{equation}

%where $\alpha\left(x,y,t\right)=\Lambda\left(y\right)+\epsilon\alpha^{'}\left(x,y,t\right)$.
 Narayanan et al. \cite{Narayananetal2002}  following Dimas and Kriger \cite{Dimas_Kiger} discussed the instability of particle laden mixing layers.  Janet et al. \cite{JANET2015106} and Patil et al. \cite{PATIL2015388} considered the fluid momentum equation as,
\begin{equation}
\frac{\partial}{\partial t}\left(\left(1-\alpha\right)u_{i}\right)+\frac{\partial}{\partial x_{k}}\left(\left(1-\alpha\right)u_{i}u_{k}\right)=-\left(1-\alpha\right)\frac{\partial p}{\partial x_{i}}+\frac{\partial}{\partial x_{k}}\left(\left(1-\alpha\right)\tau_{ik}\right)-F_{pi}.
\label{eq:Janet_Patil}
\end{equation}
Although this form agrees with the R.H.S of our equation (without modeling the closure terms),
\begin{equation}
\frac{\partial}{\partial t}\left(\left(1-\alpha\right)\left\langle u_{i}\right\rangle \right)+\frac{\partial}{\partial x_{k}}\left(\left(1-\alpha\right)\left\langle u_{i}u_{k}\right\rangle \right)=-\left(1-\alpha\right)\frac{\partial p}{\partial x_{i}}+\frac{\partial}{\partial x_{k}}\left(\left(1-\alpha\right)\left\langle \tau_{ik}\right\rangle \right)+\intop_{s^{i}}\phi_{ik}n_{k}ds^{i},
\end{equation}
the treatment of the terms $\frac{\partial}{\partial x_{k}}\left(\left(1-\alpha\right)\left\langle \tau_{ik}\right\rangle \right)$
as well as the interface integral $\intop_{s^{i}}\phi_{ik}n_{k}ds^{i}$
is not clear in their formulation. In the present work, we follow Chu and Prosperetti's \cite{CHU2016176} formulation where the two terms are combined by defining a mean shear stress for the suspension (see Eq. \ref{eq:Chu_prosperetti_form}). 

Before we proceed further with the stability analysis of planar jet, it is pertinent to discuss the migration of particles in the neighborhood of high shear regions in order to justify the zero inertia base state considered in the present study.  Martin and Meiberg \cite{martin1994accumulation} and Meiberg et al. \cite{meiburg2000vorticity} studied particle laden mixing layers numerically and found that particles strongly accumulate in the braid region of the mixing layer. They modelled the braid region as a stagnation point flow. Here, we briefly look at the arguments given by Martin and Meiberg \cite{martin1994accumulation}. We consider a stagnation point flow field given by,
\begin{equation}
\begin{aligned}
u=ax, \\
v=-ay,
\end{aligned}
\end{equation}
where $a > 0$ is the strain rate. Considering one-way coupling the governing equations for particle motion can be written as \cite{martin1994accumulation},
\begin{equation}
\begin{aligned}
\frac{du_{p}}{dt}=\frac{1}{St}\left(u-u_{p}\right), \\
\frac{dv_{p}}{dt}=\frac{1}{St}\left(v-v_{p}\right), \\
\frac{dx_{p}}{dt} = u_{p}, \\
\frac{dy_{p}}{dt} = v_{p}.
\label{MR equation}
\end{aligned}
\end{equation}  

This results in two uncoupled second order linear ODE's whose solution yields us the particle path lines,
%\begin{equation}
%\begin{aligned}
%\ddot{x}_{p}+\frac{1}{St}\dot{x}_{p}-\frac{a}{St}x_{p}=0, \\
%\ddot{y}_{p}+\frac{1}{St}\dot{y}_{p}+\frac{a}{St}y_{p}=0.
%\end{aligned}
%\end{equation}
%The solution to the above equations gives the particle path lines as, 
\begin{equation}
\begin{aligned}
x_{p}= c_{1}\exp{\lambda_{1}t}+c_{2}\exp{\lambda_{2}t}, \\
y_{p}= c_{3}\exp{\lambda_{3}t}+c_{4}\exp{\lambda_{4}t}, \\
\end{aligned}
\end{equation}
where,
\begin{equation}
\begin{aligned}
\lambda_{1}=\frac{-1+\sqrt{\left(1+4aSt\right)}}{2St}, \\
\lambda_{2}=\frac{-1-\sqrt{\left(1+4aSt\right)}}{2St}, \\
\lambda_{3}=\frac{-1+\sqrt{\left(1-4aSt\right)}}{2St}, \\
\lambda_{4}=\frac{-1-\sqrt{\left(1-4aSt\right)}}{2St},
\end{aligned}
\end{equation} 
while $c_{1}$ to $c_{4}$ are constants. We see for $St>1/4a$, that the roots are complex indicating that the particle path lines oscillates and overshoots the horizontal streamline resulting in intense accumulation of particles in the braid region which has been observed by Meiberg et al. \cite{meiburg2000vorticity}.
Depending on the value of strain rate ($a$), the upper limit of the single valuedness of the particle velocity field can be ascertained. 
%This leads us to the question of how do particles in a planar jet migrate and do particle path lines converge at any point in the flow?
Performing a similar analysis for the locally parallel flow assumption of the base flow, we show that the particles do not migrate regardless of the Stokes number, on its own, but may migrate only due to the fact that the locally parallel planar jet base flow is unstable to infinitesimal perturbations which generates perturbation velocities that guide the particles towards the jet shear layer regions.

The base state for the present study is of a laminar, steady state, locally parallel planar jet given by Tatsumi and Kakutani \cite{tatsumi_kakutani_1958} as,
%\begin{equation}
%\begin{aligned}
%u=U_{0}sech^{2}\left(\frac{y}{L}\right), \\
%v=\left(\frac{3}{4}Re_{x}\right)^{-2/3}U_{0}\left\{ \frac{2y}{L}sech^{2}\left(\frac{y}{L}\right)-tanh\left(\frac{y}{L}\right)\right\}, \\
%\end{aligned}
%\end{equation}
%where $u$ and $v$ are the jet streamwise and cross streamwise velocities respectively and the constants ,
%
%\begin{equation}
%\begin{aligned}
%U_{0}=\left(\frac{3M^{2}}{32\nu x}\right)^{1/3}, \\
%L=\left(\frac{M}{48\nu^{2}x^{2}}\right)^{-1/3}, \\
%\mbox{Local Reynolds number, } 	Re_{x}=\left(\frac{Mx}{\nu^{2}}\right)^{1/2},  \\
%\mbox{Kinematic momentum flux, } 	M=\int_{-\infty}^{\infty}u^{2}dy=\mbox{constant}. \\
%\end{aligned}
%\end{equation}
 
%We define a global Reynolds number by choosing characteristic length scale $L$ as the vortcity half thickness and velocity scale of the flow $U_{0}$, 
%\begin{equation}
%Re=\frac{U_{0}L}{\nu}=\left(\frac{9}{2}\right)^{1/3}\left(Re_{x}\right)^{2/3}.
%\end{equation}
%By taking the ratio of the velocities, we see in terms of the global Reynolds number that,
%\begin{equation}
%\frac{v}{u}=O\left(Re_{x}^{-2/3}\right)=O\left(Re^{-1}\right).
%\end{equation}
%In this limit, the vertical velocity is negligible compared to axial velocity and the variation of axial velocity with downstream location is negligible compared to the axial velocity. This justifies the locally parallel baseflow assumption resulting in the baseflow profile,
\begin{equation}
\begin{aligned}
U_{x}= U_{x}(y) = sech^{2}y, \\
U_{y} = 0. \\
\end{aligned}
\label{planar_jet_baseflow}
\end{equation}
where $U_{x}$ is the jet streamwise velocity, $U_{y}$ is the jet cross streamwise velocity which is zero for a locally parallel flow. The cross streamwise direction $y$ is non dimensionalized by a transverse length scale given by  $\frac{Q}{2U_{c}}$ where $U_{c}$ is the centreline velocity of the jet and $Q=\int_{-\infty}^{\infty}udy$, is the volume flux per unit width of the jet. Reference velocity scale, $U_{ref}$, is the jet centreline velocity $U_{c}$.
%Following Matsubara et al, \cite{matsubara_alfredsson_segalini_2020} 
%In the introduction section we discussed predicting particle migration in the case of mixing layer where the braid region is modelled as a stagnantion point flow. We saw from an Eulerian Lagrangian perspective, that for a stagnation point flow, there exists a critical Stokes number beyond which the particle path lines converge and lead to regions of intense particle accumulation. Now in the present case of 
 For planar jet, we assumed (following Saffman \cite{saffman_1962}) that the baseflow particle velocity field is equal to the baseflow fluid velocity. This implies that there cannot be any particle accumulation as far as a steady state baseflow of a planar jet is concerned. To illustrate this, let's suppose that the base state particle velocity is not equal to the base state fluid velocity. %(for instance, the assumption of unequal particle and fluid base state velocities can be found in Despirito et al, \cite{DESPIRITO20011179}). 
% For a locally parallel planar jet, we have,
%\begin{equation}
%\begin{aligned}
%u=sech^{2}y, \\
%v=0,
%\end{aligned}
%\end{equation}
Following  Martin and Meiberg \cite{martin1994accumulation}, we substitute Eq. (\ref{planar_jet_baseflow}) in Eq. (\ref{MR equation}) with the initial condition $\dot{y}_{p}=0$. (i.e. $y$-component of the particle velocity is zero), we obtain the particle path lines as,
%\begin{equation}
%\begin{aligned}
%u_{p}=\dot{x}_{p}=\frac{dx_{p}}{dt}, \\
%v_{p}=\dot{y}_{p}=\frac{dy_{p}}{dt}, \\
%\frac{du_{p}}{dt}=\frac{1}{St}\left(u-u_{p}\right)=\frac{1}{St}\left(sech^{2}y_{p}-u_{p}\right), \\
%\frac{dv_{p}}{dt}=\frac{1}{St}\left(v-v_{p}\right)=\frac{-1}{St}\left(v_{p}\right). 
%\end{aligned}
%\end{equation}  

%\begin{equation}
%\begin{aligned}
%\ddot{x}_{p}+\frac{1}{St}\dot{x}_{p}=\frac{1}{St}sech^{2}y_{p}, \\
%\ddot{y}_{p}+\frac{1}{St}\dot{y}_{p}=0.
%\end{aligned}
%\end{equation}

%\begin{equation}
%\dot{y}_{p} = Ae^{-t/St},
%\end{equation}
%and,
%\begin{equation}
%y_{p}=(-A.St)e^{-t/St}+B,
%\end{equation}
%%Initial condition at $t=0$, $\dot{y}_{p}=0$. (i,e. vertical component of the particle velocity is zero). This implies $A=0$. The solution to the above equations gives the particle path lines as, 
%Initial condition at $t=0$,  This implies $A=0$.
\begin{equation}
\begin{aligned}
x_{p}=D~St~t+ E~St~e^{-t/St} + F, \\
y_{p}=B=\mbox{Constant.}
\end{aligned}
\end{equation}
where $B$, $D$, $E$ and $F$ are constants. 
For small times $t/St<<1$, the $x$ coordinate of the particle path line, $x_{p} \sim St$ (by arbitrarily setting $F = 0$; without loss of generality).  For large times, $t/St\rightarrow\infty$, the $x-$component of the particle location varies as $x_{p}\sim t~St$,  while $y_{p}$ remains a constant for all times. This shows that for small times (time period during which linear analysis is valid), particle path lines in a locally parallel planar jet do not cross each other and hence do not result in any accumulation of particles for any Stokes number (unlike in the case of mixing layer flows which had particle path lines overshoot and oscillate about the horizontal streamline). Thus, the assumption of the zero inertia particle laden planar jet base state considered in this study, Eq.\ref{planar_jet_baseflow}, is justified. 

\section{Perturbation equations for particle laden planar flows (linearised volume averaged equations)}
\label{sec: LVAE}
In order to perform local analysis, we linearize the governing equation (volume averaged Navier Stokes equation) around a base state (baseflow) that is assumed to be steady and locally parallel, Eq. \ref{planar_jet_baseflow}. Linearized equations are obtained by writing the system of equations as a perturbation series for small amplitudes and considering the $O(\epsilon)$ equations:
\begin{equation}
\label{lsa}
\begin{aligned}
u_{x}\left(x,y,t\right)={U}_{x}\left(y\right)+\epsilon u_{x}^{'}\left(x,y,t\right)+O(\epsilon^{2}),  \\
u_{y}\left(x,y,t\right)=\epsilon u_{y}^{'}\left(x,y,t\right)+O(\epsilon^{2}),  \\
p\left(x,y,t\right)=P+\epsilon p^{'}\left(x,y,t\right)+O(\epsilon^{2}), \\   
u_{px}\left(x,y,t\right)=U_{x}\left(y\right)+\epsilon u_{px}^{'}\left(x,y,t\right)+O(\epsilon^{2}),  \\
u_{py}\left(x,y,t\right)=\epsilon u_{py}^{'}\left(x,y,t\right)+O(\epsilon^{2}),  \\
\alpha\left(x,y,t\right)=\Lambda\left(y\right)+\epsilon \alpha{'}\left(x,y,t\right)+O(\epsilon^{2}),  \\
\end{aligned}
\end{equation}
 where $\Lambda$ is the baseflow concentration field and all the primed quantities indicate the respective perturbed fields. 
\newline
Continuity equation for the carrier fluid is given by, 
\begin{equation}
\frac{\partial\alpha^{'}}{\partial t}-\left(1-\Lambda\right)\left(\frac{\partial u_{x}^{'}}{\partial x}+\frac{\partial u_{y}^{'}}{\partial y}\right)+U_{x}\frac{\partial\alpha^{'}}{\partial x}+u_{y}^{'}\frac{d\Lambda}{dy}=0.
\end{equation}
%\newpage
Continuous phase momentum equation in the $x$ direction is given by,
\[
\left(1-\Lambda\right)\left(\frac{\partial u_{x}^{'}}{\partial t}\right)-U_{x}\left(\frac{\partial\alpha^{'}}{\partial t}\right)+\left(1-\Lambda\right)U_{x}\frac{\partial u_{x}^{'}}{\partial x}-U_{x}^{2}\frac{\partial\alpha^{'}}{\partial x}+\left(1-\Lambda\right)\frac{dU_{x}}{dy}u_{y}^{'}
\]
\[
-U_{x}\left\{ \frac{d\Lambda}{dy}\left(u_{py}^{'}-u_{y}^{'}\right)+\Lambda\left(\frac{\partial u_{px}^{'}}{\partial x}+\frac{\partial u_{py}^{'}}{\partial y}\right)\right\}=-\left(1-\Lambda\right)\frac{\partial p^{'}}{\partial x}
\]
\begin{align}
&+\frac{1}{Re}\left\{ \left(1-\Lambda\right)^{2}\left(\frac{\partial^{2}u_{x}^{'}}{\partial x^{2}}+\frac{\partial^{2}u_{x}^{'}}{\partial y^{2}}\right)-\left(1-\Lambda\right)U_{x}\left(\frac{\partial^{2}\alpha^{'}}{\partial x^{2}}+\frac{\partial^{2}\alpha^{'}}{\partial y^{2}}\right)-2\left(1-\Lambda\right)\alpha^{'}\frac{d^{2}U_{x}}{dy^{2}}-2\left(1-\Lambda\right)\frac{\partial\alpha^{'}}{\partial y}\frac{dU_{x}}{dy}\right.\nonumber\\&\qquad\left. {} -2\left(1-\Lambda\right)\frac{d\Lambda}{dy}\frac{\partial u_{x}^{'}}{\partial y}-\left(1-\Lambda\right)u_{x}^{'}\frac{d^{2}\Lambda}{dy^{2}}+2\frac{d\Lambda}{dy}\frac{dU_{x}}{dy}\alpha^{'}+U_{x}\frac{d^{2}\Lambda}{dy^{2}}\alpha^{'}\right\}-\frac{\gamma\Lambda}{St}\left(u_{x}^{'}-u_{px}^{'}\right).
\end{align}

The $y-$ component of the continuous phase equation is given by,
\begin{align}
\left(1-\Lambda\right)\left(\frac{\partial u_{y}^{'}}{\partial t}\right)+\left(1-\Lambda\right)U_{x}\frac{\partial u_{y}^{'}}{\partial x}&=-\left(1-\Lambda\right)\frac{\partial p^{'}}{\partial y}+
\frac{1}{Re}\left\{ \left(1-\Lambda\right)^{2}\left(\frac{\partial^{2}u_{y}^{'}}{\partial x^{2}}+\frac{\partial^{2}u_{y}^{'}}{\partial y^{2}}\right) \right.\nonumber\\
&\qquad \left. {}    -2\left(1-\Lambda\right)\frac{d\Lambda}{dy}\frac{\partial u_{y}^{'}}{\partial y}-\left(1-\Lambda\right)\frac{d^{2}\Lambda}{dy^{2}}u_{y}^{'}\right\} -\frac{\gamma\Lambda}{St}\left(u_{y}^{'}-u_{py}^{'}\right).
\end{align}

Dispersed phase momentum equation in the $x$ direction is given by,
\begin{equation}
\frac{\partial u_{px}^{'}}{\partial t}+U_{x}\frac{\partial u_{px}^{'}}{\partial x}+u_{py}^{'}\frac{dU_{x}}{dy}=\frac{1}{St}\left(u_{x}^{'}-u_{px}^{'}\right).   
\label{eq:LPE_x}
\end{equation}

Dispersed phase momentum equation in the $y$ direction is given by,
\begin{equation}
\frac{\partial u_{py}^{'}}{\partial t}+U_{x}\frac{\partial u_{py}^{'}}{\partial x}=\frac{1}{St}\left(u_{y}^{'}-u_{py}^{'}\right).    
\label{eq:LPE_y}
\end{equation}

Continuity equation for the dispersed phase is given by,
\begin{equation}
\frac{\partial\alpha^{'}}{\partial t}+\Lambda\left(\frac{\partial u_{px}^{'}}{\partial x}+\frac{\partial u_{py}^{'}}{\partial y}\right)+U_{x}\frac{\partial\alpha^{'}}{\partial x}+u_{py}^{'}\frac{d\Lambda}{dy}=0.
\end{equation}

%\subsection{Normal mode equations}
In temporal analysis we look for freely evolving disturbances that are localised in space and growing exponentially in time. Therefore, the normal mode equations can be written as, 
\begin{equation}
\label{normal mode form equations}
\begin{aligned}
u_{x}^{'}\left(x,y,t\right)=\tilde{u}_{x}\left(y\right)e^{i\left(kx-\omega t\right)},  \\
u_{y}^{'}\left(x,y,t\right)=\tilde{u}_{y}\left(y\right)e^{i\left(kx-\omega t\right)},  \\
p^{'}\left(x,y,t\right)=\tilde{p}\left(y\right)e^{i\left(kx-\omega t\right)}, \\   
u_{px}^{'}\left(x,y,t\right)=\tilde{u}_{px}\left(y\right)e^{i\left(kx-\omega t\right)},  \\
u_{py}^{'}\left(x,y,t\right)=\tilde{u}_{py}\left(y\right)e^{i\left(kx-\omega t\right)},  \\
\alpha^{'}\left(x,y,t\right)=\tilde{\alpha}\left(y\right)e^{i\left(kx-\omega t\right)}.  \\
\end{aligned}
\end{equation}
Here $\omega$ is the complex temporal eigenvalue whose real part gives the frequency, the imaginary part gives the temporal growth rate, and $k$ is the wavenumber which is real. 
The above normal mode for perturbations is substituted in the linearised Navier--Stokes equation and can be written in the matrix form as, 
\begin{equation}
A\mathbf{q}=\omega B\mathbf{q}.
\label{normal mode form}
\end{equation}

The entries of the $A$ and $B$ matrices are given in the appendix \ref{appendix_b}.

\section{Numerical method and validation}
\label{sec:Numerics}
%\subsection{Numerical method}
The system of linearised Navier--Stokes equation in the normal form posed as an eigenvalue problem can be written in the matrix form as given in Eq. (\ref{normal mode form})
where $A$ and $B$ are operators which are a function of the base state (Eq.\ref{planar_jet_baseflow}) and the wave number $k$. The temporal eigenvalue is given by $\omega$, and
$\tilde{q}={[{\tilde{u}_{x},\tilde{u}_{y},\tilde{p},\tilde{u}_{px},\tilde{u}_{py},\tilde{\alpha}}]}^{T}$ is the eigenvector.
Far field boundary conditions are imposed, i.e. all perturbations decay to zero at infinity. 

The dispersion relation is numerically solved by using a pseudospectral collocation technique employing Chebyshev polynomials (see Boyd \cite{Boyd_2013} for details). 
Owing to the fact that Chebyshev polynomials are defined in $[-1 , 1]$ and both planar jet and mixing layer flows are unbounded, it is necessary to transform the domain. Consider a transformation from an infinite domain to a domain ${[}-1,1{]}$ given by the function
$f\left(\eta\right)$,

\begin{equation}
y=f\left(\eta\right),
\end{equation}

where $\eta\in\left[-1,1\right]$ is the Chebyshev coordinate and $y$ is the physical coordinate (jet cross streamwise direction). We choose algebraic mapping function,
\begin{equation}
f\left(\eta\right)=\frac{b\eta}{\sqrt{1-\eta^{2}}},
\end{equation}

The parameter $b$ controls the number of points in the shear layer. For the cases presented here it is found that $b = 6$ is sufficient. The derivatives in the physical coordinate ($y$) are transformed in the Chebyshev coordinate ($\eta$) as following,

\begin{equation}
    \begin{aligned}
    \frac{d}{dy}=\frac{1}{f'}\frac{d}{d\eta}, \\
    \frac{d^{2}}{dy^{2}}=\frac{1}{f'^{2}}\frac{d^{2}}{d\eta^{2}}-\left(\frac{f''}{f'^{3}}\right)\frac{d}{d\eta},    
    \end{aligned}
    \label{eq:Chebyshev_differentiation_matrices}
\end{equation}

where $\frac{d}{d\eta}$ and $\frac{d^{2}}{d\eta^{2}}$ are the Chebyshev differentiation matrices.

%\subsection{Validation}
%\label{sec:mixing_layer}
The streamwise velocity profile for a locally parallel base state given by Narayanan et al. \cite{Narayananetal2002} is used for the purpose of validation. If $U_{1}$ and $U_{2}$ are the free stream velocities at $y\rightarrow\infty$ and $y\rightarrow-\infty$,  respectively, the streamwise velocity profile is given by,  
\begin{equation}
U_{x}\left(y\right)=1+\lambda \tanh{y},
\label{baseflow vel}
\end{equation}
where $\lambda=\frac{\Delta U}{2\overline{U}}$, with $\overline{U}=\frac{U_{1}+U_{2}}{2}$ and 
$\Delta U=U_{1}-U_{2}$, is the velocity difference parameter which is taken to be equal to $1$ as given in Narayanan et al. \cite{Narayananetal2002}. The mean profile for the particle volume fraction is assumed to have a form similar to that of the mean velocity profile, following Narayanan et al. \cite{Narayananetal2002},
\begin{equation}
\Lambda\left(y\right)=\overline{\alpha}\left(1+\lambda_{p}\tanh{(l_{p}y)}\right),
\label{baseflow_concentration_mixing_layer}
\end{equation}
where $\overline{\alpha}$ is a constant taken to be $5\times10^{-4}$ so that the concentration is always in the dilute regime. $\lambda_{p}=0$, yields uniform particle concentration, $\lambda_{p}=\pm1$, yield flows loaded with particles only in one of the two streams, and $l_{p}$ is the parameter that controls the steepness of the non uniformity.
\begin{figure*}
	\centering 
	\includegraphics[width=0.8\textwidth]{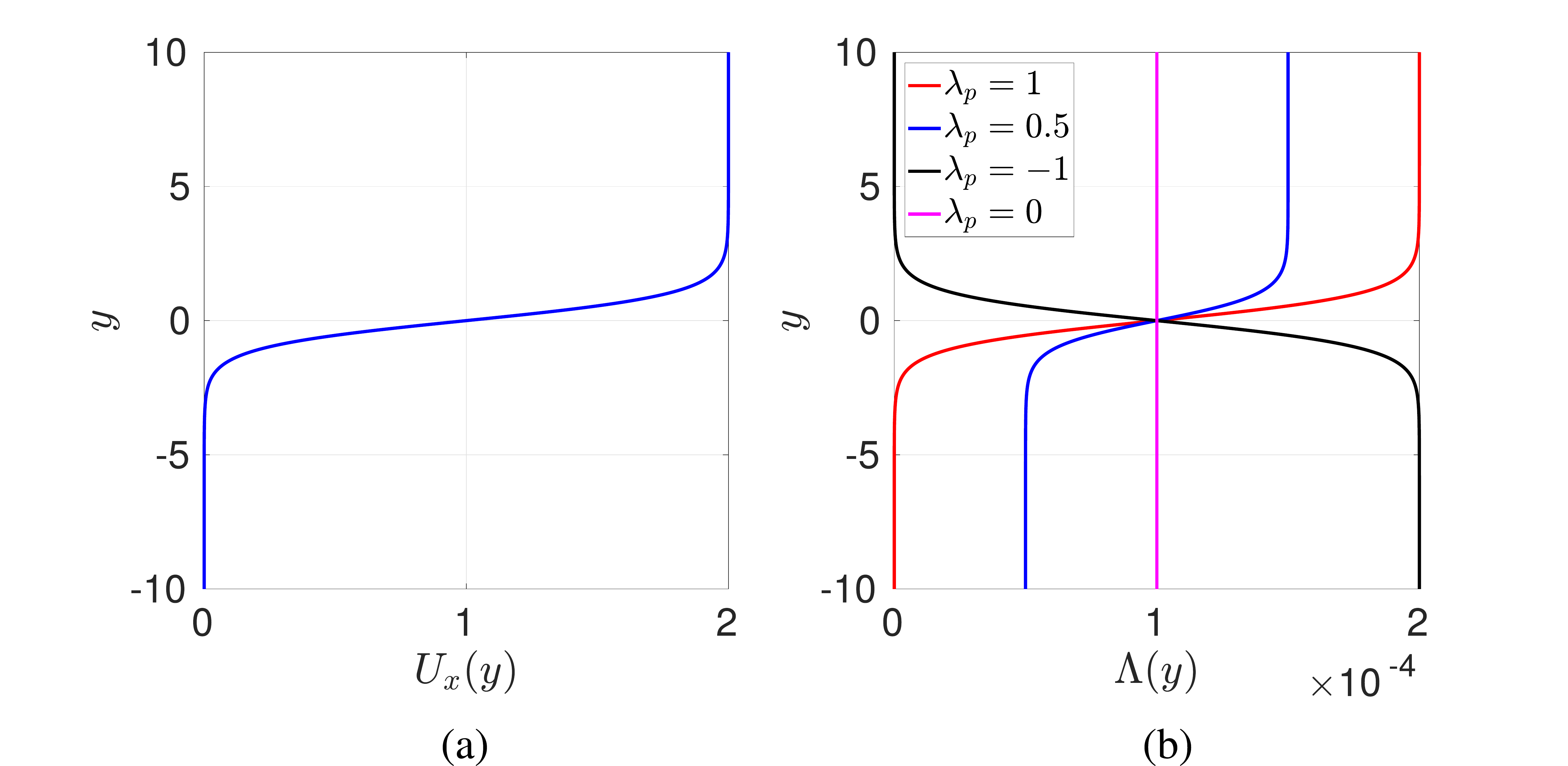}
	\caption{(a) shows base state profile for the streamwise component ($U_{x}=1+\tanh{y}$) as a function of the cross streamwise direction($y$). (b) shows base flow concentration profile as a function of the cross streamwise direction($y$) given by Eq.  (\ref{baseflow_concentration_mixing_layer}). The parameter $\lambda_{p}=0$ yields uniform particle concentration, $\lambda_{p}=\pm1$ yield flows loaded with particles only in one of the two streams.    }
	\label{mixing_layer_base_state}
\end{figure*}

\begin{figure*}
	\centering 
	\includegraphics[width=0.84\textwidth]{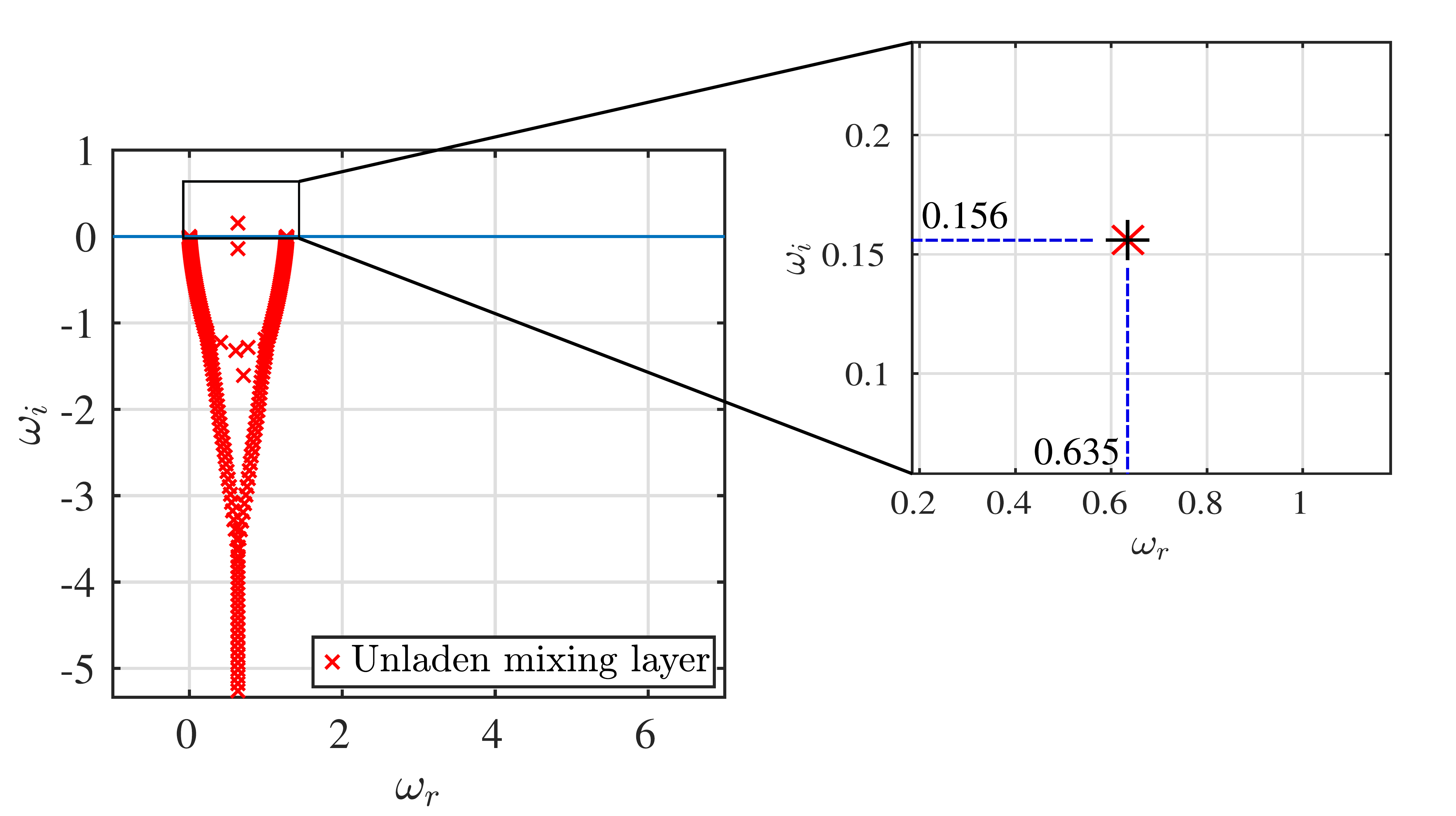}
	\caption{Eigenspectra of unladen mixing layer at $Re=250$, $k=0.635$ marked by $\color{red}\times$. Results from Narayanan et al. \cite{Narayananetal2002}, marked by $\color{black}+$, are superposed on the figure. The unstable eigenmode corresponds to a growth rate $\omega_{i}=0.156$ and frequency $\omega_{r}=0.635$.   }
	\label{unladen_mixing_layer_spectra}
\end{figure*}

\begin{figure*}
	\centering 
	\includegraphics[width=1\textwidth]{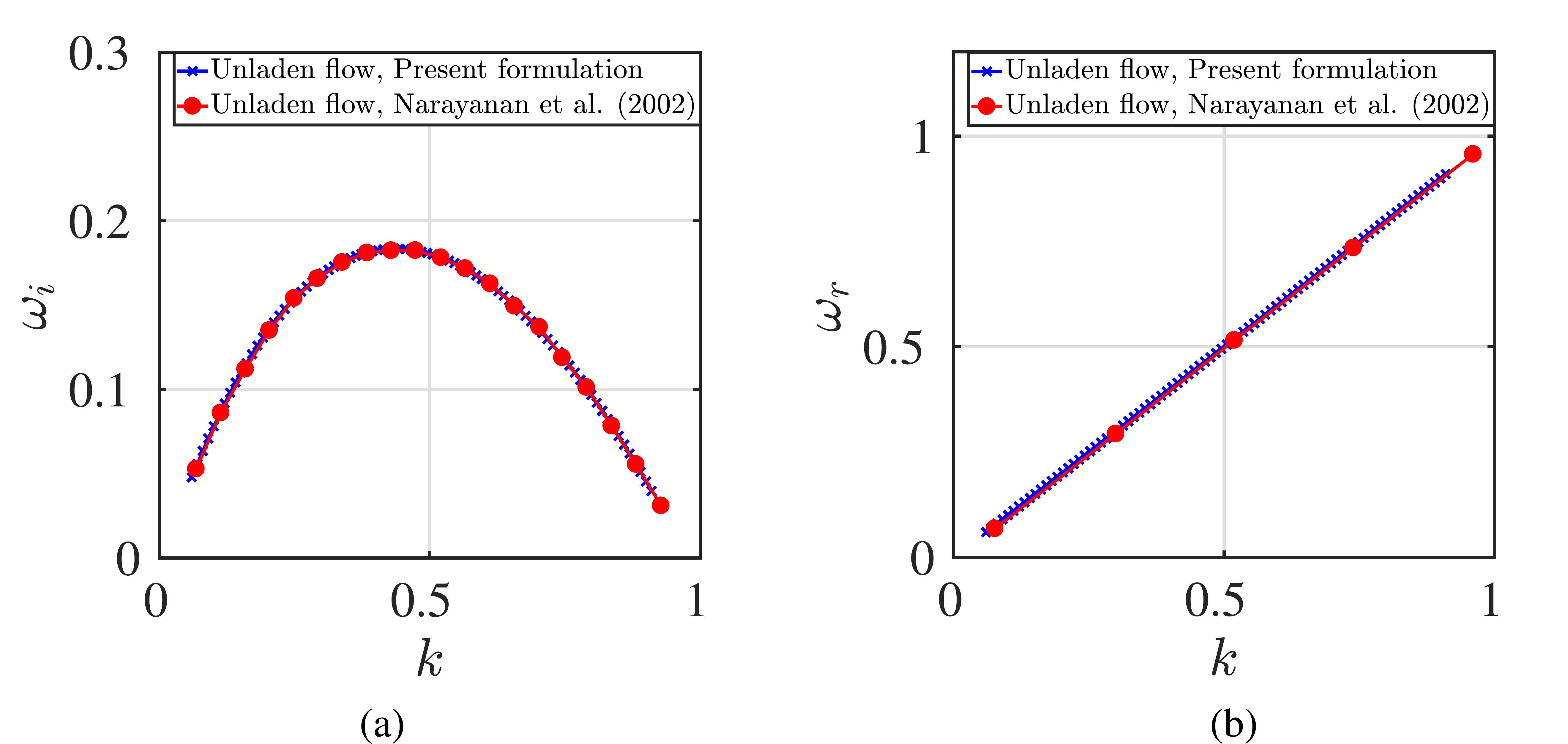}
	\caption{Validation of the present formulation for the unladen mixing layer with Narayanan et al. \cite{Narayananetal2002} at $Re=250$. The  wave speed of the disturbance $c_{r}=1$ i.e, $\omega_{r}=k$. }
	\label{unladen_mixing_layer_dispersion}
\end{figure*}

In order to validate the present formulation, we compare the unladen mixing layer, $\overline{\alpha} = 0$, eigenspectra and dispersion with the single phase locally parallel mixing layer eigenspectra and dispersion relation. 
We let the volume concentration to zero and switch off the terms containing the Stokes number in Eq.\ref{normal mode form} (see appendix \ref{appendix_b} for the terms). The corresponding eigenspectra is shown in figure (\ref{unladen_mixing_layer_spectra}). The domain height in the cross streamwise direction is truncated such that there are no oscillations in the eigenmodes and its derivatives at the boundaries. The agreement between our reduced unladen system of equations and the unladen system given by Narayanan et al. \cite{Narayananetal2002} is shown in figure \ref{unladen_mixing_layer_spectra}. The unstable mode (KH mode) is tracked for a range of wavenumbers. Peak growth rate occurs at $k=0.4344$ with a maximum growth of $\omega_{i}=0.1835$.  Figure \ref{unladen_mixing_layer_dispersion} shows that the wave speed of the disturbance is  $c_{r}=1$, i.e, $\omega_{r}=k$.  %The baseflow velocity profile is given by Eq. (\ref{baseflow vel}).   %$U_{x}\left(y\right)=1+tanh\left(y\right)$.
If the reference frame is travelling with unit velocity (mean velocity, $\overline{U}$), then  the wave speed becomes equal to that of the reference frame and thus would appear as a stationary wave. 
%Fig(\ref{unladen_mixing_layer_eigenmodes}) shows the eigenmodes of the unladen mixing layer. We see that for the streamwise component velocity that the upper and lower streams ($y>0$ and $y<0$) have unequal opposite perturbation velocities that indicates an early sign of roll up of the shear layer. This is more visible in the out of plane vorticity contour where the early signature of roll up of the shear layer is seen. The roll up process in the linear regime is also reported by Michalke \cite{MICHALKE1984159}. 

\begin{figure*}
	\centering 
	\subfigure[]{\includegraphics[width=0.66\textwidth]{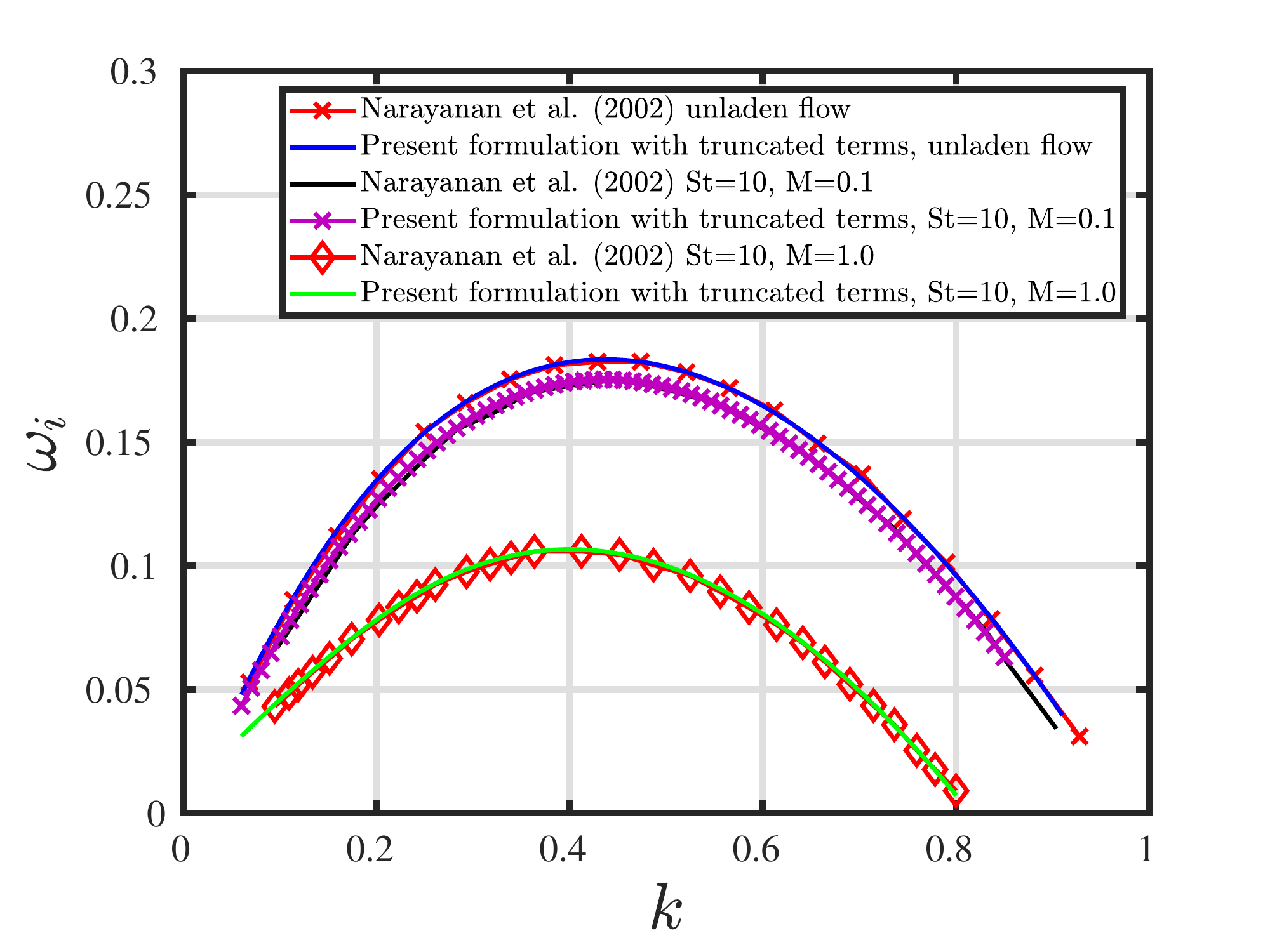}} 
	\subfigure[]{\includegraphics[width=0.7\textwidth]{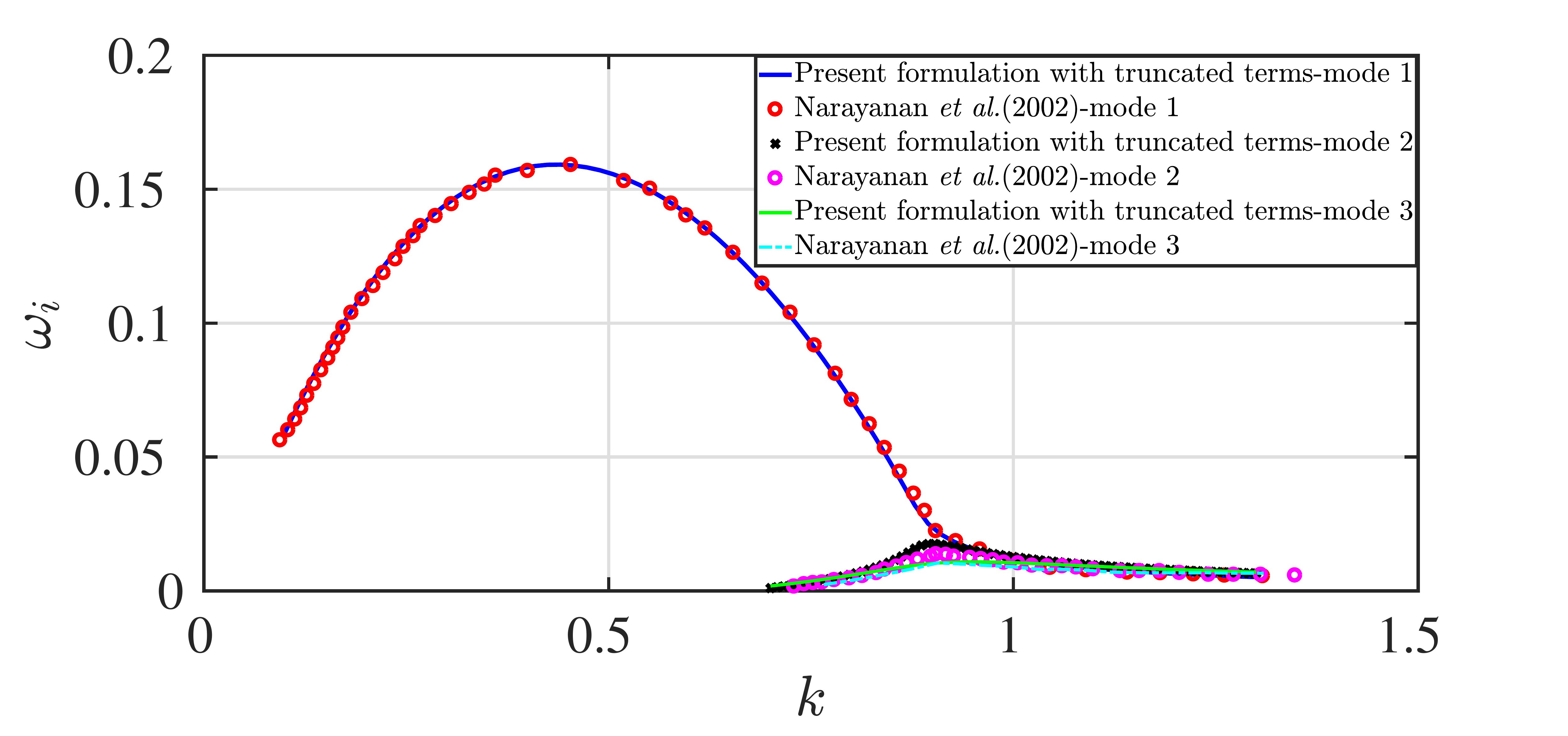}} 
	\caption{(a) Validation of constant particle loading formulation by reducing the present set of governing equations to that of Narayanan et al. \cite{Narayananetal2002} (labelled here as present formulation with truncated terms). The above cases are at intermediate Stokes number of $St=10$, $Re=250$ and $\gamma=10^3$ for uniform particle loadings of $M=0.1$ and $1.0$ (in both upper and lower streams) along with the unladen case. (b) Validation of the variable particle loading formulation. For the above cases, $l_{p}=10$, $\lambda_{p}=0.5$, $Re=250$, $St=100$, $\gamma=5000$, $\overline{\alpha}=5\times10^{-4}$. Baseflow particle concentration is a function of the cross streamwise direction and is given by Eq. (\ref{baseflow_concentration_mixing_layer}). }
	\label{particle_mixing_layer_validation}
\end{figure*}

\begin{figure}
	\centering
	%\subfigure[]{\includegraphics[width=0.46\textwidth]{st_10_M1.eps}}
	\subfigure[]{\includegraphics[width=0.45\textwidth]{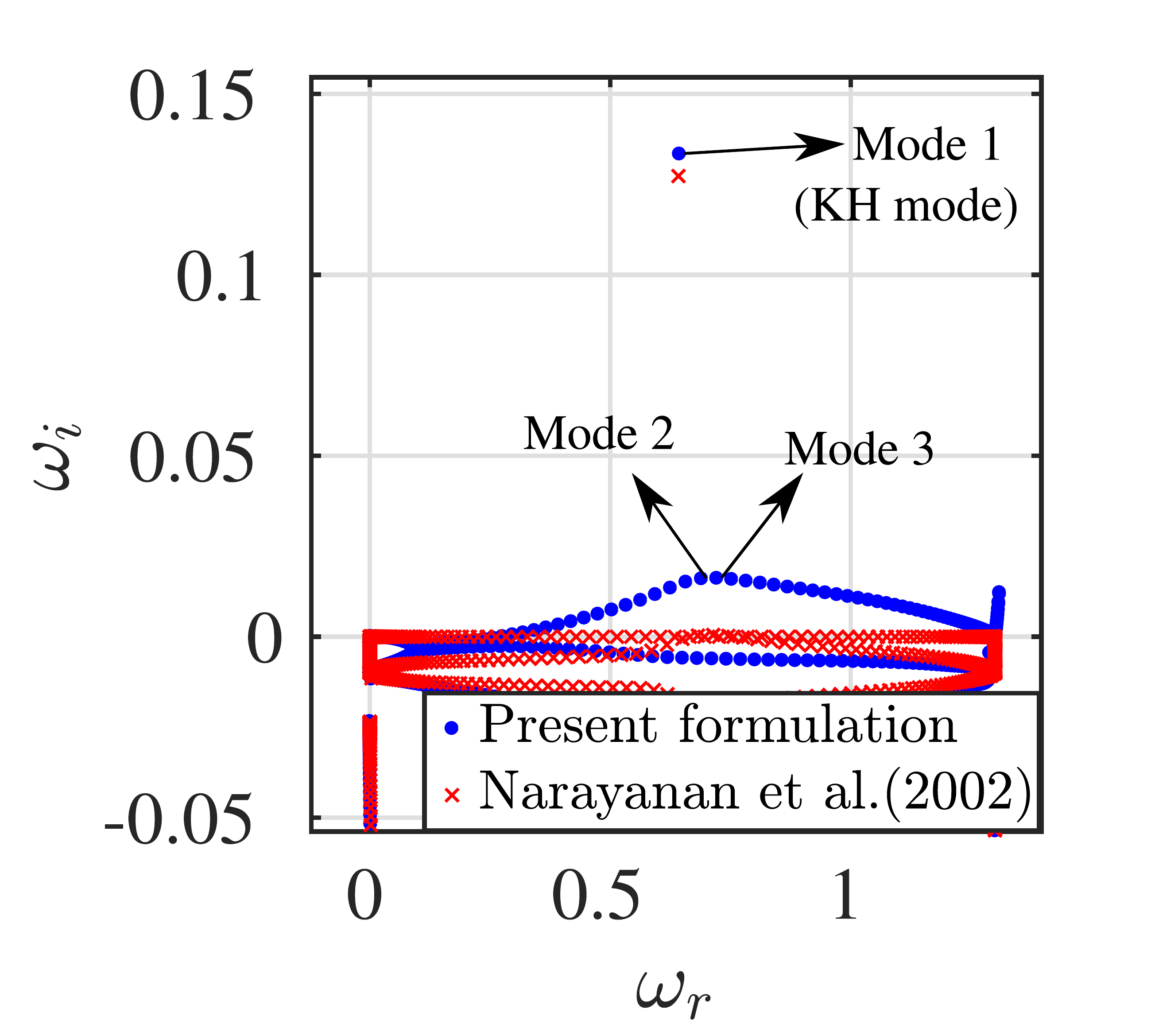}}
	\subfigure[]{\includegraphics[width=0.45\textwidth]{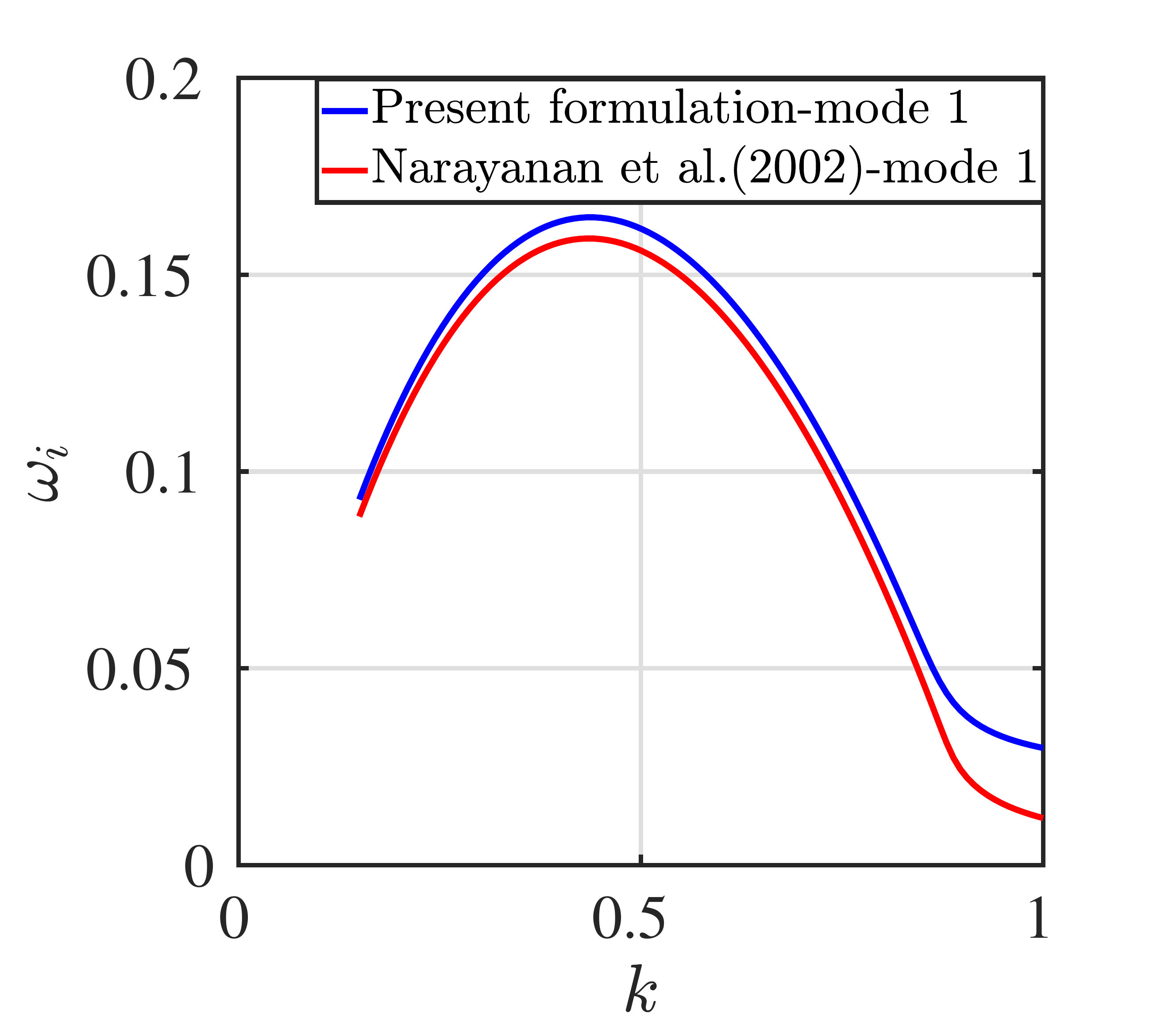}}
	\subfigure[]{\includegraphics[width=0.45\textwidth]{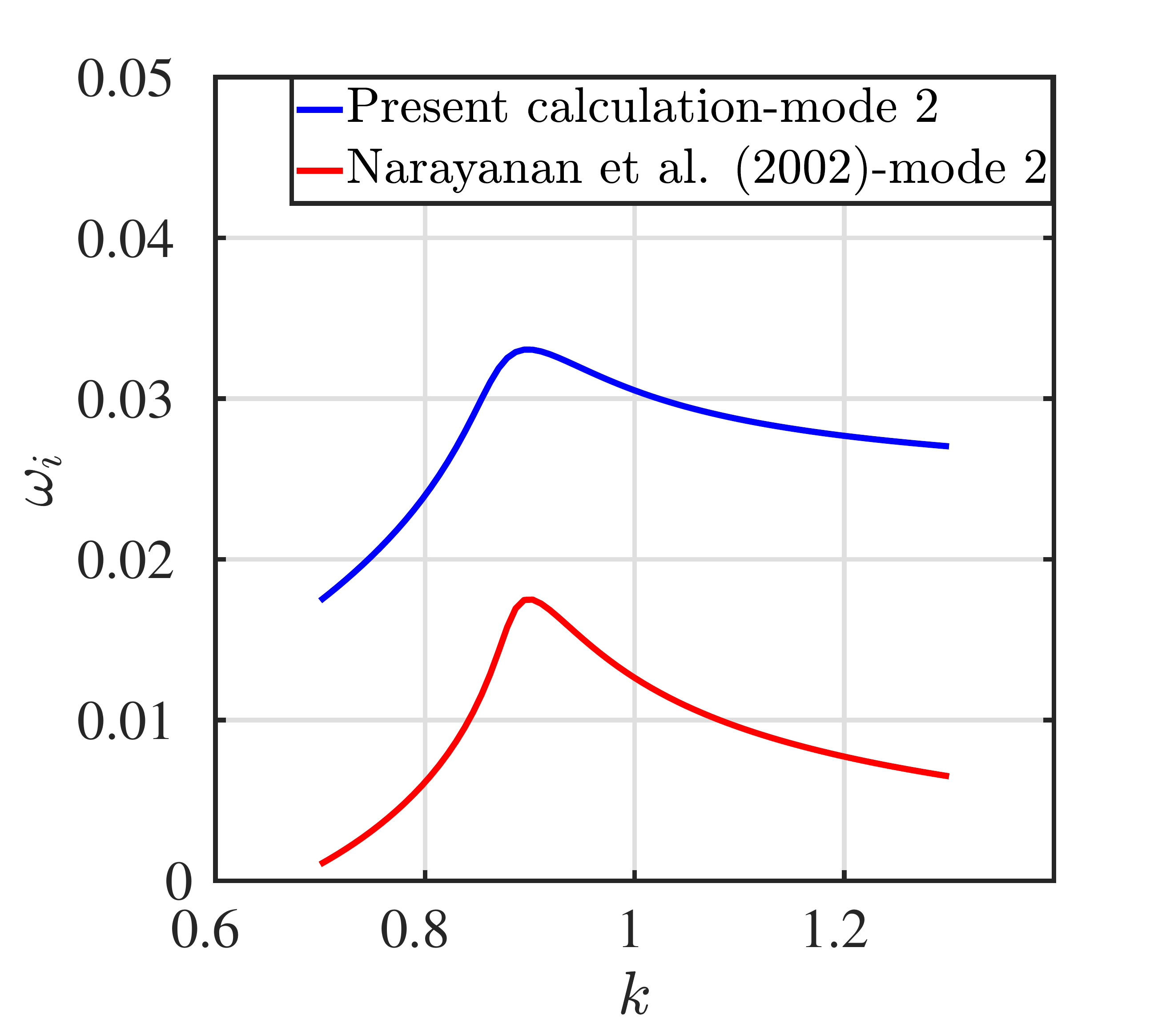}}
	\caption{Comparison of temporal growth rate from the present formulation with Narayanan et al. \cite{Narayananetal2002} for large stokes number $St=100$. $l_{p}=10$, $\lambda_{p}=0.5$, $Re=250$, $\gamma=5000$ and $\overline{\alpha}=5\times10^{-4}$, at $k=0.65$ (a) Shows the eigenspectra with mode 1 (KH mode) and additional modes $2$ and $3$. (b) Shows the difference in growth rates of mode 1. (c) Shows the growth rates of mode 2 using both the formulation. In the present case we have additional terms in the equations that contribute to the increased growth rate of both the modes. Mode 2 seems to occur over a slightly wider range of wave numbers in the present case compared to Narayanan et al. \cite{Narayananetal2002}.}
		\label{vpl_mixing_layer_comparison}
\end{figure}

For the case of uniform particle loading (uniform concentration in both upper and lower streams of the mixing layer) at Stokes number $St=10$ (intermediate Stokes number) and $Re=250$ is shown in figure (\ref{particle_mixing_layer_validation} a). A good match between Narayanan et al. \cite{Narayananetal2002} and the reduced formulation obtained by switching off the  additional terms in the present formulation. The terms present in Narayanan et al. \cite{Narayananetal2002} are indicated by the black colored terms while the additional terms in the present formulation are colored red as shown in appendix(\ref{appendix_b}). At intermediate Stokes number of 10, the effect of particles is to increase the drag force thus reducing the growth rate compared to the unladen mixing layer.  Figure (\ref{particle_mixing_layer_validation} b) shows the match between our reduced formulation and Narayanan et al. \cite{Narayananetal2002} for the case of non uniform loading at large Stokes number ($St=100$) and particle to gas density ratio $\Lambda=5000$. The  parameter $\lambda_{p}$ for the case is 0.5, indicating that the upper stream is more loaded than the lower stream (see figure {\ref{mixing_layer_base_state}b}). At large Stokes number and large density ratio there are additional unstable modes (modes 2 and 3) when the particle loading is non uniform (see Figure~\ref{vpl_mixing_layer_comparison}a). Additional modes differ in growth rates only by a small amount as shown in the (figure \ref{vpl_mixing_layer_comparison}c). Evans et al. \cite{evans1994} and  Narayanan et al. \cite{Narayananetal2002} referred to these as additional modes analogous to Holmboe instability (Ortiz et al. \cite{Ortiz_2002}) which occurs in density stratified mixing layers consisting of a pair of interfacial waves growing at the same rate but travelling in opposite directions.  

\newpage
The major difference in the present work and the work of Narayanan et al. \cite{Narayananetal2002} is in the mathematical formulation of the problem. One striking difference is the treatment of the viscous term. Consider the volume averaged fluid momentum equation in the present formulation,

\begin{equation}
\frac{\partial}{\partial t}\left(\left(1-\alpha\right)u_{i}\right)+\frac{\partial}{\partial x_{k}}\left(\left(1-\alpha\right)u_{i}u_{k}\right)=-\left(1-\alpha\right)\frac{\partial p}{\partial x_{i}}+\frac{\left(1-\alpha\right)}{Re}\frac{\partial^{2}}{\partial x_{k}^{2}}\left(\left(1-\alpha\right)u_{i}\right)-\frac{\gamma}{St}\alpha\left(u_{i}-u_{pi}\right).
\label{fluid_mom}
\end{equation}

Equation considered by Narayanan et al. \cite{Narayananetal2002} is 
\begin{equation}
\frac{\partial u_{i}}{\partial t}+u_{j}\frac{\partial u_{i}}{\partial x_{j}}=-\frac{\partial p_{i}}{\partial x}+\frac{1}{Re}\left(\frac{\partial^{2}u_{i}}{\partial x_{j}\partial x_{j}}\right)+\frac{\lambda\gamma\alpha}{St}\left(u_{pi}-u_{i}\right).
\label{error_narayanan}
\end{equation}

The expression for the diffusive flux in 
Eq. (\ref{error_narayanan}) did not consider volume concentration of the fluid which in the present formulation is $(1-\alpha)$. This leads to having a higher value of viscous force computed by Narayanan thus resulting in a reduced growth rate. This is visible in figure (\ref{vpl_mixing_layer_comparison}a,b,c), for variable particle loading where we see a noticeable change in the temporal growth rates of the KH mode as well as mode $2$. In the present case, mode 2 appears to be spread out over larger range of wave numbers and also with a slightly higher growth rate.   

%\begin{figure}
%	\centering
%	\subfigure[]{\includegraphics[width=0.46\textwidth]{st_10_M1.eps}}
%	\subfigure[]{\includegraphics[width=0.45\textwidth]{st_100_M1_present_unladen.eps}}
%	%\subfigure[]{\includegraphics[width=0.45\textwidth]{mode_1_comparison_st_100%_G5000_alpha_bar_5e_4.eps}}
%	%\subfigure[]{\includegraphics[width=0.45\textwidth]{mode_2_comparison_st_100%_G5000_alpha_bar_5e_4.eps}}
%	\caption{Comparison of temporal growth rate for constant particle loading from the present formulation with Narayanan et al. \cite{Narayananetal2002}. The cases shown above are (a) $St=10$, (b) $St=100$.  In the present case we have additional terms in the equations that contribute to the increased growth rate of both the modes.}
%		\label{23}
%\end{figure}

%\newpage
\section{Results and discussions}
\label{sec: results}
We perform local temporal stability analysis of both the unladen and the particle laden planar jet using the present formulation. In the previous section we have already discussed the effect of ignoring certain terms (present formulation with truncated terms) against the growth rates obtained using a formulation with all the terms present in the context of a locally parallel particle laden mixing layers. Before we discuss the effect of addition of particles on the stability of the jet, we briefly describe the stability of the unladen jet below. 
\subsection{Unladen planar jet}
\begin{figure*}[!ht]
	\centering 
%	\subfigure[]{\includegraphics[width=0.35\textwidth]{base_state.eps}}
%	\subfigure[]{\includegraphics[width=0.55\textwidth]{unladen_re_250_grid_converge.pdf}}
	\includegraphics[width=1.1\textwidth]{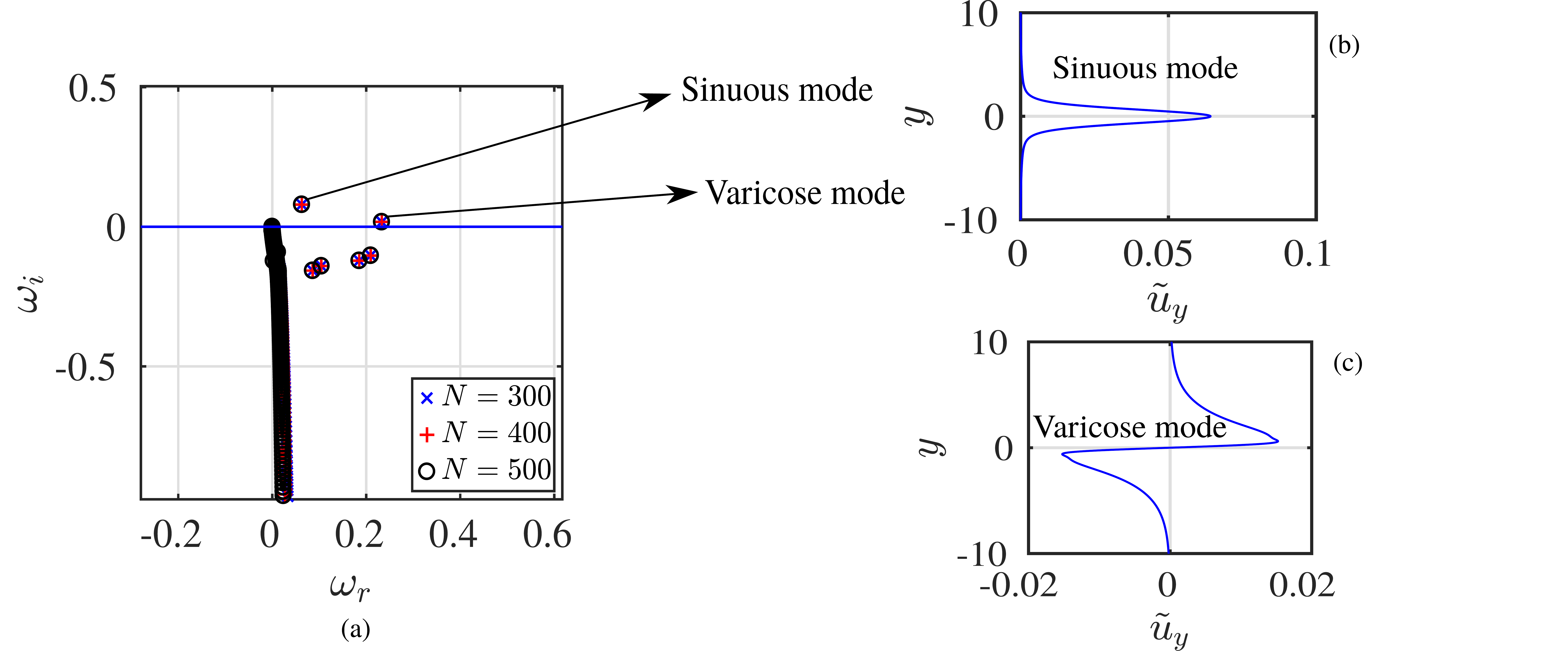}
	\caption{(a) shows  unladen planar jet spectrum for $Re=250$, $k=0.3$. The spectrum was  obtained by solving the system of equations by reducing the system in the absence of particles. There are two unstable eigenmodes : sinuous and varicose mode. Grid convergence for the two unstable modes for $N = 300,400,500$ are shown. (b) shows the sinuous mode (symmetric mode) with cross streamwise component $\tilde{u}_{y}(-y)=\tilde{u}_{y}(y)$. (c) Varicose mode (antisymmetric mode) with $\tilde{u}_{y}(-y)=-\tilde{u}_{y}(y)$. }
	\label{unladen spectra}
\end{figure*}

Figure(\ref{unladen spectra}) shows the eigenspectra for the unladen planar jet at $Re=250$, obtained by solving the system of equations by reducing the equations in the absence of particles. There two unstable modes namely sinuous and varicose modes. The two unstable  modes in figure (\ref{unladen spectra}a) correspond to even and odd eigenfunctions respectively. Figure(\ref{unladen spectra}b) and figure(\ref{unladen spectra}c) shows the sinuous and varicose mode shapes respectively. The sinuous mode is characterised by the cross streamwise perturbation being an even function $u^{'}_{y}(x,-y,t)=u^{'}_{y}(x,y,t)$, while the varicose mode  being an odd function,  $u^{'}_{y}(x,-y,t)=-u^{'}_{y}(x,y,t)$. They are allowed solutions of the Rayleigh equation (can also be shown for Orr Sommerfeld equation) provided, the base flow velocity $U_{x}$, is an even function and the boundary conditions are symmetric. Under these conditions it can be shown that odd and even eigenfunctions are allowed solutions of the Rayleigh equation. Now in terms of velocity potential, we have the Rayleigh equation, 

\begin{equation}
\tilde{\phi}^{''}-\left(k^{2}+\frac{U_{x}^{''}}{U_{x}-c}\right)\tilde{\phi}=0,
\label{eq:unladen_Rayliegh}
\end{equation}

or 
\begin{equation}
\mathcal{L}\tilde{\phi}=0,
\end{equation}

where $\mathcal{L}=\frac{d^{2}}{dy^{2}} - (k^2 + \frac{U_{x}^{''}}{U_{x}-c}))$, subject to boundary conditions $\tilde{\phi}\left(y\right)\rightarrow0$ as  $y\rightarrow\pm\infty$. 

Replacing $y$ with $-y$ and using the fact that the base flow velocity is an even function we have, 
\begin{equation}
\mathcal{L}\tilde{\phi}\left(-y\right)=\mathcal{L}\tilde{\phi}\left(y\right) = 0.    
\end{equation}

\begin{figure*}[!ht]
	\centering 
	\includegraphics[width=0.9\textwidth]{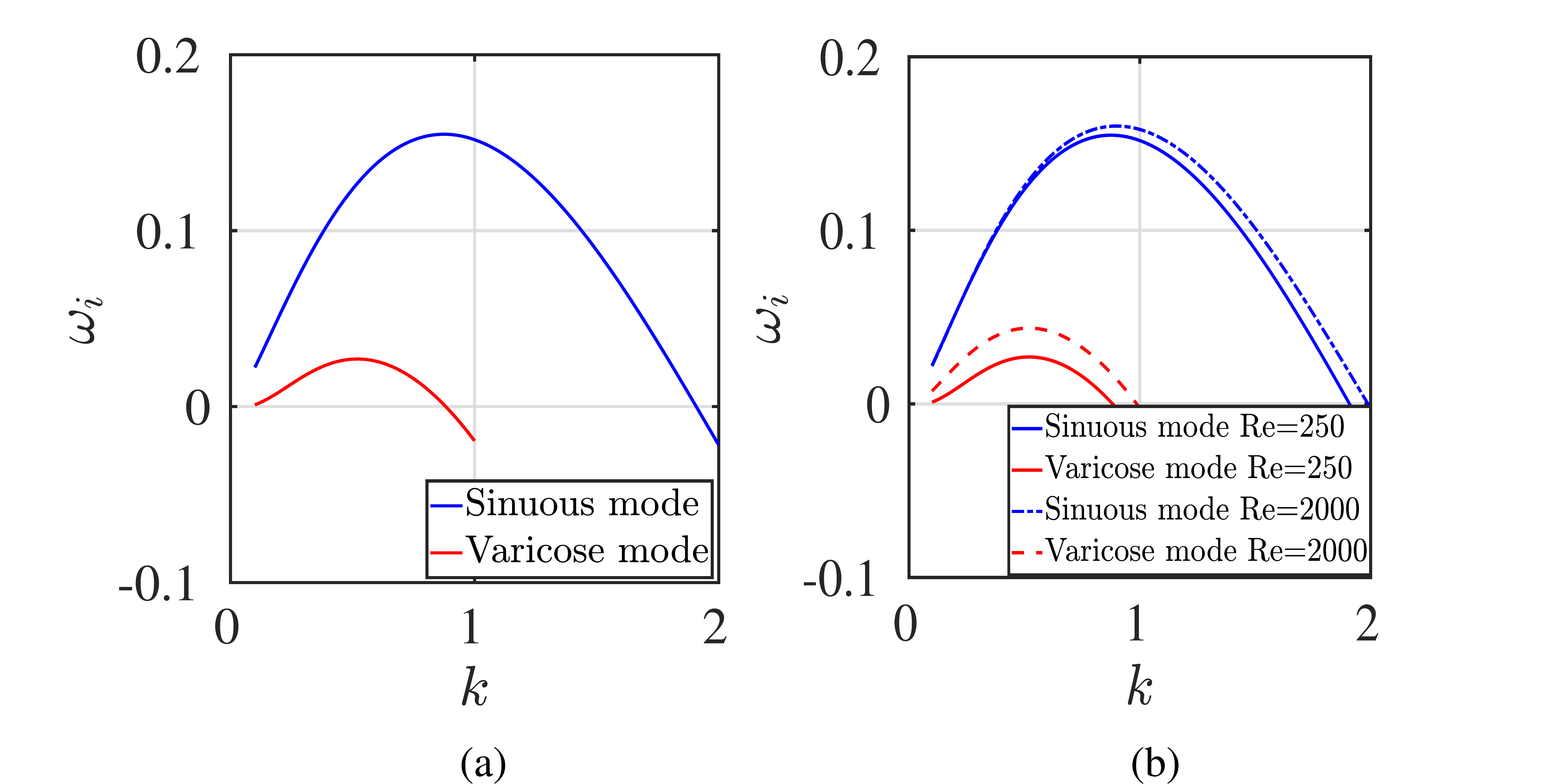}
	\caption{(a) Growth rates for unladen planar jet at $Re_{jet} = 250$ for both sinuous and varicose modes. Sinuous mode is more unstable than the varicose mode and for a larger range of wavenumbers. (b) Growth rates at $Re_{jet} = 250, 2000$. A higher $Re_{jet}$ implies smaller viscous effects and therefore at a smaller $Re_{jet}$, shorter waves are smoothed to a greater extent which decreases their growth rate .}
	\label{unladen_sin_var}
\end{figure*}

Both even and odd eigen functions satisfy the Rayleigh equation provided the baseflow is an even function and the boundary conditions are symmetric as in case of the locally parallel planar jet where the baseflow profile is an even function $(U_{x}(y)= sech^{2}y)$ and perturbations decaying to zero at infinity. While in the case of a locally parallel mixing layer base flow velocity given by ($U_{x}(y)=1+\tanh{y}$), is not an even function and therefore odd and even eigenfunctions are not allowed solutions of the Rayleigh equation.  Figure (\ref{unladen_sin_var}a) shows their growth rates at $Re_{jet}=250$.  As Reynolds number increases (see figure \ref{unladen_sin_var}b), maximum growth rate increases especially at large wavenumbers. This is expected as large wavenumber implies disturbances of smaller wavelengths (shorter waves) and they are smoothed (damped) by viscosity and increasing Reynolds number decreases this effect resulting in higher growth rate. However the varicose mode is more sensitive to changes in Reynolds number.
%and streamwise perturbation being an even function $u^{'}_{x}(x,-y,t)=u^{'}_{x}(x,y,t)$.  The sinuous and varicose modes shown in figures(\ref{unladen sinuous}) and (\ref{unladen varicose}) are qualitatively similar to the experimental results of Matsubara et al, \cite{matsubara_alfredsson_segalini_2020}. (See figures 15 and 16 in Matsubara et al, \cite{matsubara_alfredsson_segalini_2020}). They studied fully developed turbulent planar jet experimentally. Streamwise and Cross stream wise velocities were measured and the mean was compared with the $Sech^{2}y$ profile that we have used. The jet was periodically forced to study the coherant structures associated with the introduction of perturbation. They performed a weakly non parallel (Velocities are slowly varying in the stream wise direction). From figures 15 and 16 in Matsubara et al, \cite{matsubara_alfredsson_segalini_2020}) we see that their spatial evolution of the sinous mode matches qualitatively well with our local analysis upto small distances downstream of the jet exit. Their experiments too confirmed  the fact that sinuous modes were dominant compared to varicose modes which is revealed from our local analysis.

%\begin{figure*}[!ht]
%	\centering 
%	\includegraphics[width=0.7\textwidth]{sinuous_varicose.pdf}
%%	\includegraphics[width=0.7\textwidth]{sinuous_unladen_vorticity.pdf}
%	\caption{Shows the sinuous mode (symmetric mode) with streamwise component ((a) and (b)) where  $\tilde{u}_{x}(-y)=-\tilde{u}_{x}(y)$ and cross streamwise component ((c) and (d)) where $\tilde{u}_{y}(-y)=\tilde{u}_{y}(y)$.}
%	\label{unladen sinuous}
%\end{figure*}

%
%\begin{figure*}[!ht]
%	\centering 
%	\includegraphics[width=0.8\textwidth]{unladen_varicose_mode_Re_250.pdf}
%	\caption{Varicose mode (antisymmetric mode) with $\tilde{u}_{x}(-y)=\tilde{u}_{x}(y)$;  $\tilde{u}_{y}(-y)=-\tilde{u}_{y}(y)$.}
%	\label{unladen varicose}
%\end{figure*}

%\begin{figure*}[!ht]
%	\centering 
%	\includegraphics[width=0.7\textwidth]{unladen_sinuous_varicose_re_250_2000_N600.pdf}
%	\caption{Dispersion relation curves for unladen jet at $Re_{jet} = 250, 2000$. A higher $Re_{jet}$ implies smaller viscous effects and therefore at a smaller $Re_{jet}$, shorter waves are smoothened to a greater extent which decreases their growth rate as seen above.}
%	\label{high_re_unladen}
%\end{figure*}

\newpage
\subsection{Addition of particles at low Stokes number ($St=10^{-5} - 0.1$) : Physical mechanism of instability}\label{sec:low_St}
We now study the effect of addition of particles on the stability of the jet. Let $L$ be the length scale and $U^{*}$ be the velocity scale of the base flow. 
%Following Saffman \cite{saffman_1962}, we assume that the wavelength of the most dominant disturbance is of the order of $L$ (length scale of the flow).
Writing the linearised particle phase equation in vector form (from Eq. (\ref{eq:LPE_x}),(\ref{eq:LPE_y})) we have,
\begin{equation}
\frac{\partial \boldsymbol{u_{p}}^{'}}{\partial t}+U_{p}\frac{\partial \boldsymbol{u_{p}}^{'}}{\partial x}+u_{py}^{'}\frac{dU_{x}}{dy}\boldsymbol{e_{x}}=\frac{1}{St}\left(\boldsymbol{u^{'}}-\boldsymbol{u_{p}}^{'}\right)
\end{equation}   
The L.H.S  is  $O\left(\frac{U^{*}u_{p}^{'}}{L}\right)$ and the R.H.S is $O\left( \frac{u^{'}-u_{p}^{'}}{\tau}\right)$. 
\begin{equation}
\frac{U^{*}u_{p}^{'}}{L}\sim \frac{\left(u^{'}-u_{p}^{'}\right)}{\tau}
\end{equation}

\begin{figure}[!ht]
	\centering
	%	\subfigure[]{\includegraphics[width=0.7\textwidth]{spectra_st_1e_minus_5_lambda_1e_minus_4.pdf}}
	\includegraphics[width=0.95\textwidth]{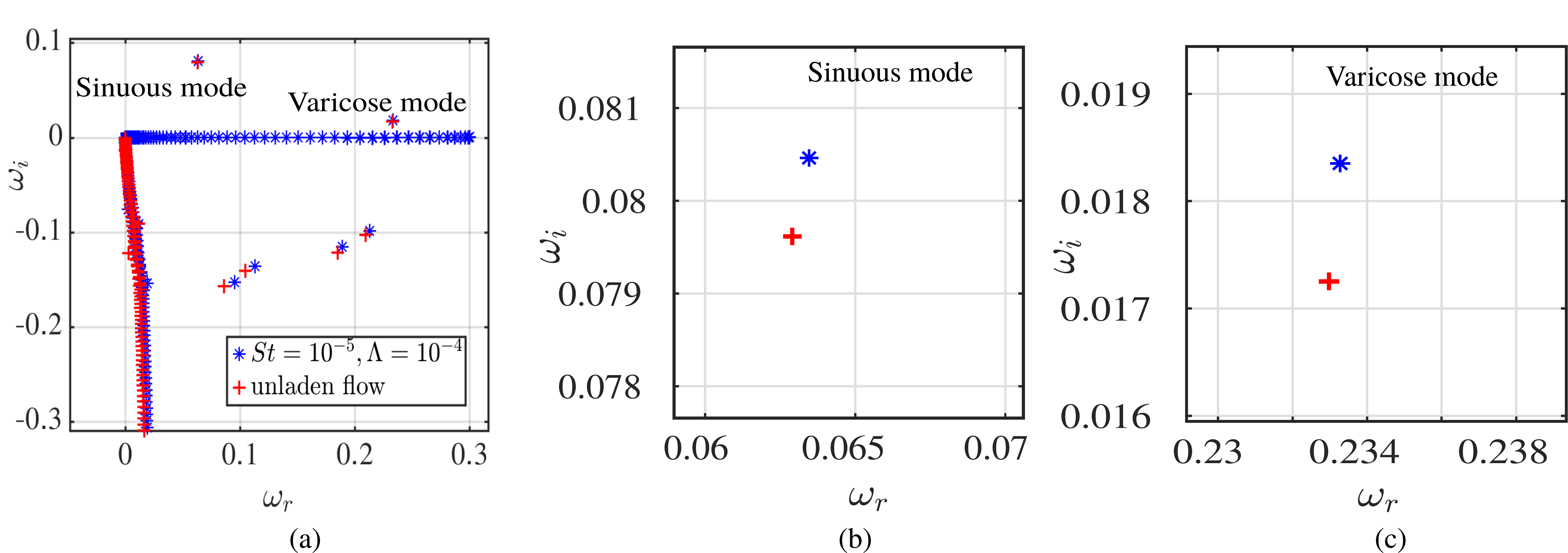}
	%	\subfigure[]{\includegraphics[width=0.35\textwidth]{st_1e_minus_5_M1.eps}}
	\caption{(a) Low Stokes number planar jet spectrum showing  two unstable eigenmodes : sinuous and varicose mode at $St=10^{-5}, \Lambda=10^{-4}$ compared with the sinuous and varicose modes of the unladen flow at $k=0.3$. (b,c) Zoomed in view showing growth rate of sinuous mode with particles at low Stokes number greater than of unladen flow and similarly, the varicose mode with particles is more unstable than the unladen varicose mode.  Particle concentration $\Lambda=10^{-3}$ (one order higher) at the same wavenumber. The difference in the growth rates between unladen and particle laden sinuous and varicose modes are more clearly seen at higher particle  concentration. }
	\label{spectra_st_1e_minus_5} 
\end{figure}

\begin{figure}
	\centering
		\includegraphics[width=1.0\textwidth]{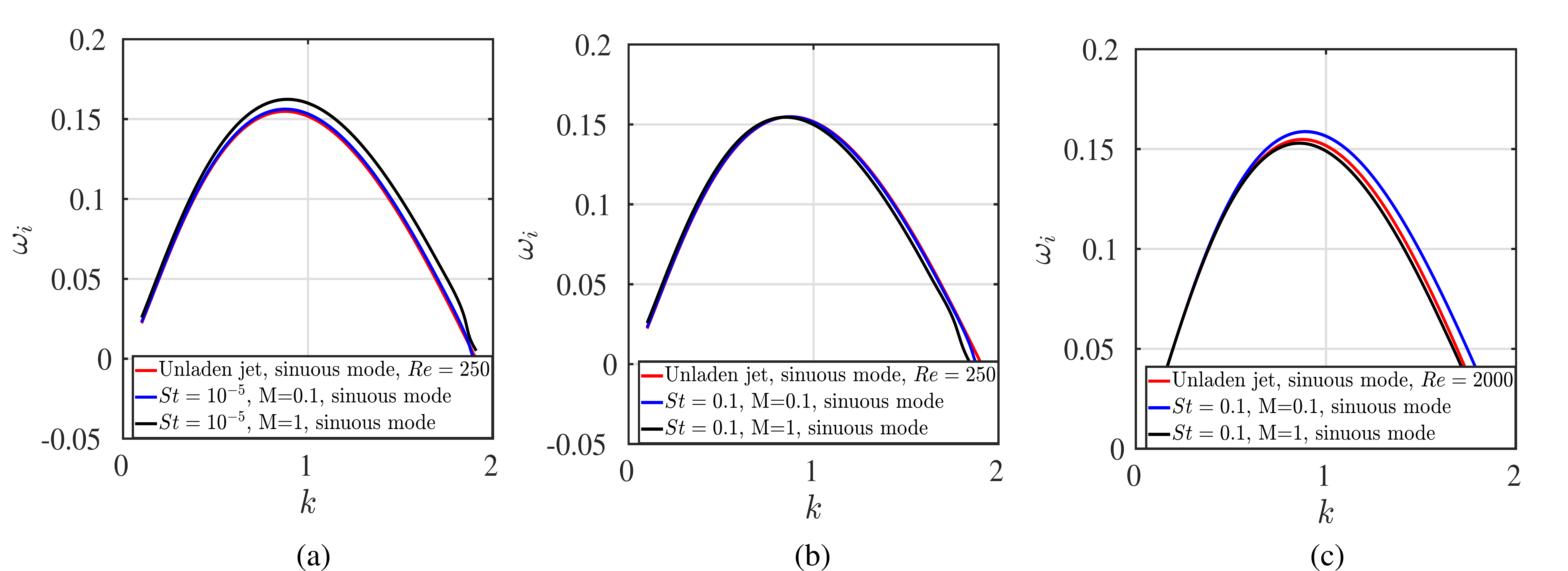}
	\caption{Temporal dispersion curves for sinuous modes at Stokes numbers of $10^{-5}$ and $0.1$ for different particle concentrations ($\Lambda=10^{-3}, 10^{-4}$) compared against the unladen flow dispersion curves for different Reynolds number $Re=250$ and $Re=2000$. (a) Shows growth rates at $St=10^{-5}$ and $Re=250$ at $\Lambda=10^{-3}, 10^{-4}$. Particle laden growth rates are greater than the unladen growth rates owing to negligible drag and lower viscous dissipation. (b) Shows growth rates at $St=0.1$ and $Re=250$ at $\Lambda=10^{-3}, 10^{-4}$. Sinuous modes with particles are more unstable than the unladen ones in the lower $k$ region while at larger $k$, the growth rates almost match the unladen ones.  (c) Shows growth rates at $St=0.1$ and $Re=2000$ at $\Lambda=10^{-3}, 10^{-4}$.} 
	\label{Re_250_st_1e_minus_1_comparison}  
\end{figure}

For the baseflow, the particle relaxation time is zero from our assumption $\boldsymbol{U}=\boldsymbol{U}_{p}$ (zero inertia flow). The condition $\tau \ll \frac{L}{U^{*}}$, is when the relaxation time of the disturbance field is much smaller than the convective time scale of the baseflow. We refer to this as the low Stokes number regime ($St\ll 1$).
Figure(\ref{spectra_st_1e_minus_5}) shows the eigenspectra of particle laden planar jet at low Stokes number of $St=10^{-5}$. The two unstable modes are the sinuous and varicose modes which are similar to the ones seen in the case of particle free planar jets. We see that addition of particles at low Stokes numbers increases the growth rate of both the sinuous and varicose modes compared to the unladen case. This behaviour can be explained as in the limit of small Stokes number, the Stokes drag is negligible compared to the viscous stress per unit volume as the perturbed particle velocity field responds to the fluid velocity field almost instantaneously. Meanwhile small Stokes number implies smaller particle size ($St\sim d^{2}_{p}$ where $d_{p}$ is the diameter of the particle). This has a small effect of the viscosity of the particle laden suspension but the density of the suspension exceeds the density of the unladen flow. This results in a smaller kinematic viscosity of the suspension compared to the unladen flow due to the presence of fine particles, resulting in smaller viscous damping effect on the jet, thus leading to an increased growth rate. This was first discussed by Saffman \cite{saffman_1962} in the context of dusty gas flows in a channel. However at very high Reynolds number (greater than 2000) and  Stokes number of $St=0.1$, particles have stabilizing effect, while at moderate $Re=250$ at $St=0.1$,  particle laden flow is more unstable than the unladen flow as shown in figure (\ref{Re_250_st_1e_minus_1_comparison}). This is better understood by considering the linearized fluid momentum equations. The $x$ component equation reads,
\[
    \left(1-\Lambda\right)\left(\frac{\partial {u_{x}}^{'}}{\partial t}\right)-U_{x}\left(\frac{\partial\alpha^{'}}{\partial t}\right)+\left(1-\Lambda\right)U_{x}\frac{\partial {u_{x}}^{'}}{\partial x}-{U_{x}}^{2}\frac{\partial\alpha^{'}}{\partial x}+\left(1-\Lambda\right)\frac{d U_{x}}{dy}{u_{y}}^{'}
\]
\[
-U_{x}\left\{\frac{d\Lambda}{dy} \left(u_{py}^{'}-u_{y}^{'}\right)+\Lambda\left(\frac{\partial u_{px}^{'}}{\partial x}+\frac{\partial u_{py}^{'}}{\partial y}\right)\right\}=-\left(1-\Lambda\right)\frac{\partial p^{'}}{\partial x}
\]
\begin{align}
&+\frac{1}{Re}\left\{ \left(1-\Lambda\right)^{2}\left(\frac{\partial^{2}u_{x}^{'}}{\partial x^{2}}+\frac{\partial^{2}u_{x}^{'}}{\partial y^{2}}\right)-\left(1-\Lambda\right)U_{x}\left(\frac{\partial^{2}\alpha^{'}}{\partial x^{2}}+\frac{\partial^{2}\alpha^{'}}{\partial y^{2}}\right)-2\left(1-\Lambda\right)\alpha^{'}\frac{d^{2}U_{x}}{dy^{2}}-2\left(1-\Lambda\right)\frac{\partial\alpha^{'}}{\partial y}\frac{dU_{x}}{dy}\right.\nonumber\\&\qquad\left. {} -2\left(1-\Lambda\right)\frac{d\Lambda}{dy}\frac{\partial u_{x}^{'}}{\partial y}-\left(1-\Lambda\right)u_{x}^{'}\frac{d^{2}\Lambda}{dy^{2}}+2\frac{d\Lambda}{dy}\frac{dU_{x}}{dy}\alpha^{'}+U_{x}\frac{d^{2}\Lambda}{dy^{2}}\alpha^{'}\right\}-\frac{\gamma\Lambda}{St}\left(u_{x}^{'}-u_{px}^{'}\right)
\end{align}
%\raisebox{.5pt}{\textcircled{\raisebox{-.9pt} {1}}}
Denoting the convective term as,
\[
\mbox{Term \raisebox{.5pt}{\textcircled{\raisebox{-.9pt} {1}}}}= \left(1-\Lambda\right)U_{x}\frac{\partial u_{x}^{'}}{\partial x}-U_{x}^{2}\frac{\partial\alpha^{'}}{\partial x}+\left(1-\Lambda\right)\frac{dU_{x}}{dy}u_{y}^{'}-U_{x}\left\{\frac{d\Lambda}{dy} \left(u_{py}^{'}-u_{y}^{'}\right)+\Lambda\left(\frac{\partial u_{px}^{'}}{\partial x}+\frac{\partial u_{py}^{'}}{\partial y}\right)\right\}
\]

Viscous diffusion term,
\begin{align*}
\mbox{Term \raisebox{.5pt}{\textcircled{\raisebox{-.9pt} {2}}}}= &\frac{1}{Re}\left\{ \left(1-\Lambda\right)^{2}\left(\frac{\partial^{2}u_{x}^{'}}{\partial x^{2}}+\frac{\partial^{2}u_{x}^{'}}{\partial y^{2}}\right)-\left(1-\Lambda\right)U_{x}\left(\frac{\partial^{2}\alpha^{'}}{\partial x^{2}}+\frac{\partial^{2}\alpha^{'}}{\partial y^{2}}\right)-2\left(1-\Lambda\right)\alpha^{'}\frac{d^{2}U_{x}}{dy^{2}}\right.\nonumber\\&\qquad\left. {} -2\left(1-\Lambda\right)\frac{\partial\alpha^{'}}{\partial y}\frac{dU_{x}}{dy}-2\left(1-\Lambda\right)\frac{d\Lambda}{dy}\frac{\partial u_{x}^{'}}{\partial y}-\left(1-\Lambda\right)u_{x}^{'}\frac{d^{2}\Lambda}{dy^{2}}+2\frac{d\Lambda}{dy}\frac{dU_{x}}{dy}\alpha^{'}+U_{x}\frac{d^{2}\Lambda}{dy^{2}}\alpha^{'}\right\}    
\end{align*}

Drag force term,
\[
\mbox{Term \raisebox{.5pt}{\textcircled{\raisebox{-.9pt} {3}}}}=\frac{\gamma\Lambda}{St}\left(u_{x}^{'}-u_{px}^{'}\right)
\]

$y$ component of the continuous phase equation reads,
\begin{align}
\left(1-\Lambda\right)\left(\frac{\partial u_{y}^{'}}{\partial t}\right)+\left(1-\Lambda\right)U_{x}\frac{\partial u_{y}^{'}}{\partial x}&=-\left(1-\Lambda\right)\frac{\partial p^{'}}{\partial y}+
\frac{1}{Re}\left\{ \left(1-\Lambda\right)^{2}\left(\frac{\partial^{2}u_{y}^{'}}{\partial x^{2}}+\frac{\partial^{2}u_{y}^{'}}{\partial y^{2}}\right) \right.\nonumber\\
&\qquad \left. {}    -2\left(1-\Lambda\right)\frac{d\Lambda}{dy}\frac{\partial u_{y}^{'}}{\partial y}-\left(1-\Lambda\right)\frac{d^{2}\Lambda}{dy^{2}}u_{y}^{'}\right\} -\frac{\gamma\Lambda}{St}\left(u_{y}^{'}-u_{py}^{'}\right)
\end{align}

$y$ component of the convective term  
\[
\mbox{Term \raisebox{.5pt}{\textcircled{\raisebox{-.9pt} {4}}}} = \left(1-\Lambda\right)U_{x}\frac{\partial u_{y}^{'}}{\partial x} 
\]

$y$ component of the viscous diffusion term,
\[
\mbox{Term \raisebox{.5pt}{\textcircled{\raisebox{-.9pt} {5}}}}= \frac{1}{Re}\left\{ \left(1-\Lambda\right)^{2}\left(\frac{\partial^{2}u_{y}^{'}}{\partial x^{2}}+\frac{\partial^{2}u_{y}^{'}}{\partial y^{2}}\right)  -2\left(1-\Lambda\right)\frac{d\Lambda}{dy}\frac{\partial u_{y}^{'}}{\partial y}-\left(1-\Lambda\right)\frac{d^{2}\Lambda}{dy^{2}}u_{y}^{'}\right\}
\]

$y$ component of the drag force,
\[
\mbox{Term \raisebox{.5pt}{\textcircled{\raisebox{-.9pt} {6}}}}= \frac{\gamma\Lambda}{St}\left(u_{y}^{'}-u_{py}^{'}\right)
\]   

\begin{figure}[!ht]
	\centering
	\includegraphics[width=0.65\textwidth]{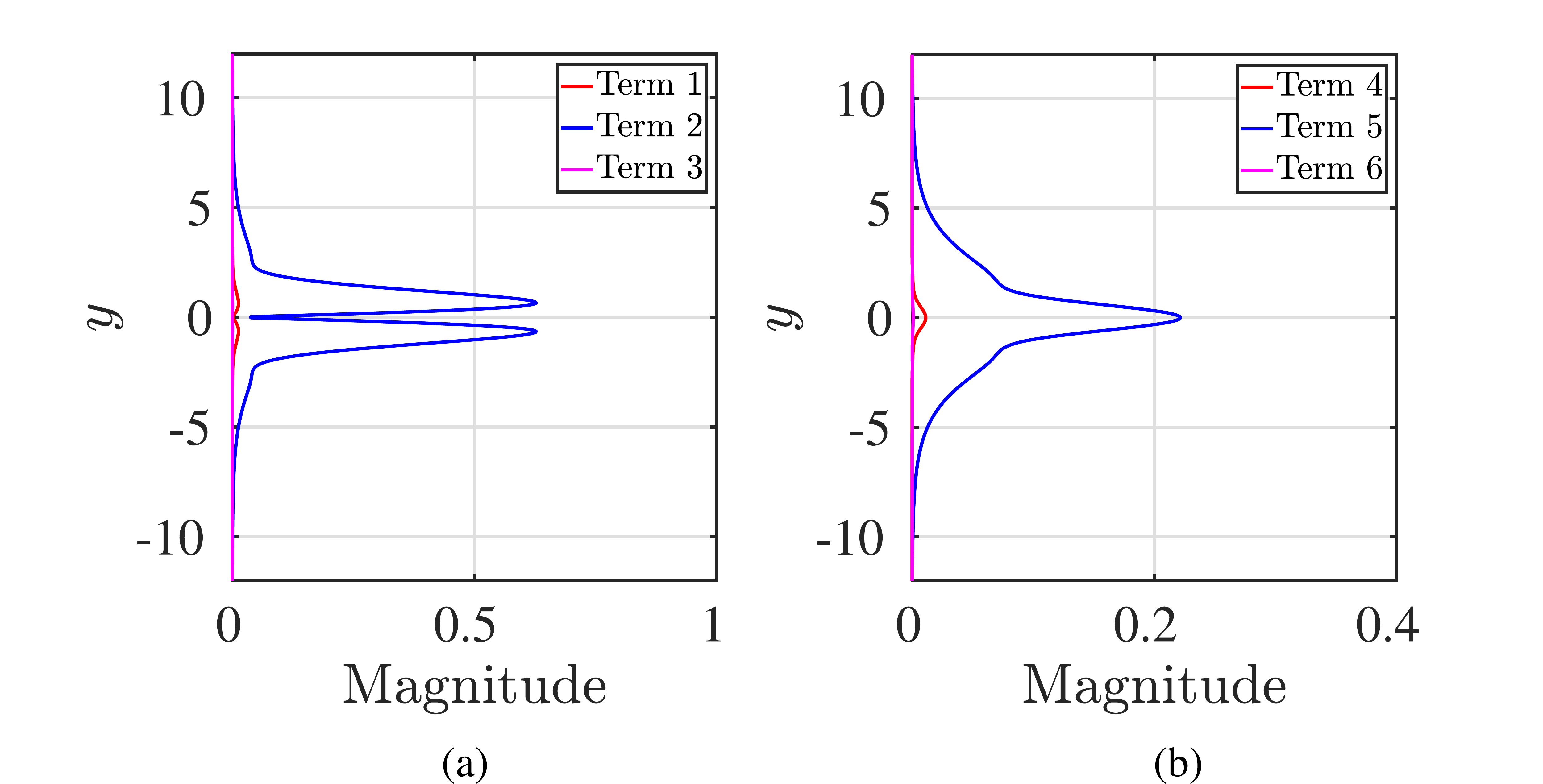}
    \includegraphics[width=0.65\textwidth]{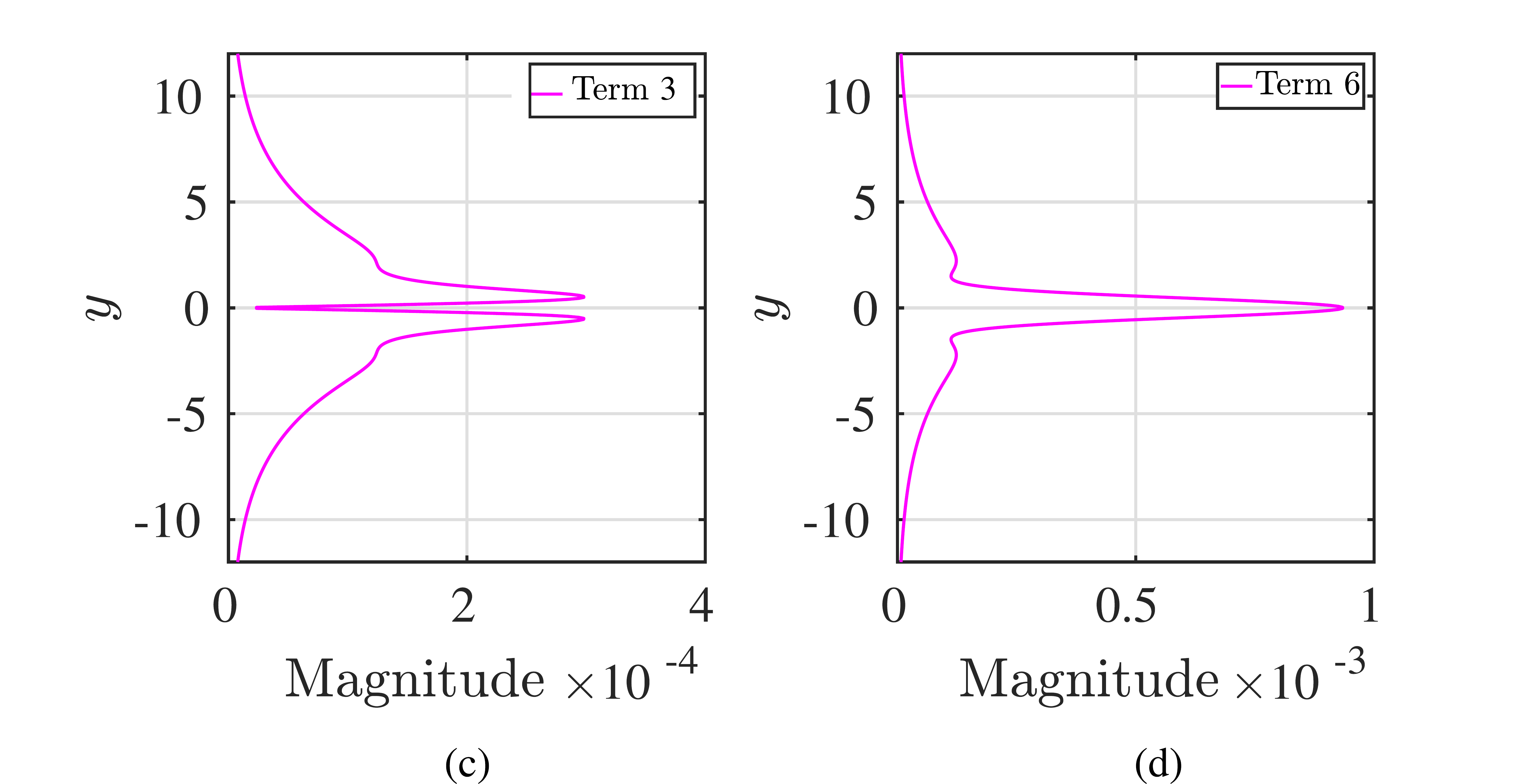}
	\caption{ Term 1 is the $x$ component of the convective term. Term 2 is the $x$ component of the viscous diffusion term and term 3 is the $x$ component of the drag force. Terms $4,5$ and $6$ are the y components of convective, viscous diffusion and drag force respectively. Terms $3$ and $6$ are separately plotted to indicate the difference in magnitude between the drag force terms with the other terms. Calculations shown are performed at $St=10^{-5}$, $\Lambda=10^{-4}$, $Re=250$ and $k=0.3$.  } 
	\label{Re_250_st_1e_minus_5_budget}  
\end{figure}

\begin{figure}[!ht]
	\centering
	\includegraphics[width=0.8\textwidth]{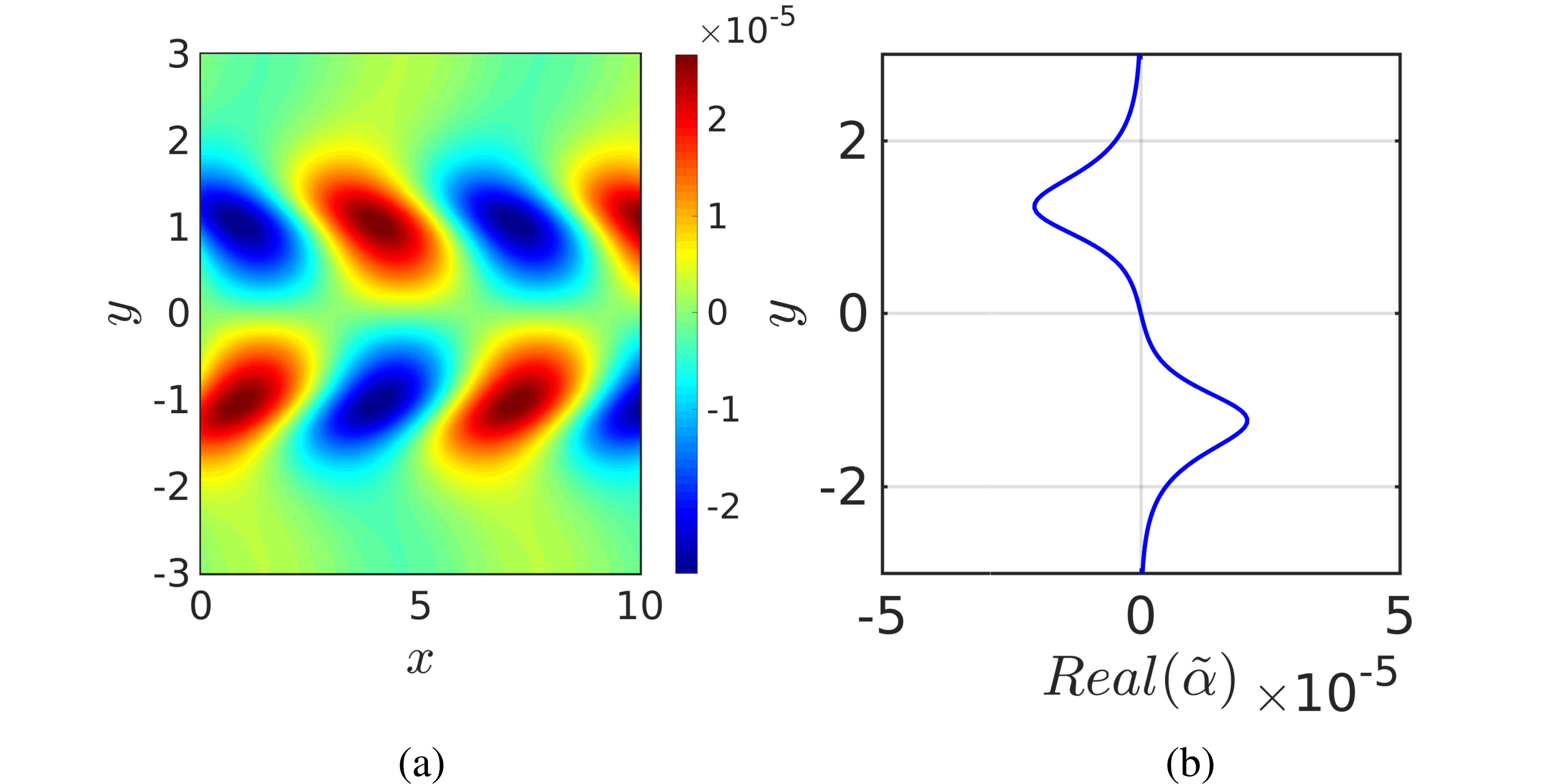} 
	\includegraphics[width=0.8\textwidth]{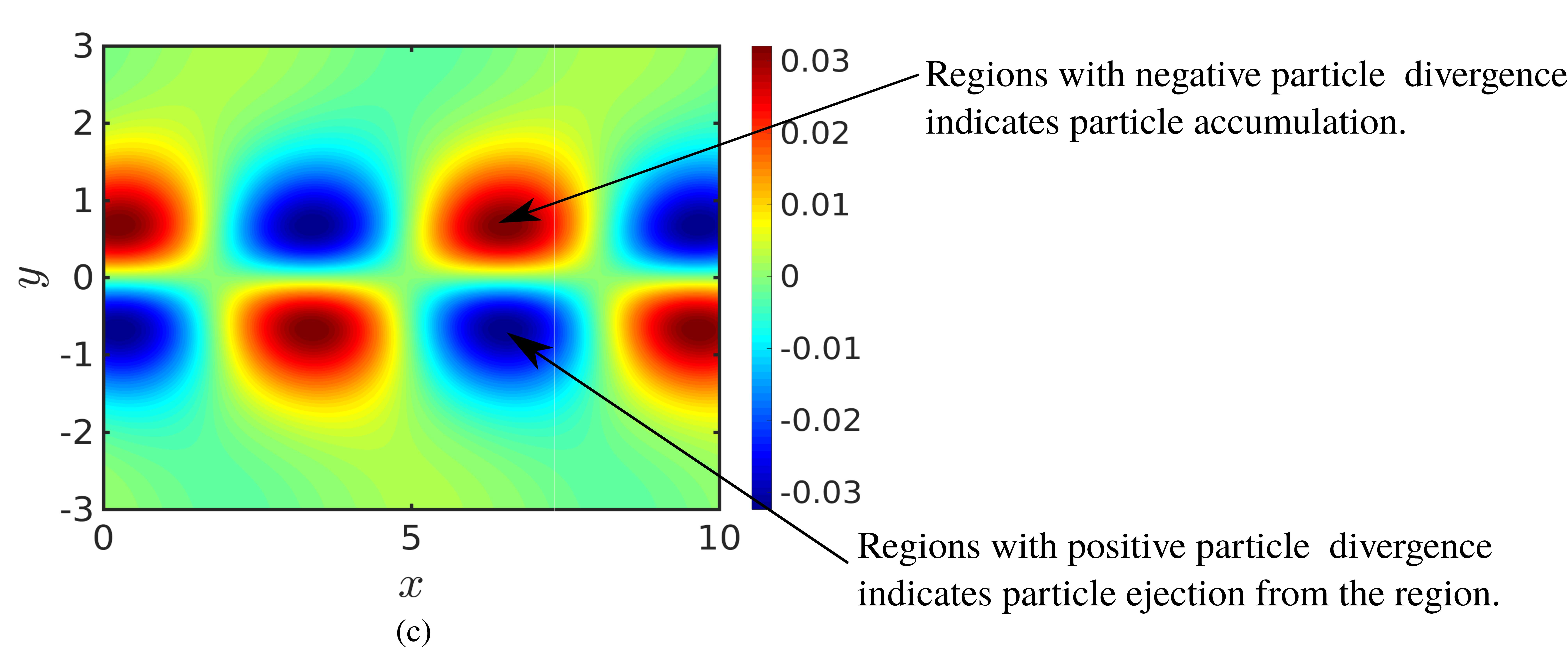}
	\caption{(a) Perturbed particle concentration field at low Stokes number ($St=10^{-5}$) for the sinuous mode. (b) As expected, the deviations from the mean concentration occurs in the shear layers. (c) Perturbed particle divergence field for sinuous mode at low stokes number $St = 0.1$. The regions of negative divergence indicate regions where particles tend to accumulate and are expelled from the positive regions. Sinuous mode is more unstable than the varicose mode, so this sort of particle accumulation is expected.} 
	\label{phi} 
\end{figure}

In figure(\ref{Re_250_st_1e_minus_5_budget}), we see that at low Stokes number ($St=10^{-5}$) the  magnitude of Stokes drag term (term \raisebox{.5pt}{\textcircled{\raisebox{-.9pt} {3}}} and term \raisebox{.5pt}{\textcircled{\raisebox{-.9pt} {6}}}) is smaller compared to that of viscous terms (term \raisebox{.5pt}{\textcircled{\raisebox{-.9pt} {2}}} and term \raisebox{.5pt}{\textcircled{\raisebox{-.9pt} {5}}}) by three orders. As stated earlier at low Stokes number, Stokes drag is negligible compared to viscous diffusion term.  At low Stokes numbers, particle velocity quickly adjusts to the local fluid flow field. From figure (\ref{phi} a,b), we see that particles laden uniformly, tends to accumulate in the upper shear layer (shear layer at $y>0$) and lower shear layer ($y<0$), alternatively as the jet flaps in the transverse direction due to the fact that sinuous modes are more unstable than varicose modes. This is more clearly illustrated in figure(\ref{phi} c,d) showing the regions with $\nabla\cdot u_{p}<0$ is where the particles tend to accumulate and particles are expelled from regions $\nabla\cdot u_{p}>0$. Varicose mode is the one where the vertical perturbation velocity is an odd function (or antisymmetric) ie., $\tilde{u}(-y)=-\tilde{u}(y)$.    
  From the analysis presented in section \ref{sec:formulation}, we see that even if we do not consider Saffman's \cite{saffman_1962} assumption of zero inertia flow, the locally parallel planar jet baseflow in itself would not have led to particle migration. Intuitively this would make sense, as since the flow is locally parallel, we do not expect the particles to migrate in the $y$ direction, which is what we see from the  analysis presented in section \ref{sec:formulation}. However particles do migrate due to the fact that planar jet profile ($Sech^{2}y$) is unstable to infinitesimal perturbations and creates a flow pattern as shown in figures(\ref{phi}c) that guides the particles towards the shear layer of the jet. Particle migration is a result of the base flow profile being unstable to infinitesimal disturbances. In the context of volume averaged equations however, the reasoning is more straight forward since the definition of divergence of a vector field itself would indicate regions where particles accumulate and  get expelled from. 

%\begin{figure}[!ht]
%	\centering
%%	\subfigure[]{\includegraphics[width=0.7\textwidth]{spectra_st_1e_minus_5_lambda_1e_minus_4.pdf}}
%\includegraphics[width=0.95\textwidth]{spectra.pdf}
%%	\subfigure[]{\includegraphics[width=0.35\textwidth]{st_1e_minus_5_M1.eps}}
%	\caption{(a) Low Stokes number planar jet spectrum showing  two unstable eigenmodes : sinuous and varicose mode at $St=10^{-5}, \Lambda=10^{-4}$ compared with the sinuous and varicose modes of the unladen flow at $k=0.3$. (b,c) Zoomed in view showing growth rate of sinuous mode with particles at low Stokes number greater than of unladen flow and similarly, the varicose mode with particles is more unstable than the unladen varicose mode.  Particle concentration $\Lambda=10^{-3}$ (one order higher) at the same wavenumber. The difference in the growth rates between unladen and particle laden sinuous and varicose modes are more clearly seen at higher particle  concentration. }
%	\label{spectra_st_1e_minus_5} 
%\end{figure}

\newpage
\clearpage
\subsubsection{Addition of particles at intermediate Stokes number ($St=1,10$) : Physical mechanism of instability}\label{sec:intermediate_st}
\begin{figure}
	\centering
	\subfigure[]{\includegraphics[width=0.28\textwidth]{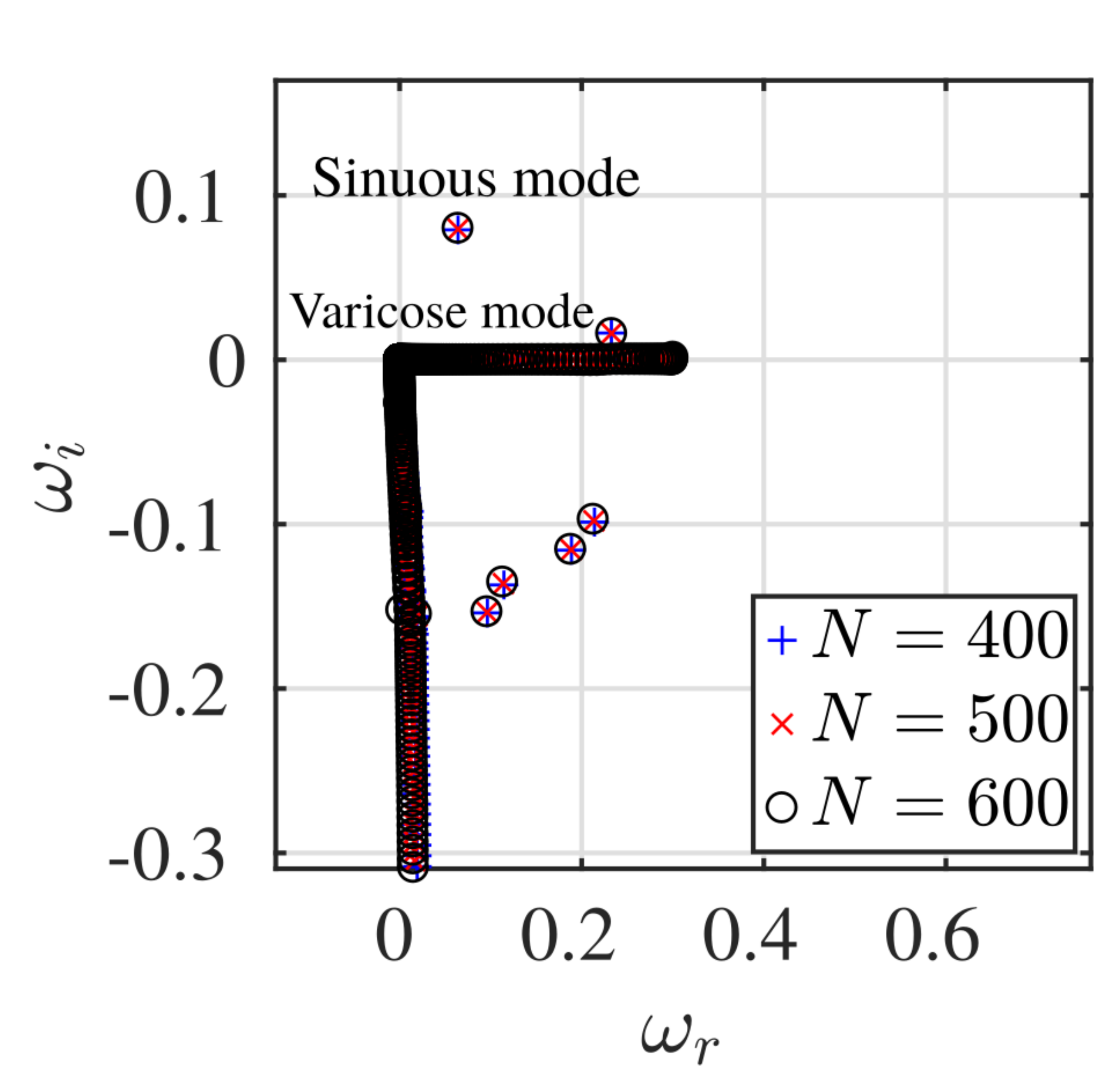}}
    \subfigure[]{\includegraphics[width=0.3\textwidth]{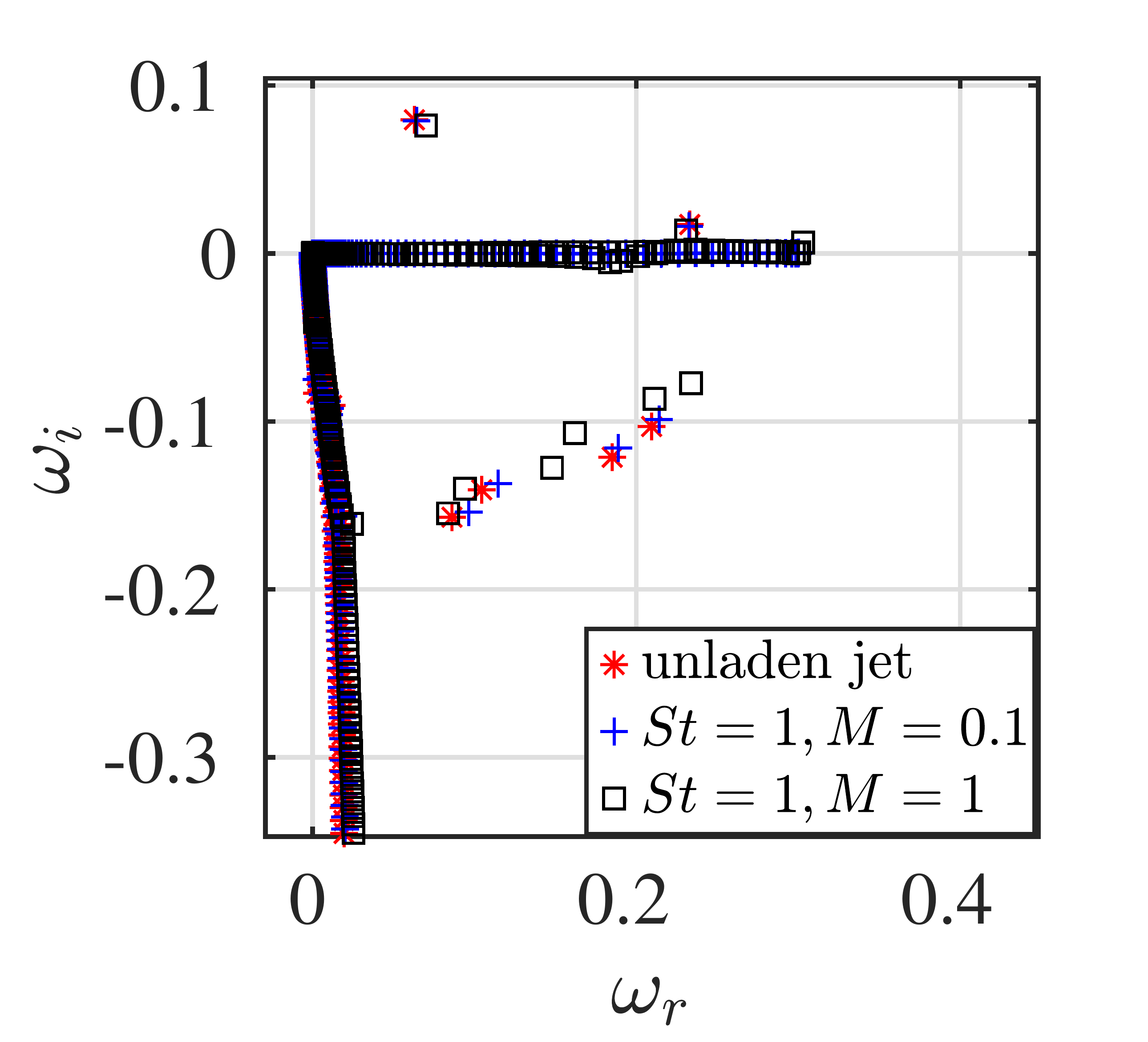}}
	\subfigure[]{\includegraphics[width=0.3\textwidth]{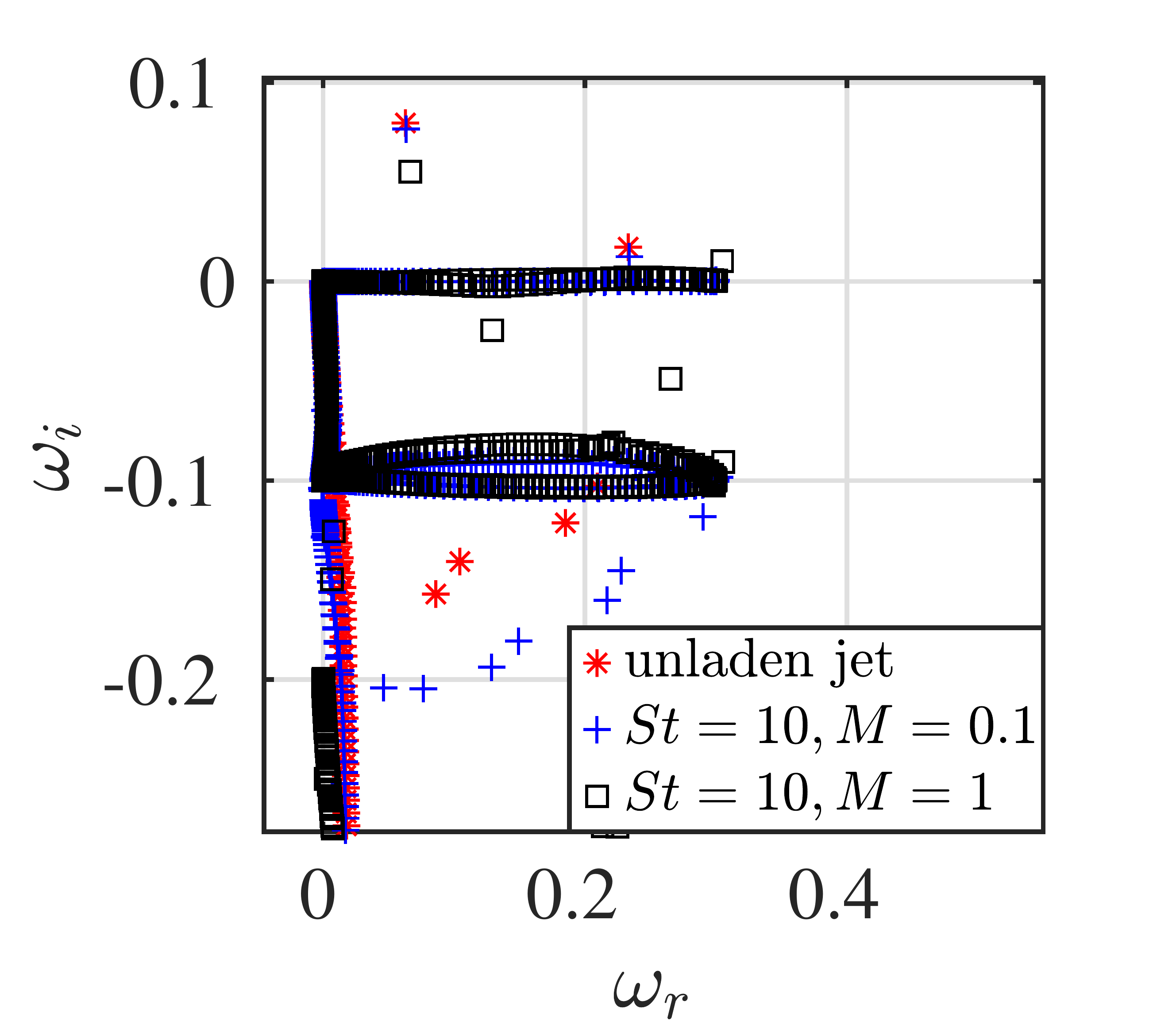}}
	\caption{(a) Eigenspectra convergence for $St=1$, at $k=0.3$. 
	(b) Intermediate Stokes number planar jet spectrum showing  two unstable eigenmodes : sinuous and varicose mode at $St=1$, $\Lambda=10^{-4}$ and $\lambda=10^{-3}$ in comparison with the sinuous and varicose modes of the unladen flow at $k=0.3$. (c) Eigenspectra for $St=10$ at $k=0.3$ showing the stable varicose mode at particle loading $M=1$. Varicose mode with particles is more unstable than the unladen varicose mode. }
	\label{spectra_st_1} 
\end{figure}

\begin{figure}[!ht]
	\centering
	\includegraphics[width=0.85\textwidth]{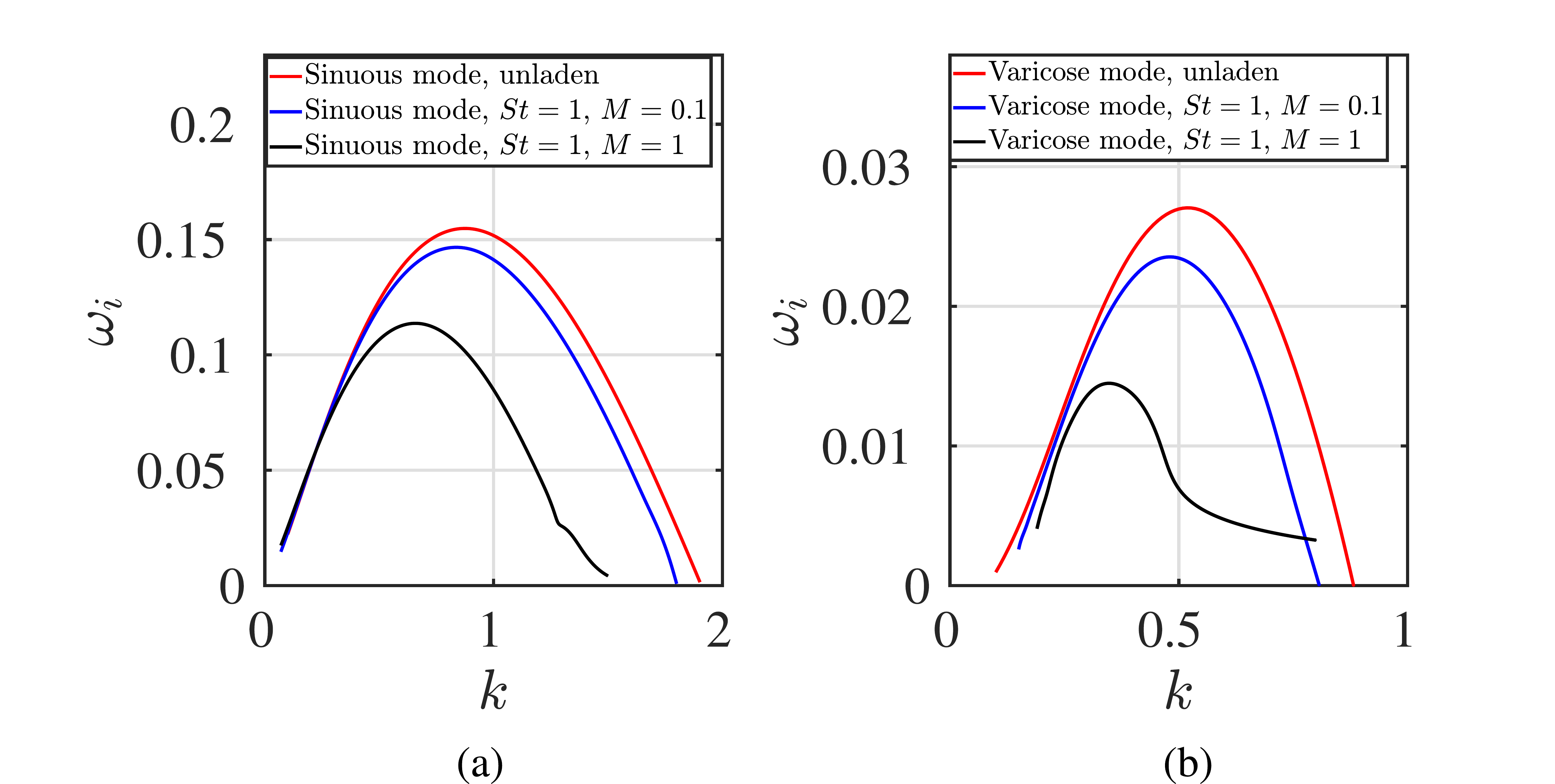}
	\includegraphics[width=0.85\textwidth]{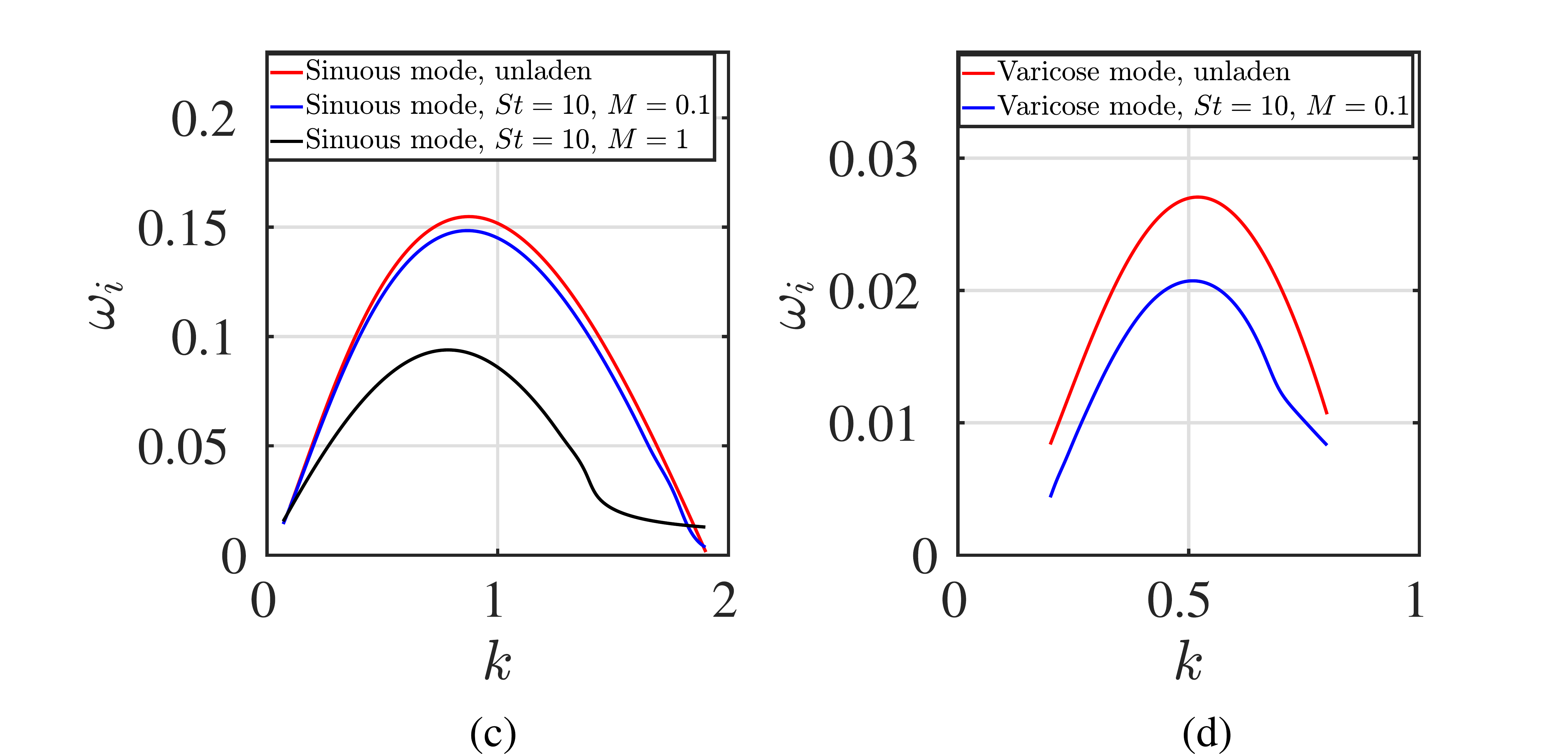}
	\caption{Intermediate Stokes number planar jet with growth rates of the sinuous and varicose mode. (a) shows $St=1, \Lambda=10^{-4}$ and $10^{-3}$ compared with the sinuous modes of the unladen flow. Growth rate of sinuous mode with particles at intermediate Stokes number is much smaller than of unladen flow. (b) Varicose modes at  $St=1, \Lambda=10^{-4}$ and $10^{-3}$. Varicose mode with particles is more stable than the unladen varicose mode and unstable over a much smaller wave number. Varicose modes have much smaller growth rates compared to sinuous modes. (c)  $St=10, \Lambda=10^{-4}$ and $10^{-3}$ compared with the sinuous modes of the unladen flow. The growth rate of sinuous mode with particles is much smaller than of unladen flow. (d) shows $St=10$, $\Lambda=10^{-4}$ and $10^{-3}$. Varicose mode with particles is unstable only at smaller mass loading (for $M<1$). For $M=1.0$, the varicose modes are stabilized. }
	\label{growthrate_st_1_10} 
\end{figure}

For intermediate Stokes number ($St=1$), the perturbed particle relaxation time ($\tau$) is of the order the convective time scale of the baseflow. Figure(\ref{spectra_st_1} a) shows the eigenspectra for $St=1$ at $k=0.3$, with mass loading  of $0.1$. Just as in the case of unladen jet, we see the unstable  sinuous and varicose modes. We see that both the sinuous and varicose modes have converged but there are a few  unstable modes which have very small growth rates compared to the sinuous mode which are not important to the present study. We see that these are modes with very slow convergence, i.e, their growth rate changes by a very small amount as the grid is refined (see \ref{spectra_st_1}a, inset).  Figure (\ref{spectra_st_1} b) shows the eigenspectra for $St=1$, with mass loading $M=\gamma\Lambda=0.1$ and $1$ at $k=0.3$ compared with that of the unladen jet eigenspectra. At $St=10$, however the varicose mode becomes stable as seen in figure (\ref{spectra_st_1} c) while the sinuous mode remains unstable. 
%Unlike in the small Stokes number regime, here we see a noticeable difference in the fluid and particle disturbance velocity field due to increase in the effective viscosity of the suspension. 
Figure (\ref{growthrate_st_1_10}) shows the growth rate for the sinuous and varicose modes modes for $St=1$ and $10$. Figure (\ref{growthrate_st_1_10} a) shows the growth rate at $St=1$, for the sinuous mode at different particle loading. Particle loading as mentioned previously given by $M=\gamma\Lambda$, is varied by keeping the density ratio constant ($\gamma=10^{3}$) and varying the particle volume fraction $\Lambda=10^{-4}, 10^{-3}$. For both particle loading at small wavenumbers (longer waves), the growth rates are very nearly equal to the growth rate of the unladen planar jet. For higher wavenumbers, growth rate with particles is lower than the unladen case. This is because perturbations of higher wavenumbers (shorter waves) are smoothed by the action of viscosity and results in decreased growth rate. We see that with increase the particle loading, Stokes drag increases, decreasing the growth rate. Figure (\ref{growthrate_st_1_10} b) shows the growth rate for the varicose mode for both particle loading mentioned above. Compared to the sinuous mode, the varicose mode has a significantly smaller growth rate as seen with the unladen case. The difference here is that the varicose modes have a higher particle concentration compared to the sinuous mode. This reduces the fluid inertia and leads to smaller growth rates compared to the unladen jet. Further at higher wavenumbers (shorter waves) the difference in growth rate is more drastic due to the fact that viscosity dampens out the shorter waves. Figure (\ref{growthrate_st_1_10} c) shows the growth rate at $St=10$, for the sinuous mode for particle loading of $0.1$ and $1.0$. At small wavenumbers, for $St=10$ and particle loading $M=1.0$, the growth rates are smaller than suspensions with $M=0.1$ and the unladen jet. Also noteworthy is the fact that for $St=10$ and $M=1.0$, the varicose mode is stabilized as shown in figure (\ref{growthrate_st_1_10} d). This is expected that as we increase the Stokes number, the effect of addition of particles is to stabilize the flow. 

%\begin{figure}[!ht]
%	\centering
%	\includegraphics[width=0.75\textwidth]{St_1_convergence.pdf}
%	%	\includegraphics[width=0.45\textwidth]{st_10_Re_250_eigenspectra.eps}
%	\caption{Eigenspectra convergence at $St=1$, $k=0.3$. Presence of marginaly stable spurious modes visible as they donot converge as the number of Chebyshev nodes are increased. These are mainly visible in the intermediate Stokes number regime.  }
%	\label{converge_spectra_st_1} 
%\end{figure}

%\begin{figure}
%	\centering
%	\includegraphics[width=0.83\textwidth]{fluid_inertia_comp.pdf}
%	\caption{Fluid inertia comparison for the varicose mode for particle loading $M=0.1,1$ with varicose mode of unladen jet. Varicose modes have a higher particle concentration and a smaller fluid inertia compared to the unladen jet. } 
%	\label{Re_250_st_1_varicose_fluid_inertia}  
%\end{figure}

%\begin{figure}[!ht]
%	\includegraphics[width=1.0\textwidth]{M1_e1_st_1_Re_250_sinuous_mode_total.pdf}
%	\includegraphics[width=1.0\textwidth]{M1_e1_st_1_Re_250_varicose_mode_total.pdf}
%	\caption{Eigenmode shapes of the sinuous and varicose modes at Intermediate Stokes number $St=1$ and  $\Lambda=10^{-4}$ at  $k=0.3$. (a), (b) and (c) shows the mode shapes of the sinuous modes while (d), (e) and (f) shows the mode shapes of the varicose modes. (a) shows the mode cross streamwise disturbance velocities of the fluid and particles while (b) shows the jet streamwise disturbance velocities of the fluid and particles. (c) shows the perturbed particle volume fraction field and at either ends of the jet shear, the volume fraction field peaks. (d), (e) and (f) shows the cross streamwise, streamwise and particle volume fraction fields respectively for the varicose mode. Particle concentration is higher in the varicose mode than the sinuous mode.  }
%	\label{st_1_eigenmodes} 
%\end{figure}

\newpage
\clearpage
\subsubsection{Addition of particles at large Stokes number ($St=100$) : Physical mechanism of instability}\label{sec:large_st}
We first asses the validity of the formulation at large Stokes number. Stokes number is defined as the ratio of particle relaxation time $\left(\tau=\frac{2}{9}\frac{a^2\rho_{s}}{\mu}\right)$ to the flow convective time scale $\left(t_{c}=\frac{L}{U_{0}}\right)$, where the characteristic length scale $L$ is given by $\frac{Q}{2U_{0}}$ with $Q$ being the volume flux per unit width of the jet and $U_{0}$ being the centreline velocity of the jet. Stokes number in terms of volume fraction is given by
\begin{equation}
St=\frac{1}{18}\Lambda^{2/3}\gamma Re.
\end{equation}
Maximum allowed concentration for dilute suspension is $\Lambda=10^{-3}$ (one part in thousand is usually  considered dilute enough. See Elghobashi \cite{Elghobashi1991} and \cite{Elghobashi1994} for particle concentration values for different regimes). Putting $\gamma=10^3$, $Re=250$ yields, $St\sim O(100)$. This is consistent with the estimates of Narayanan et al.  \cite{Narayananetal2002} for largest Stokes number allowed for the case of particle laden mixing layers. Moreover we have already shown that for small times (time period during which linear analysis is valid), particle path lines in a locally parallel planar jet do not cross each other and hence do not result in accumulation of particles.  We note from the formulation that the density ratio $\left(\gamma=\frac{\rho_s}{\rho_f}\right)$ has to be large so that added mass, Basset force can be ignored for which we set $\gamma=1000$ atleast. To study large Stokes number behaviour (at constant $Re$), we see  that Stokes number can be increased by increasing density ratio or by increasing the particle size ($St\sim a^2$). We discuss both the routes at large Stokes number below
%\newpage
\subsubsection{Large Stokes number by increasing particle size} \label{sec:large_st_increasing particle_size}
\begin{figure}[!ht]
	\includegraphics[width=1\textwidth]{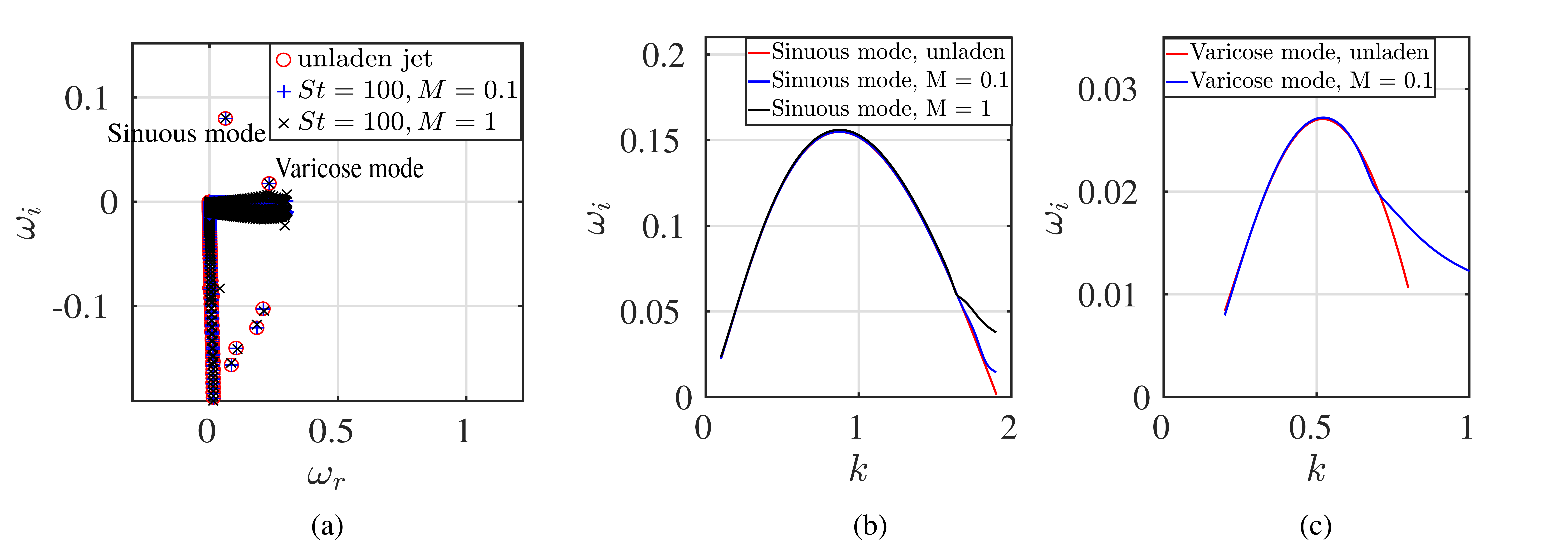}
	\caption{(a) Eigenspectra showing sinuous and varicose modes at large Stokes number  $St=100$, for different particle loading of $M=0.1, 1$ and $Re=250$, against the unladen flow by increasing particle size. Large Stokes numbers achieved by increasing particle size (particle diameter $d_p$) is accompanied by a reduction in the number of particles. The effect of reducing the number of particles is far more drastic than the increase in particle diameter resulting in the growth rates of particle laden jet similar to the unladen jet as seen in (b) and (c).} 
	\label{st_100_dispersion_relation_size}  
\end{figure}

Figure (\ref{st_100_dispersion_relation_size}) shows the growth rates for different particle loading at large Stokes numbers by increasing the particle diameter while keeping the same volume fraction for a given particle loading. We see that the growth rates with particles is nearly the same as that of the unladen flow. Compared to the intermediate Stokes number regime, the growth rate has increased as Stokes number is increased to $100$. This is due to the fact that any increase in diameter of the particle is accompanied by a decrease in the number of particles. This results in smaller dissipation compared to increasing Stokes number by increasing density ratio. This decrease in the number of particles has an overwhelming destabilization effect compared to the stabilizing effect due to increased diameter of the particle and thus the growth rates are closer to the unladen growth rate.

\newpage
\subsubsection{Large Stokes number by increasing density ratio} \label{sec:large_St_increasing density ratio} 
\begin{figure}[!ht]
	\centering
	\includegraphics[width=0.98\textwidth]{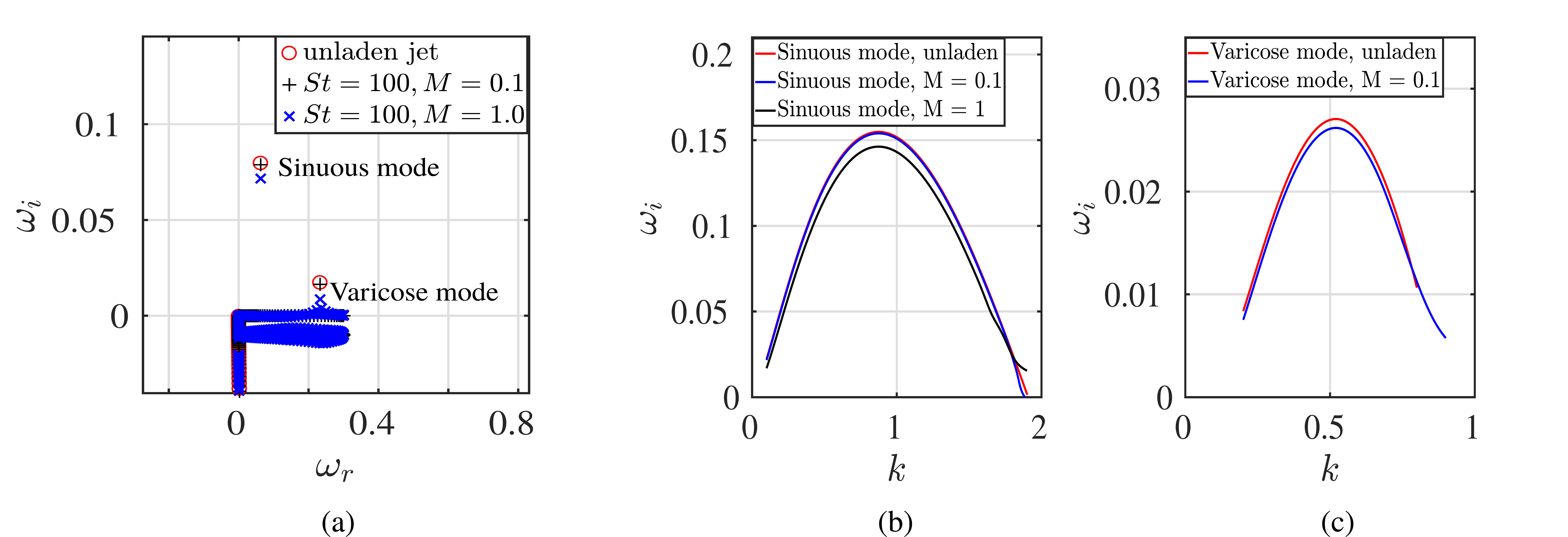}
	\caption{(a) Eigenspectra showing sinuous and varicose modes at large Stokes number  $St=100$, for different particle loading of $M=0.1, 1$ and $Re=250$, against the unladen flow by increasing density ratio $\left(St\sim\gamma\right)$. Increasing density ratio leads to the reduction in growth rates compared to unladen flow as seen in (b) and (c).} 
	\label{st_100_dispersion_relation_density_ratio}  
\end{figure}

\begin{figure}[!ht]
	\centering
	\subfigure[]{\includegraphics[width=0.33\textwidth]{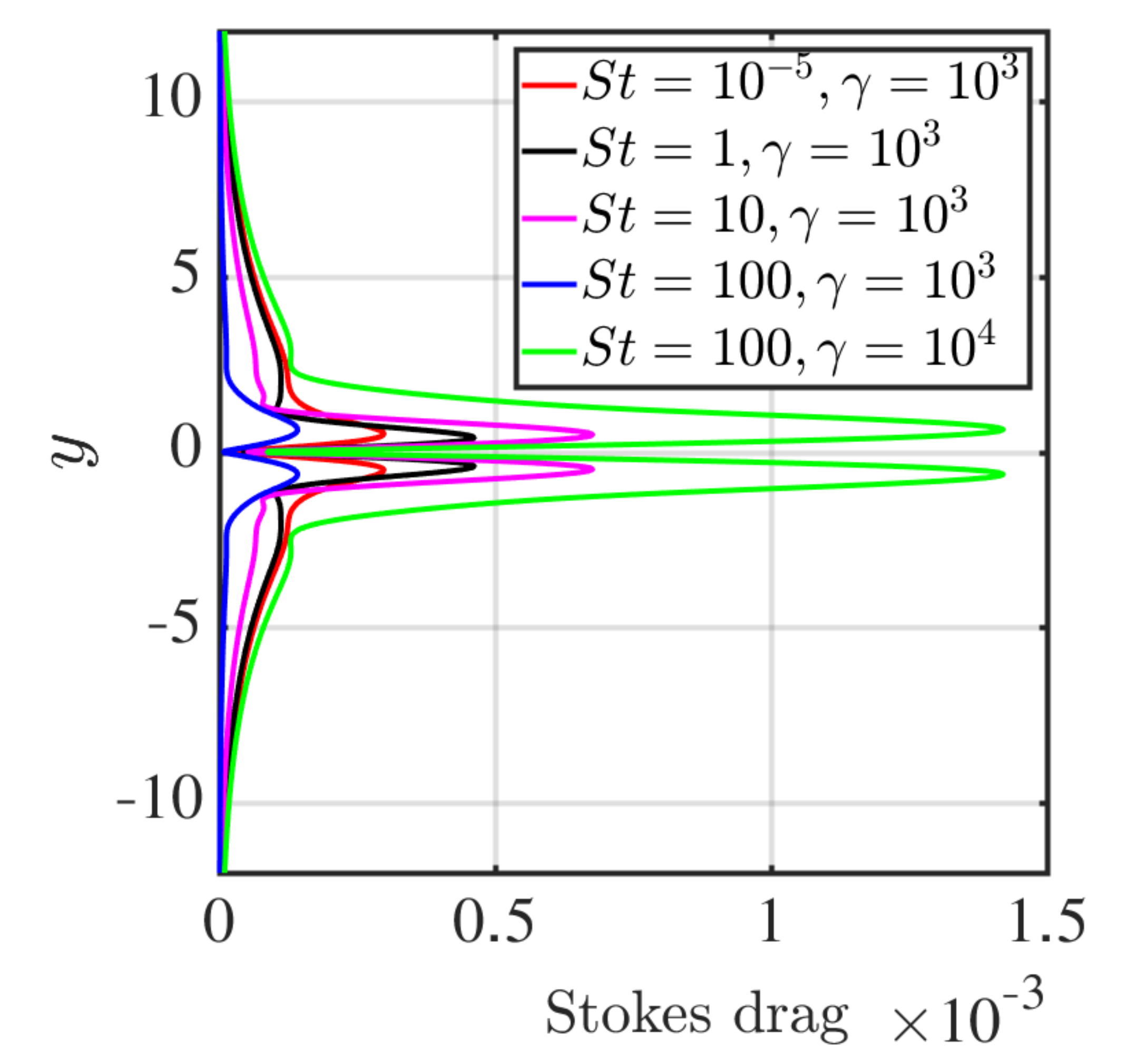}}
	\subfigure[]{\includegraphics[width=0.35\textwidth]{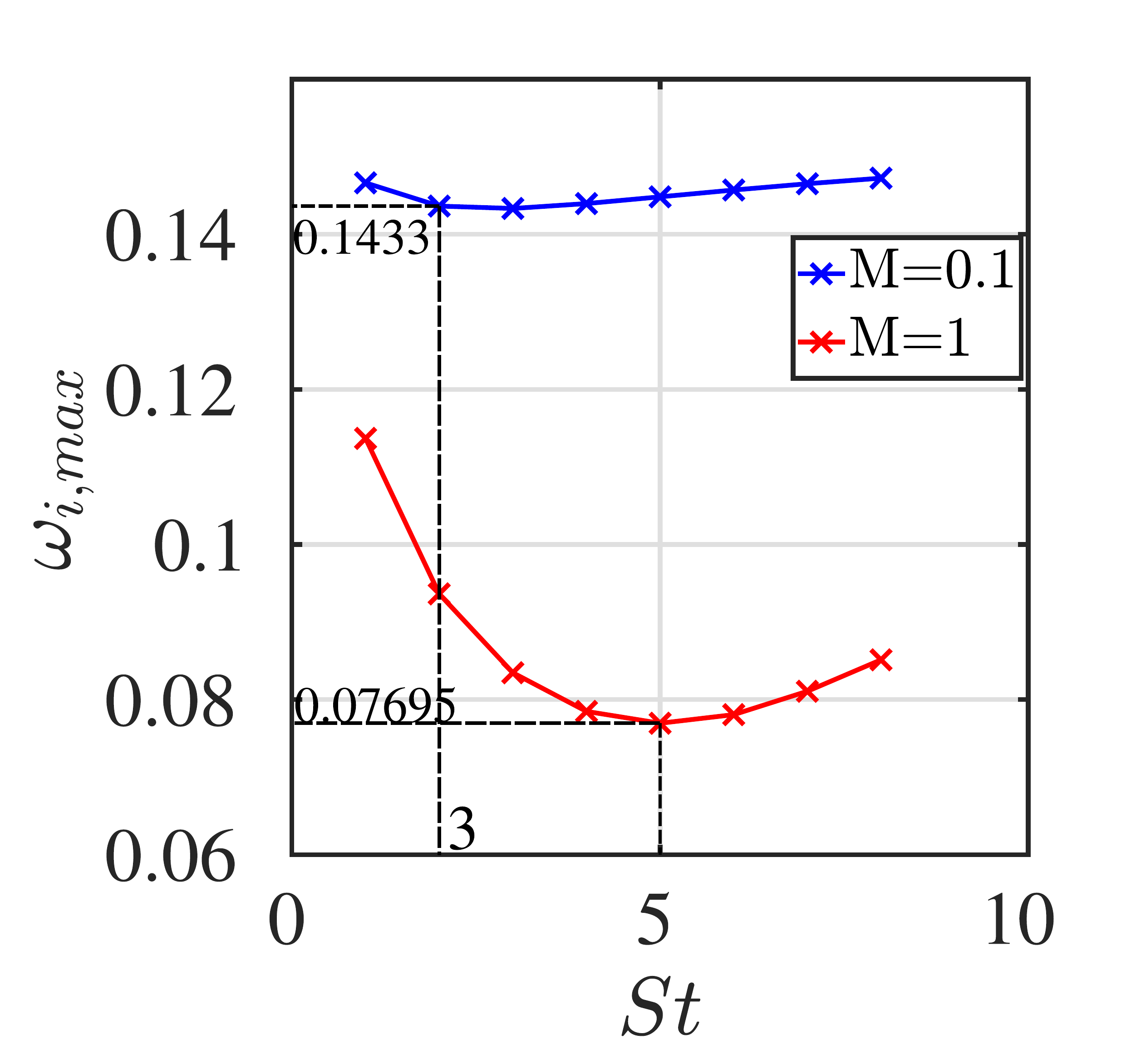}}
	\caption{(a) Stokes drag for various Stokes numbers for constant particle loading. For low Stokes number of $St=10^{-5}$ and $\gamma=10^{3}$, Stokes drag is low (given by red line). As Stokes number increases ($St = 1$ and $10$), Stokes drag increases (given by black line and magenta line respectively). Increasing Stokes number to 100 by increasing particle size decreases the Stokes drag (given by blue line) while for the same Stokes number increasing density ratio ($10^{4}$ in this case), Stokes drag increases. (b) Shows the variation of maximum growth rate for the sinuous mode in the intermediate Stokes number regime ($St=1$ to $10$). The particles have stabilizing action in this regime. For particle loading of $M=\gamma\Lambda=0.1$, growth rate is minimum at $St=3$ and for particle loading of $M=\gamma\Lambda=1.0$, growth rate is minimum at $St=5$.} 
	\label{stokes_drag_sinuous}  
\end{figure}

Figure (\ref{st_100_dispersion_relation_density_ratio}) shows the growth rates for different particle loading at large Stokes number with increasing density ratio. We see that addition of particles at large Stokes number by increasing the density ratio leads to reduction in the growth rate compared to the unladen flow. The reduction in the growth rate is due to the fact that there is dissipation of energy on account of larger response times to fluctuations of the baseflow. Figure (\ref{stokes_drag_sinuous}a) shows the Stokes drag comparison for different Stokes number regimes. For low Stokes number, the drag force is small and as Stokes number increases the Stokes drag increases (for the intermediate range). At large Stokes number, increasing the Stokes number by increasing the size of the particle, reduces the drag force. While increasing the Stokes number by increasing density ratio increases the drag force. Addition of particles has the most stabilizing effect at the intermediate Stokes numbers as shown in fig(\ref{stokes_drag_sinuous}b). Peak growth rate for different particle loading in the intermediate regime is plotted against Stokes numbers. For mass loading of $0.1$ (Concentration $\Lambda=10^{-4}$,$\gamma=10^{3}$) growth rate is the lowest for $St=3$.  While at mass loading of unity, lowest growth rate occurs at $St=5$.   

%\begin{figure}[!ht]
%	\centering
%	\includegraphics[width=0.4\textwidth]{critical_St.pdf}
%	\caption{} 
%	\label{critical_st}  
%\end{figure}

\newpage
\subsection{Non uniform particle laden planar jet}
\label{sec:vpl}
\begin{figure}[!ht]
	\centering
	\includegraphics[width=0.7\textwidth]{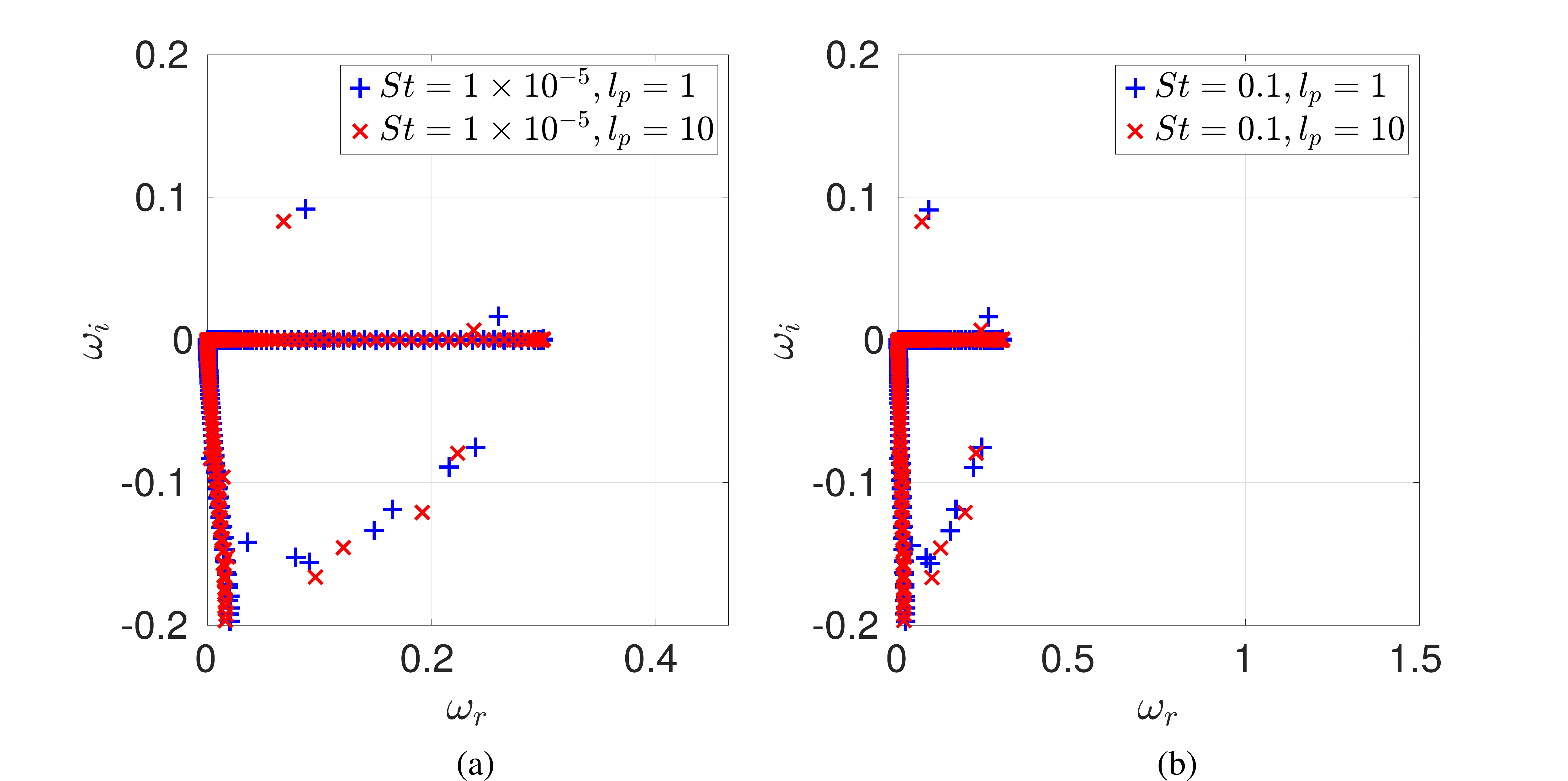}
	\includegraphics[width=0.7\textwidth]{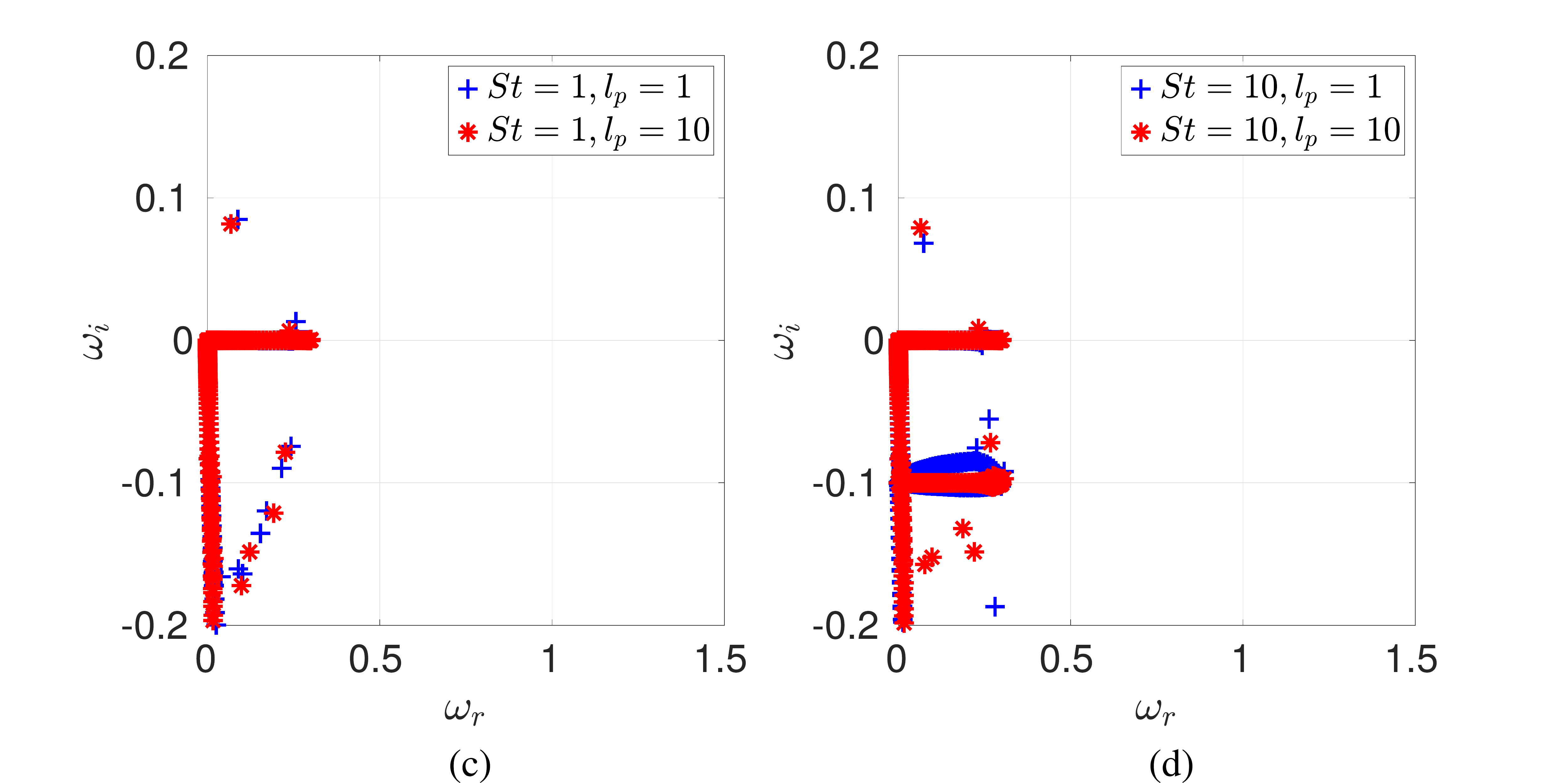}
	\includegraphics[scale=0.25]{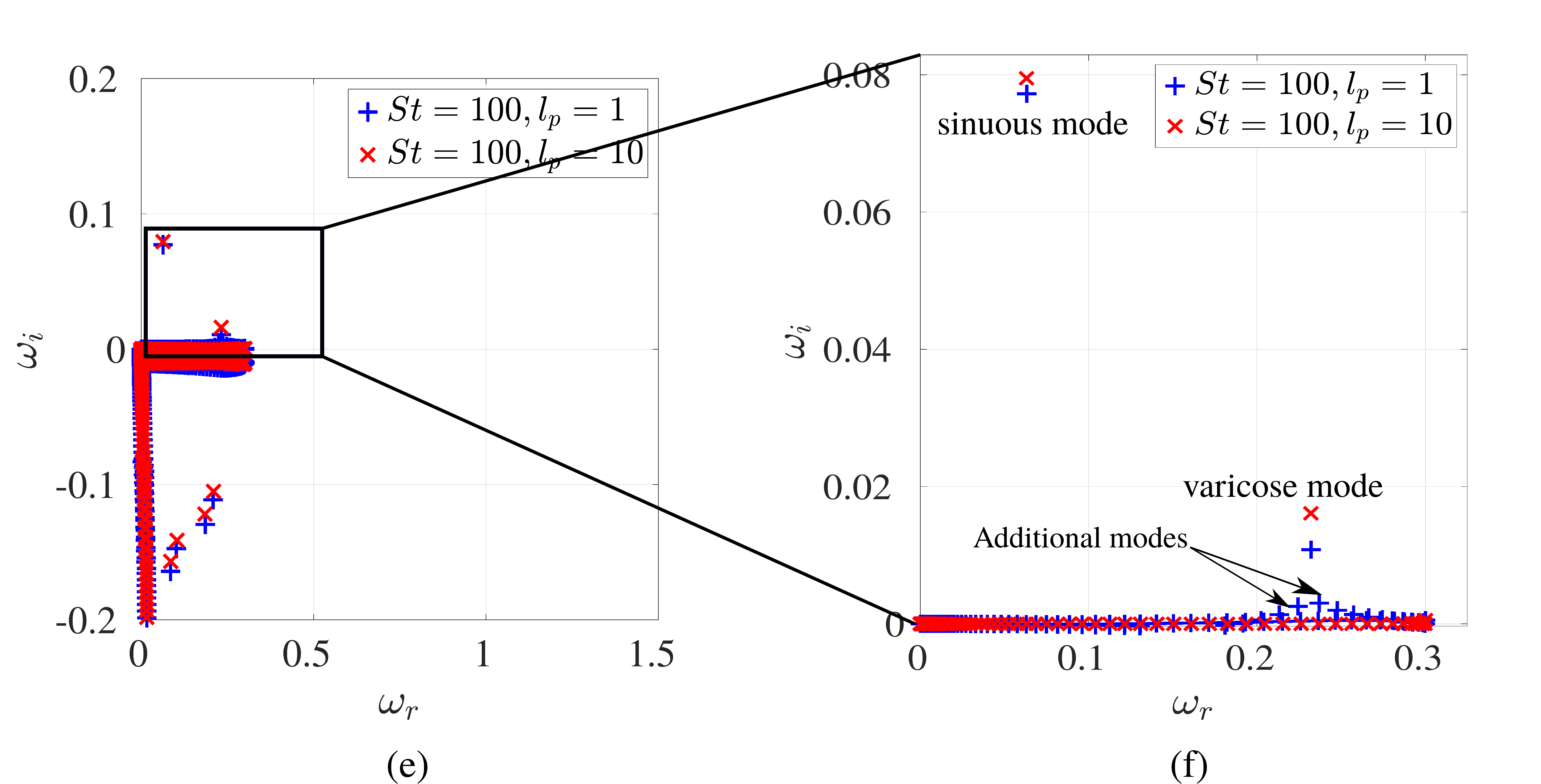}
	\caption{Variable particle loading growth rates for the sinuous modes with  steepness parameter $l_{p}=1,10$. For small Stokes number $St=1\times 10^{-5}$, $St=0.1$ (as in fig (a), (b)) we see sinuous mode and varicose mode just as in the case of constant particle loading. In the intermediate Stokes number regime ((c) and (d)) at $St=10$ and $l_{p}=10$, we see only the sinuous mode and the varicose mode is damped, while for $l_{p}=1$ both the sinuous and varicose modes are unstable. At large Stokes number ((e) and (f)), apart from the sinuous and varicose modes, we see additional set of unstable modes.} 
	\label{vpl_st_spectra}  
\end{figure}

\begin{figure}
	\centering 
	\includegraphics[width=0.8\textwidth]{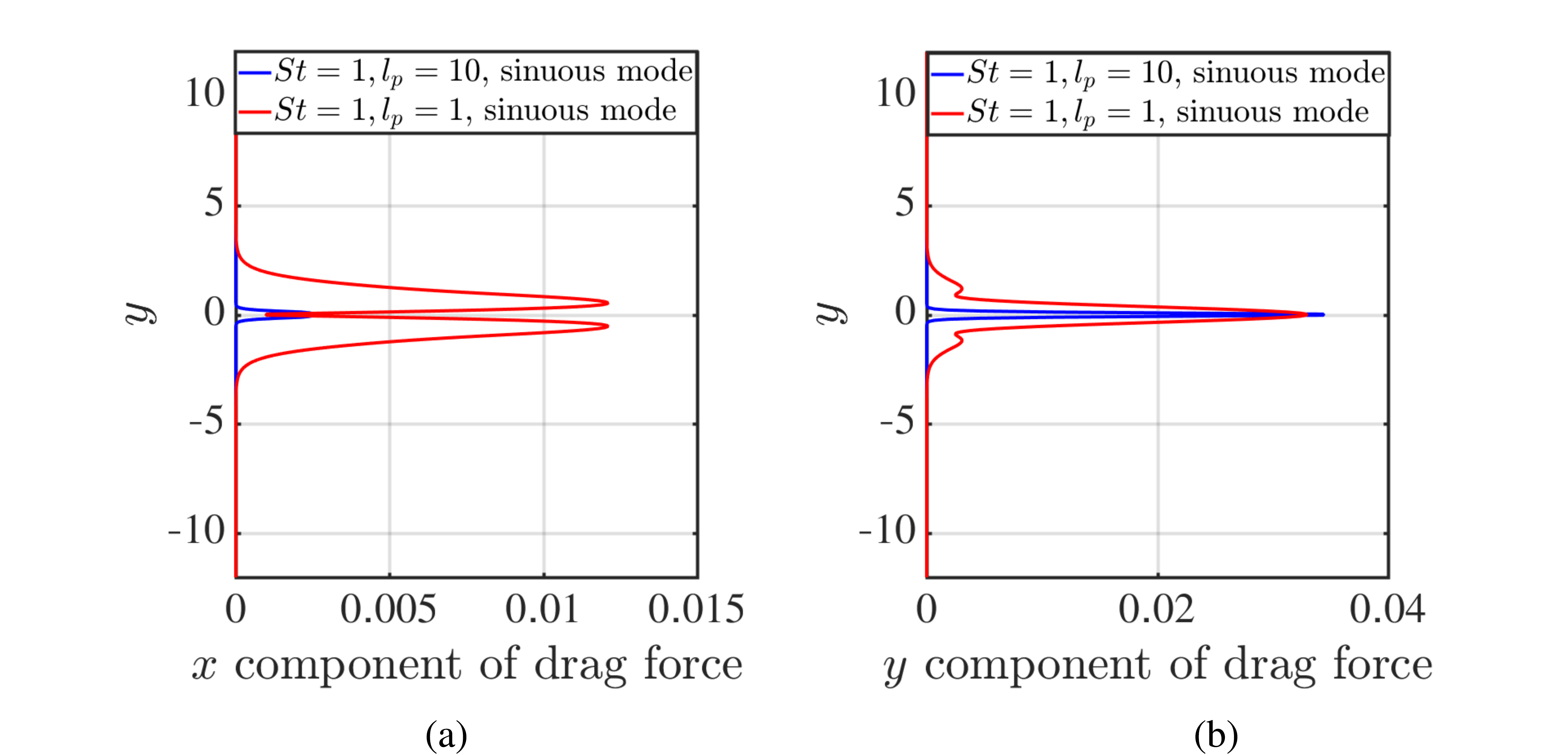}
	\includegraphics[width=0.75\textwidth]{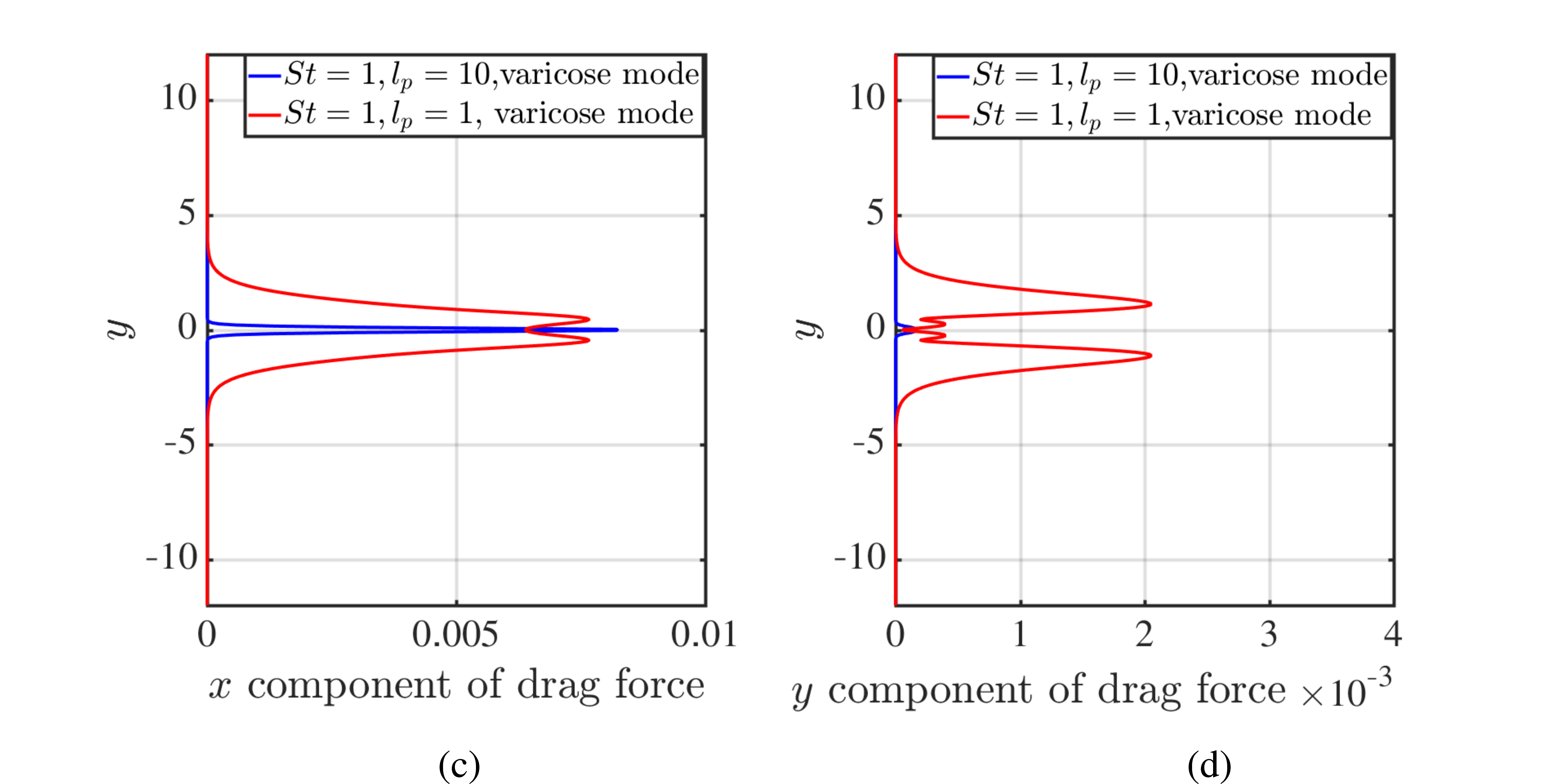}
	\caption{Drag force comparison at $St=1$ for $l_p=1$ and $l_p=10$ at $k=1$. For the sinuous mode, Stokes drag is greater for $l_p=1$ compared to $l_p=10$ resulting in reduced growth rate for $l_p=1$. For varicose mode, $l_p=10$ has a higher drag which reduces the growth rate. The increase in drag force slows down the bulging of the jet.} 
	\label{vpl_sinuous_St_1_drag} 
\end{figure}

We consider the base state particle concentration to be a function of the cross streamwise jet direction given by,
\begin{equation}
\Lambda(y)=\overline{\alpha}sech^{2}(l_{p}y),
\label{planar_jet_concentration}
\end{equation} 
\begin{figure}[!ht]
	\centering
	\includegraphics[width=0.65\textwidth]{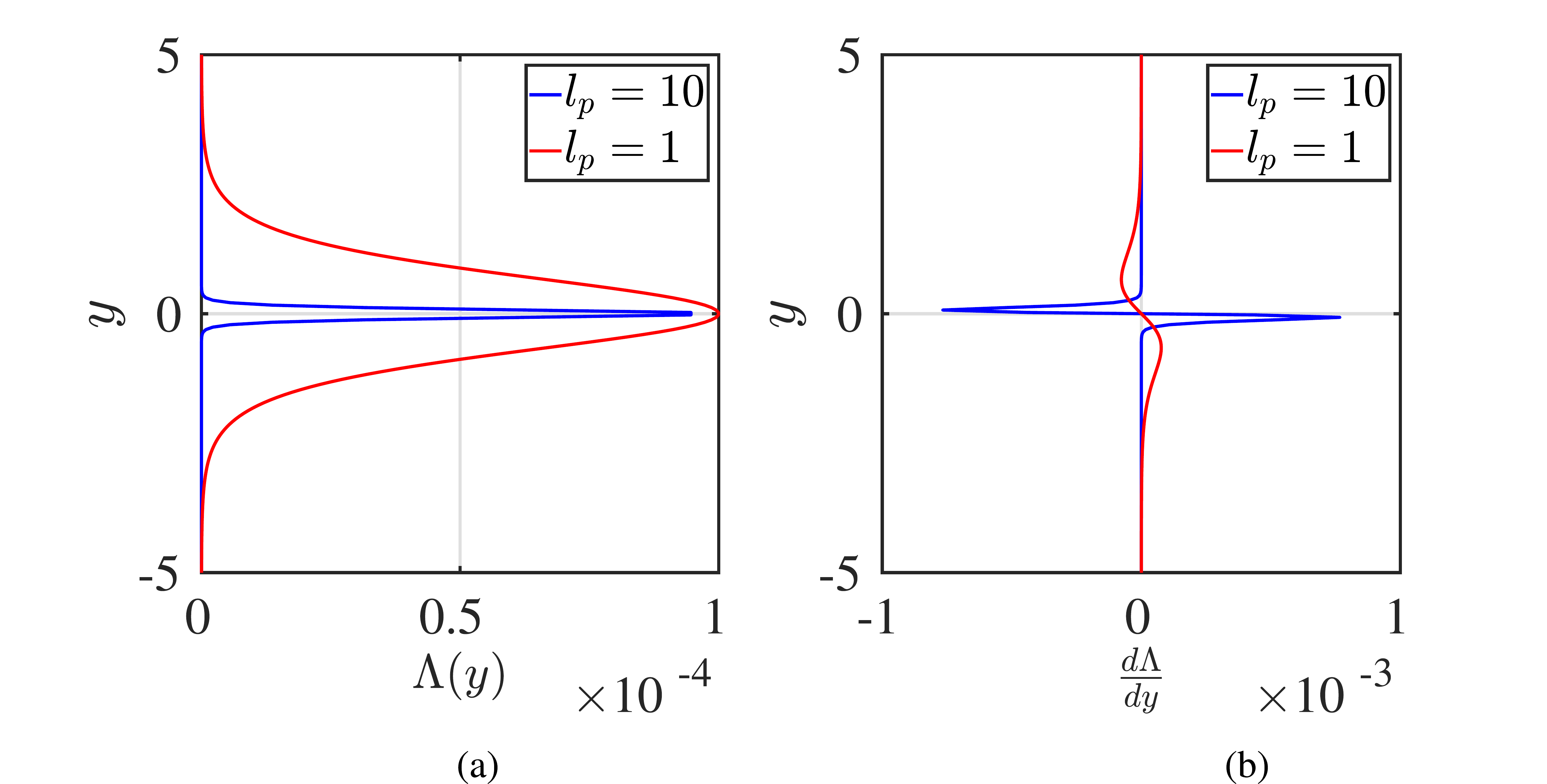}
	\caption{Base state concentration profile $\Lambda$ as a function of the cross streamwise jet direction given by Eq. (\ref{planar_jet_concentration}) with $\overline{\alpha}=10^{-4}$. The steepness parameter $l_{p}$ is taken as $1$ and $10$. Higher value of $l_{p}$ results in larger concentration gradient.} 
	\label{basestate_lp}  
\end{figure}
where $\overline{\alpha}$ is a constant taken to be $10^{-4}$ to ensure the concentration is dilute. The parameter $l_{p}$ controls the steepness of the concentration as shown in figure (\ref{basestate_lp}). Particle concentration is maximum at the centreline and decays to zero outside the shear layer. Parameter $l_p=10$, results in particle concentration confined to the neighbourhood of the jet centreline, while $l_p=1$, results in a profile that is less steeper and decays to zero gradually. Figure (\ref{vpl_st_spectra}) shows the eigenspectra with variable particle loading at $k=0.3$, for steepness parameter $l_p=1$ and $10$. For small Stokes number $St=1\times 10^{-5}$ and $St=0.1$, growth rate for $l_p=1$ is greater than that of $l_p=10$ which is true for both the sinuous and the varicose modes. For $l_{p}=1$, the growth rate with particles exceed the unladen jet similar to the constant loading case (see appendix (\ref{CPL_app}) for growth rate curves for different values of the steepness parameter). For $l_{p}=10$, the growth rates are smaller than the unladen jet. 
In the intermediate Stokes number regime at $St=1$ and $10$, addition of particles leads to reduction in temporal growth rates for both $l_{p}=1$ and $10$. For the sinuous mode, the growth rate for $l_p=1$, is smaller than that of $l_p=10$. This is because of the increased drag force for $l_p=1$ profile as shown in figure (\ref{vpl_sinuous_St_1_drag}). At higher wavenumbers, the reduction in the growth rate is more prominent due to the fact that viscosity dampens (smears) out shorter waves (higher wavenumber fluctuations) more effectively than the longer waves. From figure (\ref{vpl_st_spectra}) for $St=10$ and $l_p=1$, we see that the varicose mode is stabilized. Previously we observed that for constant particle loading, that at $St=10$, the varicose mode is stabilized. This is seen in variable particle loading too.  For the varicose modes, the growth rate at intermediate Stokes for $l_p=10$, is more damped compared to a base flow concentration profile with $l_p=1$.  This is reflected in the increased drag force for $l_p=10$, in the vicinity of the centreline owing to that fact that a higher value of steepness parameter implies higher concentration near the jet centreline. This increase in the drag force dampens the bulging of the mode. 

The large Stokes number limit ($\sim100$) behaviour is similar to that observed for constant particle loading.  As discussed previously, increasing Stokes number by increasing the density ratio, leads to a reduction in growth rate of the modes as particles have insufficient time to respond to fluctuations in the base state. In the increasing size scenario, the number density of the particles has to reduce to ensure dilute suspension. This results in smaller dissipation compared to increasing Stokes number by increasing particle size.  This is shown in figure (\ref{vpl_st_spectra}). For the varicose mode, the behaviour is similar to the sinuous mode, in that the drag force decreases as the steepness parameter increases. Increase in the steepness parameter results in a higher concentration near the jet centreline while at the shear layers the concentration is lower and therefore the bulging of the shear is less damped than in the case of $l_p=1$, making it more unstable.  

\begin{figure}[!ht]
	\includegraphics[scale=0.2]{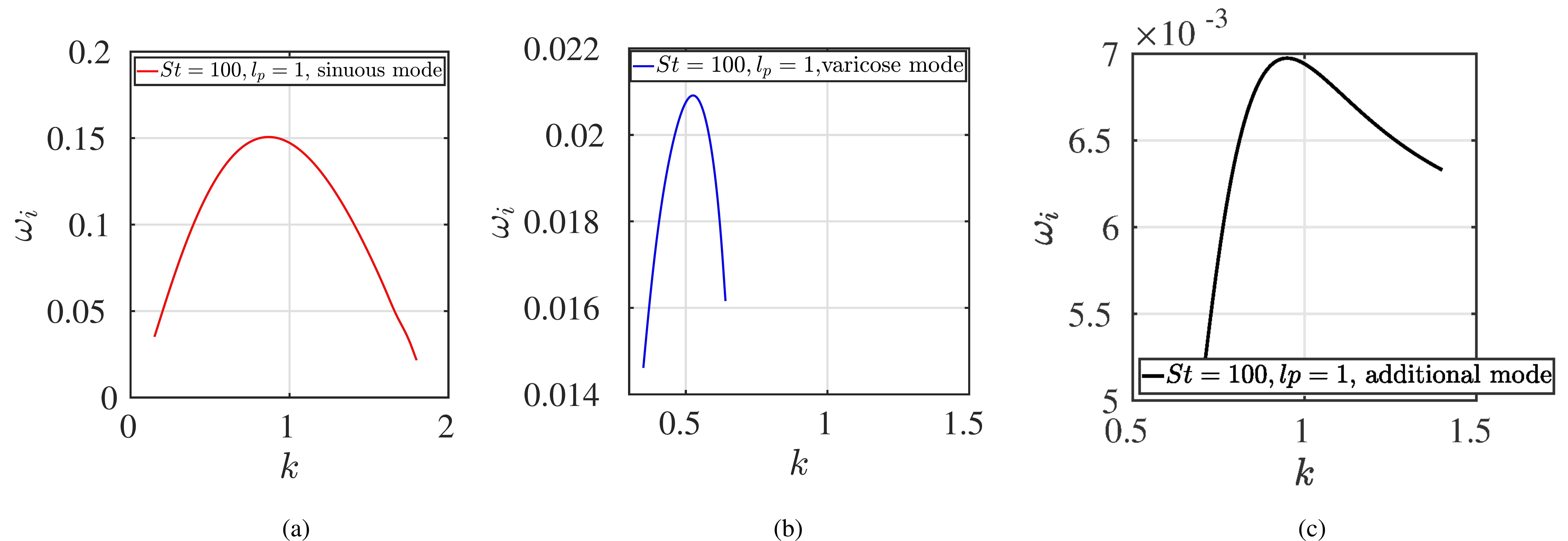}
	\caption{Growth rate vs wavenumber for (a) sinuous mode, (b) varicose mode and (c) additional mode. The additional mode has a significantly smaller growth rate compared to sinuous and varicose modes.} 
	\label{vpl_additional_modes_growth_rates}  
\end{figure}

\begin{figure}
	\centering
	\includegraphics[width=1\textwidth]{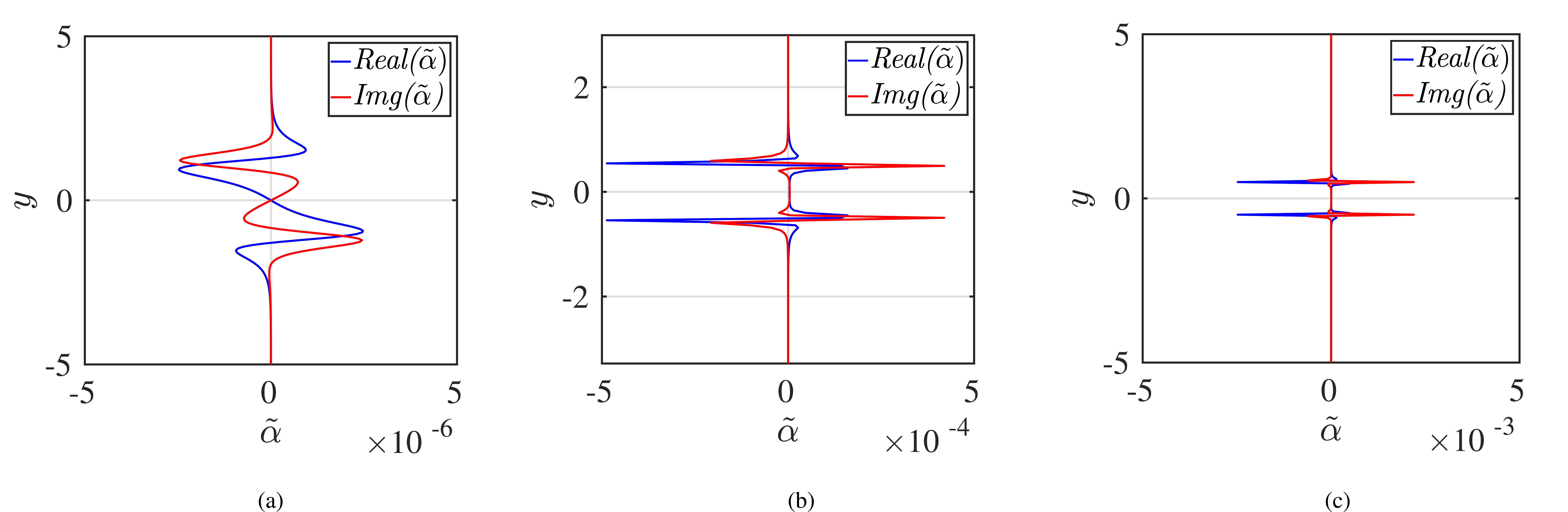}
	\caption{Particle concentration eigen mode shapes for (a) sinuous mode, (b) varicose mode and (c) additional mode at $St=100$, $k=0.3$, $l_p=1$, $\gamma=10^{4}$. The magnitude of the fluctuations in the sinuous mode and the varicose mode differs by two orders while for the additional mode, the particle concentrations are three orders greater than sinuous and an order of magnitude greater than the varicose mode.} 
	\label{vpl_phi_modes}  
\end{figure}

\begin{figure}[!ht]
	%	\centering
	\includegraphics[width=0.85\textwidth]{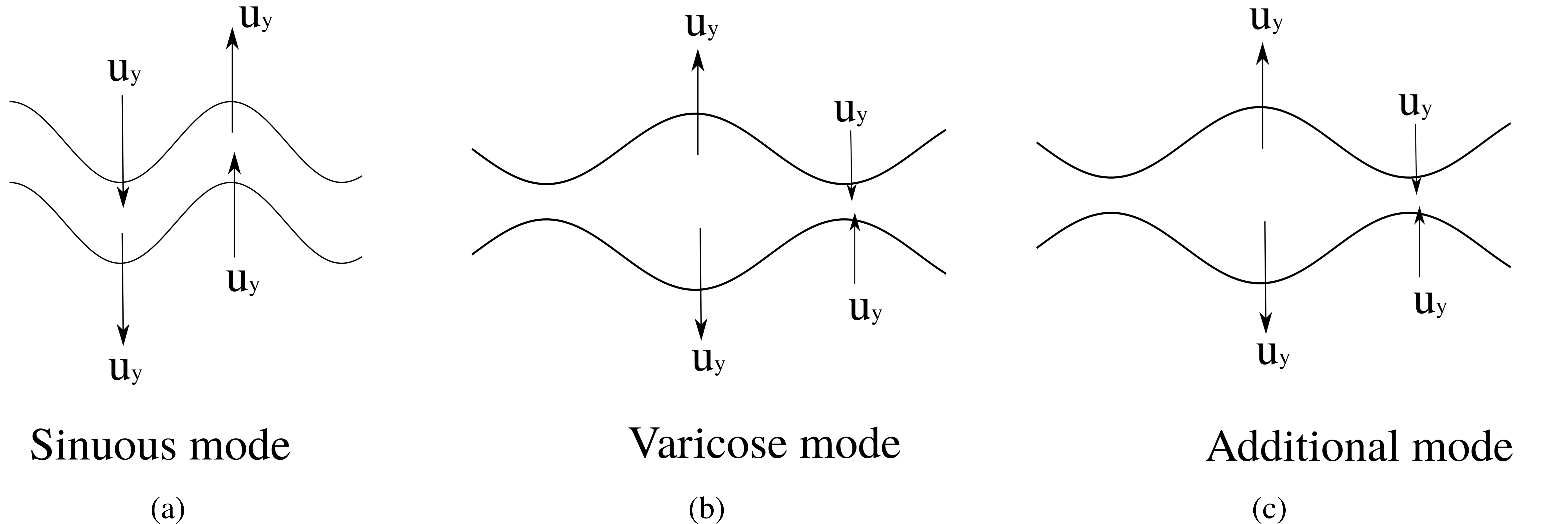}
	\caption{Pictorial representation of sinuous, varicose and the additional modes. (a) Sinuous mode are symmetric modes  with $\tilde{u}_{y}(-y)=\tilde{u}_{y}(y)$. (b) Varicose modes are antisymmetric modes  with $\tilde{u}_{y}(-y)=-\tilde{u}_{y}(y)$. (c) The additional mode is similar to the varicose mode in that they are antisymmetric modes but has a greater disturbance particle concentration.} 
	\label{vpl_additional_mode_shapes}  
\end{figure}

In the context of particle laden mixing layer with non uniform loading,  Evans et al. \cite{evans1994},  Narayanan et al. \cite{Narayananetal2002} (reproduced in the validation section in the present work) and  Giacomo et al. \cite{Giacomo_2015}  reported an additional set of unstable modes at large Stokes number and high density ratios which have very small growth rates and attributed them to Holmboe instability. In the present case of particle laden planar jets with non uniform loading at large Stokes numbers (via the increasing density ratio path) for $l_p=1$, we see additional modes apart from the sinuous and varicose modes as previously seen in figure (\ref{vpl_st_spectra}e,f) . In the present work we consider only the most unstable mode among the set of unstable additional modes. These modes are only seen in the case of variable particle loading suggesting that they are distinct from Shear layer instability.  The additional modes are very similar to the varicose modes in that the eigen solutions are antisymmetric (cross streamwise velocity perturbations are odd functions while the streamwise fluctuations are even functions). However they differ from the varicose modes in that the particle concentrations are an order of magnitude larger than the varicose mode as shown in figures (\ref{vpl_phi_modes}) and (\ref{vpl_additional_mode_shapes}). At large Stokes numbers, the convective flow of the continuous phase (fluid phase) does not have sufficient time to influence the motion of particles as the particle takes a longer time to respond to the flow.  Further, as seen in figure (\ref{vpl_additional_modes_growth_rates}), the growth rate of the additional modes are significantly lower than the sinuous and varicose mode. The disturbance particle concentration increases in the shear layers which locally increases the effective viscosity of the suspension and dampens out the vortices in the shear layer.

\vspace{1cm}
\newpage
\section{Conclusions}
\label{sec:conclusion}
 We explore the linear local stability characteristics of a planar jet laden with particles. The present formulation consists of terms in addition to the ones in Narayanan et al. \cite{Narayananetal2002}. The closure term arising from volume averaging is neglected assuming the suspension to be dilute and particle Reynolds number to be very small. Linear temporal analysis carried out for the unladen planar jet shows two unstable modes viz sinuous and varicose modes which are symmetric and antisymmetric modes (flapping and bulging modes) respectively. Sinuous mode is more unstable compared to the varicose mode. These modes are found to qualitatively similar to the experimental results of Matsubara et al. \cite{matsubara_alfredsson_segalini_2020} valid for short distances downstream beyond which non parallel effects begin to dominate. Addition of particles (uniform particle loading) at low Stokes numbers (fine particles) results in growth rates greater than that of the unladen jet. This is because of the fact that at low Stokes numbers, Stokes drag is negligible compared to viscous diffusion.  Addition of particles results in decrease in effective kinematic viscosity making the suspension more unstable than the unladen jet. This has been observed by Saffman \cite{saffman_1962} in the context of dusty channel flow. Xie et al. \cite{xie_09} arrived at the same conclusion using the equation proposed by Saffman for the same velocity profile that is used in the present study (Bickley jet). We observe similar behavior for the varicose mode for small Stokes numbers. However for large Reynolds numbers ($\geq2000$) and Stokes number of $0.1$, particles have a stabilizing effect. In the intermediate Stokes number regimes, addition of particles have a stabilizing effect on both sinuous and varicose modes. Calculations at $St=10$, shows that the varicose mode is damped. Increasing the Stokes number further can be done via two paths namely, increasing particle size path and increasing density ratio path. At $St=100$, both sinuous and varicose modes show increasing growth rates almost equal to the unladen jet. This is due to the fact that increase in diameter of the particle is accompanied by a decrease in the number of particles. This decrease in the number of particles overpowers the stabilizing effect of increased diameter of the particle. However increasing density ratio has a stabilizing effect on the flow.  
   
The effect of non uniform particle loading is studied by incorporating the parameter $l_p$ which indicates the steepness of the non uniformity of the base state particle concentration. Particle concentration is maximum at the centreline and decays to zero outside the shear layer. Higher value of $l_p$ implies more particles at the jet centreline and rapidly decays to zero. For small Stokes numbers, growth rate for smaller $l_p$ values is greater than that of a more steeper profile, which is true for both the sinuous and varicose modes. For $l_{p}=1$, the growth rate with particles exceed the unladen jet (similar to the constant loading case), while for a higher $l_p$ value, growth rates are smaller than the unladen jet. At large Stokes numbers (via the increasing density ratio path) additional modes apart from the sinuous and varicose modes are seen.
These modes are only seen in the case of variable particle loading and thus could be thought of as concentration driven instability.  The additional modes are very similar to the varicose modes in that the eigen solutions are antisymmetric, however they differ from the varicose modes in that the particle concentrations are an order of magnitude larger than the varicose mode.

\section*{Acknowledgement}
The author is grateful to Prof. Santosh Hemchandra (Department of Aerospace Engg, IISc Bangalore) for useful discussions on volume averaging and the interpretation of the linear stability results.
%\newpage
\appendix
\section{Appendix}
\label{appendix_a}
\subsection{Volume averaging }
 Volume average of any quantity $\mathbf{q}^{(j)}$, where $j$ indicates
the phase 1,2 is defined as

\begin{equation}
\label{vol_avg}
\left\langle \mathbf{q}^{(j)}\left(\mathbf{x},t\right)\right\rangle =\frac{1}{V}\int_{V}\mathbf{q}\left(\mathbf{x}+\boldsymbol{\xi},t\right)\gamma^{(j)}\left(\mathbf{x}+\boldsymbol{\xi},t\right)dV_{\xi}.
\end{equation}

$V=V_{1}+V_{2}$ is the total volume and $V_{1},V_{2}$ are the volume occupied by the individual phases. Fig(\ref{coordinate system}) shows the coordinate system used. The coordinates $(x_{1},x_{2},x_{3})$ is the global coordinate, which contains the averaging volume whose center is at a distance $\mathbf{x}$ from the origin of the global coordinate axes. The local coordinate $\boldsymbol{\xi}$ is used to specify the coordinates within the averaging volume. 

Mathematically, the two phases separated by the interface surface $S^{(i)}$ is expressed as a Dirac delta function. Consider the Heaviside function, in one dimension,

\begin{equation}
H\left(x_{1}-a\right)=\begin{cases}
0 & , x_{1} < a \\
1 & , x_{1} \geqslant a.
\end{cases}
\end{equation}

\begin{equation}
\frac{dH\left(x_{1}-a\right)}{dx_{1}}=\delta\left(x_{1}-a\right)
\end{equation}
where $\delta\left(x_{1}-a\right)$ is the Dirac delta function centered at $a$.
Consider the function $\gamma^{(j)}\left(x_{1}\right)$ defined as

\begin{equation}
\gamma^{(j)}\left(x_{1}\right)=H\left(x_{1}-a_{0}\right)-H\left(x_{1}-a_{1}\right)+H\left(x_{1}-a_{2}\right)-H\left(x_{1}-a_{3}\right)+....
\end{equation}
Taking derivative w.r.t the coordinate $x_{1}$, we get ,

\begin{equation}
\frac{d\gamma^{(j)}\left(x_{1}\right)}{dx_{1}}=\delta\left(x_{1}-a_{0}\right)-\delta\left(x_{1}-a_{1}\right)+\delta\left(x_{1}-a_{2}\right)-\delta\left(x_{1}-a_{3}\right)+....
\end{equation}

If a unit normal vector $\mathbf{n}^{j}$ is defined which points
outward from phase 1 onto phase 2 at the interface 1-2 and $\mathbf{i}$
be the unit vector in the +ve $x_{1}$ direction,

\begin{equation}
\frac{d\gamma^{(j)}\left(x_{1}\right)}{dx_{1}}=-\sum_{k=0}^{n}\mathbf{n}^{j}\cdot\mathbf{i}\delta\left(x_{1}-a_{k}\right).
\end{equation}

We extend the argument for three dimensional case as,

\begin{equation}
\label{grad_gamma}
\nabla\gamma^{(j)}\left(\mathbf{x}\right)=-\mathbf{n}^{j}\delta\left(\mathbf{x}-\mathbf{x}^{(i)}\right),
\end{equation}

where $\mathbf{x}^{(i)}$ denotes the interface location. 

As we noted in the beginning of the paper, integrating the pde over a particular volume results in terms that are averages of the temporal and spatial derivatives of the quantity of interest. However what is usually measured are the temporal and spatial derivatives of the averages. Hence it becomes necessary to relate the averages of temporal and spatial derivatives to the temporal and spatial derivatives of the averages. These are derived below.    

\subsubsection{First averaging theorem}
Applying the definition of volume averaging (Eq. (\ref{vol_avg})) to gradient of a quantity,
\begin{equation*}
\left\langle \nabla\mathbf{q}^{(j)}\right\rangle =\frac{1}{V}\int_{V}\nabla\mathbf{q}\left(\mathbf{x}+\boldsymbol{\xi},t\right)\gamma^{(j)}\left(\mathbf{x}+\boldsymbol{\xi},t\right)dV_{\xi}.
\end{equation*}
This can be written as,
\begin{equation*}
\left\langle \nabla\mathbf{q}^{(j)}\right\rangle =\frac{1}{V}\int_{V}\nabla\left\{ \mathbf{q}\left(\mathbf{x}+\boldsymbol{\xi},t\right)\gamma^{(j)}\left(\mathbf{x}+\boldsymbol{\xi},t\right)\right\} dV_{\xi}-\frac{1}{V}\int_{V}\left\{ \mathbf{q}\left(\mathbf{x}+\boldsymbol{\xi},t\right)\nabla\gamma^{(j)}\left(\mathbf{x}+\boldsymbol{\xi},t\right)\right\} dV_{\xi}.
\end{equation*}

using Eq. (\ref{grad_gamma}) we have,

\begin{equation*}
\left\langle \nabla\mathbf{q}^{(j)}\right\rangle =\frac{1}{V}\int_{V}\nabla\left\{ \mathbf{q}\left(\mathbf{x}+\boldsymbol{\xi},t\right)\gamma^{(j)}\left(\mathbf{x}+\boldsymbol{\xi},t\right)\right\} dV_{\xi}+\frac{1}{V}\int_{V}\left\{ \mathbf{q}^{(j)}\left(\mathbf{x}+\boldsymbol{\xi},t\right)\mathbf{n}^{(j)}\delta\left(\mathbf{x}+\boldsymbol{\xi}-\mathbf{x}^{(i)}\right)\right\} dV_{\xi}.
\end{equation*}

The $\delta$ function is zero everywhere except at 1-2 interface, which leads to,

\textbf{
\begin{equation}
\boxed{\left\langle \nabla\mathbf{q}^{(j)}\right\rangle =\nabla\left\langle \mathbf{q}^{(j)}\right\rangle +\frac{1}{V}\int_{s^{(i)}\left(\mathbf{x},t\right)}\mathbf{q}^{(j)}\left(\mathbf{x}+\boldsymbol{\xi},t\right)\mathbf{n}^{(j)}ds.}
\end{equation}
}

\subsubsection{Second averaging theorem}
Applying the definition of volume averaging (Eq. (\ref{vol_avg})) to temporal derivative of a quantity,
\begin{equation}
\left\langle \frac{\partial\mathbf{q}^{(j)}}{\partial t}\right\rangle =\frac{1}{V}\int_{V}\frac{\partial\mathbf{q}\left(\mathbf{x}+\boldsymbol{\xi},t\right)}{\partial t}\gamma^{(j)}\left(\mathbf{x}+\boldsymbol{\xi},t\right)dV_{\xi}.
\end{equation}

This is can be written as,
\begin{equation}
\left\langle \frac{\partial}{\partial t}\mathbf{q}^{(j)}\right\rangle =\frac{1}{V}\int_{V}\frac{\partial}{\partial t}\left\{ \mathbf{q}\left(\mathbf{x}+\boldsymbol{\xi},t\right)\gamma^{(j)}\left(\mathbf{x}+\boldsymbol{\xi},t\right)\right\} dV_{\xi}-\frac{1}{V}\int_{V}\left\{ \mathbf{q}\left(\mathbf{x}+\boldsymbol{\xi},t\right)\frac{\partial}{\partial t}\left(\gamma^{(j)}\left(\mathbf{x}+\boldsymbol{\xi},t\right)\right)\right\} dV_{\xi}.
\label{time_avg}
\end{equation}

\begin{equation}
\frac{d}{dt}\left(\gamma^{(j)}\left(\mathbf{x}+\boldsymbol{\xi},t\right)\right)=\frac{\partial\gamma^{(j)}}{\partial t}+\frac{dx_{1}}{dt}\left(\frac{\partial\gamma^{(j)}}{\partial x_{1}}\right)+\frac{dx_{2}}{dt}\left(\frac{\partial\gamma^{(j)}}{\partial x_{2}}\right)+\frac{dx_{3}}{dt}\left(\frac{\partial\gamma^{(j)}}{\partial x_{3}}\right),
\label{eq:material_derivative_gamma}
\end{equation}

Since the derivative of $\gamma^{(j)}$ is the Dirac delta function which is non zero only at the interface, the terms $\frac{dx_{1}}{dt}$, $\frac{dx_{2}}{dt}$, $\frac{dx_{3}}{dt}$ are the velocity components of the interface located at  $\boldsymbol{x}^{(i)}$. Travelling with the interface, we have the material derivative equal to $0$. Eq. (\ref{eq:material_derivative_gamma}) becomes,

\begin{equation}
\frac{d}{dt}\left(\gamma^{(j)}\left(\mathbf{x}+\mathbf{\xi},t\right)\right)=\frac{\partial\gamma^{(j)}}{\partial t}+\mathbf{w}^{(i)}\left(\mathbf{x}+\mathbf{\xi},t\right)\cdot\nabla\gamma^{(j)}\left(\mathbf{x}+\mathbf{\xi},t\right)=0,
\label{material_gamma}
\end{equation}

where $\mathbf{w}^{(i)}$ is the velocity of the interface. From Eq. (\ref{time_avg}) and Eq. (\ref{grad_gamma}) we have,

\begin{equation}
\boxed{\left\langle \frac{\partial\mathbf{q}^{(j)}}{\partial t}\right\rangle =\frac{\partial\left\langle \mathbf{q}^{(j)}\right\rangle }{\partial t}-\frac{1}{V}\int_{s^{(i)}}\left\{ \mathbf{q}^{(j)}\mathbf{w}^{(i)}\cdot\mathbf{n}^{(j)}\right\} ds.}
\end{equation}

\subsubsection{Averaging theorem for divergence of a quantity}

Taking double contraction of the first averaging theorem with unit tensor $I$,

\textbf{
\begin{equation}
I:\left\langle \nabla\mathbf{q}^{j}\right\rangle =I:\nabla\left\langle \mathbf{q}^{j}\left(\mathbf{x},t\right)\right\rangle +I:\left(\frac{1}{V}\int_{s_{i}\left(\mathbf{x},t\right)}\mathbf{q}^{j}\left(\mathbf{x}+\xi,t\right)\mathbf{n}^{j}ds\right).
\end{equation}
}

\textbf{Note: }\textit{Double contraction of an identity tensor $I$
with a second order tensor $\left(\mathbf{u}\otimes\mathbf{v}\right)$
is given by}, 

\[
I:\left(u\otimes v\right)=\left(e_{i}\otimes e_{i}\right):\left(\mathbf{u}\otimes\mathbf{v}\right)=\left(e_{i}\cdot\mathbf{u}\right)\left(e_{i}\cdot\mathbf{v}\right)=u_{i}v_{i}=\mathbf{u}\cdot\mathbf{v}.
\]

Applying the above tensor identity,
\begin{equation*}
\left\langle \frac{\partial q_{k}}{\partial x_{l}}\left(e_{k}\otimes e_{l}\right):\left(e_{m}\otimes e_{m}\right)\right\rangle =\frac{\partial\left\langle q_{k}\right\rangle }{\partial x_{l}}\left(\left(e_{k}\otimes e_{l}\right):\left(e_{m}\otimes e_{m}\right)\right)+\frac{1}{V}\int_{s^{i}}q_{k}n_{l}\left(\left(e_{k}\otimes e_{l}\right):\left(e_{m}\otimes e_{m}\right)\right)ds^{i}.
\end{equation*}

This results in ,

\begin{equation}
\boxed{\left\langle \nabla\cdot\mathbf{q}^{(j)}\right\rangle =\nabla\cdot\left\langle \mathbf{q}^{(j)}\right\rangle +\frac{1}{V}\int_{s^{(i)}\left(\mathbf{x},t\right)}\mathbf{q}^{(j)}\left(\mathbf{x}+\boldsymbol{\xi},t\right)\cdot\mathbf{n}^{(j)}ds.}
\end{equation}

\newpage
\section{Matrix form for particle laden planar mixing layer}
\label{appendix_b}
\begin{equation}
\begin{bmatrix}A_{11} & A_{12} & 0 & 0 & 0 & A_{16}\\
A_{21} & A_{22} & A_{23} & A_{24} & A_{25} & A_{26}\\
0 & A_{32} & A_{33} & 0 & A_{35} & 0\\
A_{41} & 0 & 0 & A_{44} & A_{45} & 0\\
0 & A_{52} & 0 & 0 & A_{55} & 0\\
0 & 0 & 0 & A_{64} & A_{65} & A_{66}
\end{bmatrix}\begin{Bmatrix}\tilde{u}_{x}\\
\tilde{u}_{y}\\
\tilde{p}\\
\tilde{u}_{px}\\
\tilde{u}_{py}\\
\tilde{\alpha}
\end{Bmatrix}=\omega\begin{bmatrix}0 & 0 & 0 & 0 & 0 & {\color{red}-i}\\
i\left(1-{\color{red}\Lambda}\right) & 0 & 0 & 0 & 0 & {\color{red}-iU_{x}}\\
0 & i\left(1-{\color{red}\Lambda}\right) & 0 & 0 & 0 & 0\\
0 & 0 & 0 & i & 0 & 0\\
0 & 0 & 0 & 0 & i & 0\\
0 & 0 & 0 & 0 & 0 & i
\end{bmatrix}\begin{Bmatrix}\tilde{u}_{x}\\
\tilde{u}_{y}\\
\tilde{p}\\
\tilde{u}_{px}\\
\tilde{u}_{py}\\
\tilde{\alpha}
\end{Bmatrix}    
\end{equation}

The linearised equations are arranged in a row wise fashion as continuity equation, $x$ momentum, $y$ momentum for the continuous phase and $x$ momentum, $y$ momentum and continuity equation for the dispersed phase.
The terms marked red are the terms present in our formulation in addition to the terms given by Narayanan et al. \cite{Narayananetal2002}(terms colored black)

$A_{11}=\left(ik-{\color{red}ik\Lambda}\right)$, $A_{12}=\left(\frac{d\Lambda}{dy}+\frac{d}{dy}-{\color{red}\Lambda\frac{d}{dy}}\right)$,
$A_{16}={\color{red}-iU_{x}k}$,

$A_{21}=\left\{ \left(1-{\color{red}\Lambda}\right)ikU_{x}-\frac{\left(1-{\color{red}\Lambda}\right)^{{\color{red}2}}}{Re}\left(\frac{d^{2}}{dy^{2}}-k^{2}\right)+\frac{\Lambda\gamma}{St}-\frac{1}{Re}\left(-2\left(1-\Lambda\right)\frac{d\Lambda}{dy}\frac{d}{dy}-\left(1-\Lambda\right)\frac{d^{2}\Lambda}{dy^{2}}\right)\right\} $,

$A_{22}=\left(1-{\color{red}\Lambda}\right)\left(\frac{dU_{x}}{dy}\right) + \left({\color{red}U_{x}\frac{d\Lambda}{dy}}\right) $,
$A_{23}=\left(ik-{\color{red}ik\Lambda}\right)$, $A_{24}=-\left(\frac{\Lambda\gamma}{St}+{\color{red}ikU_{x}\Lambda}\right)$,
\newline
$A_{25}=-\left({\color{red}U_{x}\frac{d\Lambda}{dy}+U_{x}\Lambda\frac{d}{dy}}\right)$ ,

$A_{26}={\color{red}-ikU_{x}^{2}-\frac{1}{Re}\left\{ -\left(1-\Lambda\right)U_{x}\left(\frac{d^{2}}{dy^{2}}-k^{2}\right)-2\left(1-\Lambda\right)\frac{d^{2}U_{x}}{dy^{2}}-2\left(1-\Lambda\right)\frac{dU_{x}}{dy}\frac{d}{dy}+2\frac{dU_{x}}{dy}\frac{d\Lambda}{dy}+U_{x}\frac{d^{2}\Lambda}{dy^{2}}\right\} }$,

$A_{32}=\left\{ \left(1-{\color{red}\Lambda}\right)ikU_{x}-\frac{\left(1-{\color{red}\Lambda}\right)^{{\color{red}2}}}{Re}\left(\frac{d^{2}}{dy^{2}}-k^{2}\right)+\frac{\Lambda\gamma}{St}-\frac{1}{Re}\left(-2\left(1-\Lambda\right)\frac{d\Lambda}{dy}\frac{d}{dy}-\left(1-\Lambda\right)\frac{d^{2}\Lambda}{dy^{2}}\right)\right\} $,

$A_{33}=\left(1-{\color{red}\Lambda}\right)\frac{d}{dy}$, $A_{35}=-\frac{\Lambda\gamma}{St}$, 

$A_{41}=-\frac{1}{St}$, $A_{42}=\left(ikU_{x}+\frac{1}{St}\right)$,
$A_{44}=\left(ikU_{x}+\frac{1}{St}\right)$,

$A_{52}=-\frac{1}{St}$, $A_{55}=\left(ikU_{x}+\frac{1}{St}\right)$,

$A_{64}=ik\Lambda$, $A_{65}=\left(\frac{d\Lambda}{dy}+\Lambda\frac{d}{dy}\right)$,
$A_{66}=ikU_{x}.$ 

\newpage
\section{Non uniform particle loading growth rates for different values of the steepness parameter}
\label{CPL_app}
\begin{figure}[!ht]
    \centering
	\includegraphics[width=0.73\textwidth]{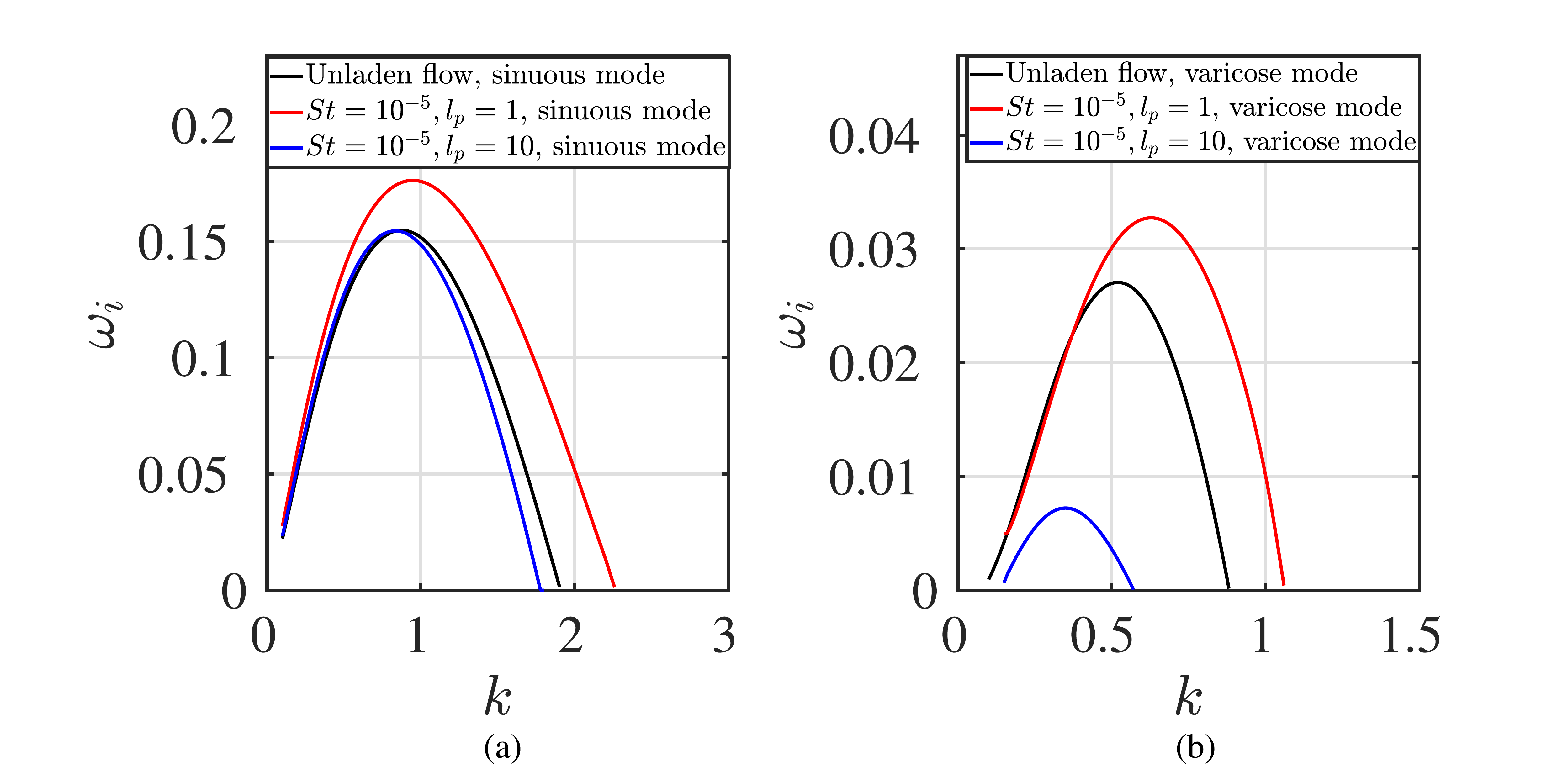}
	\includegraphics[width=0.73\textwidth]{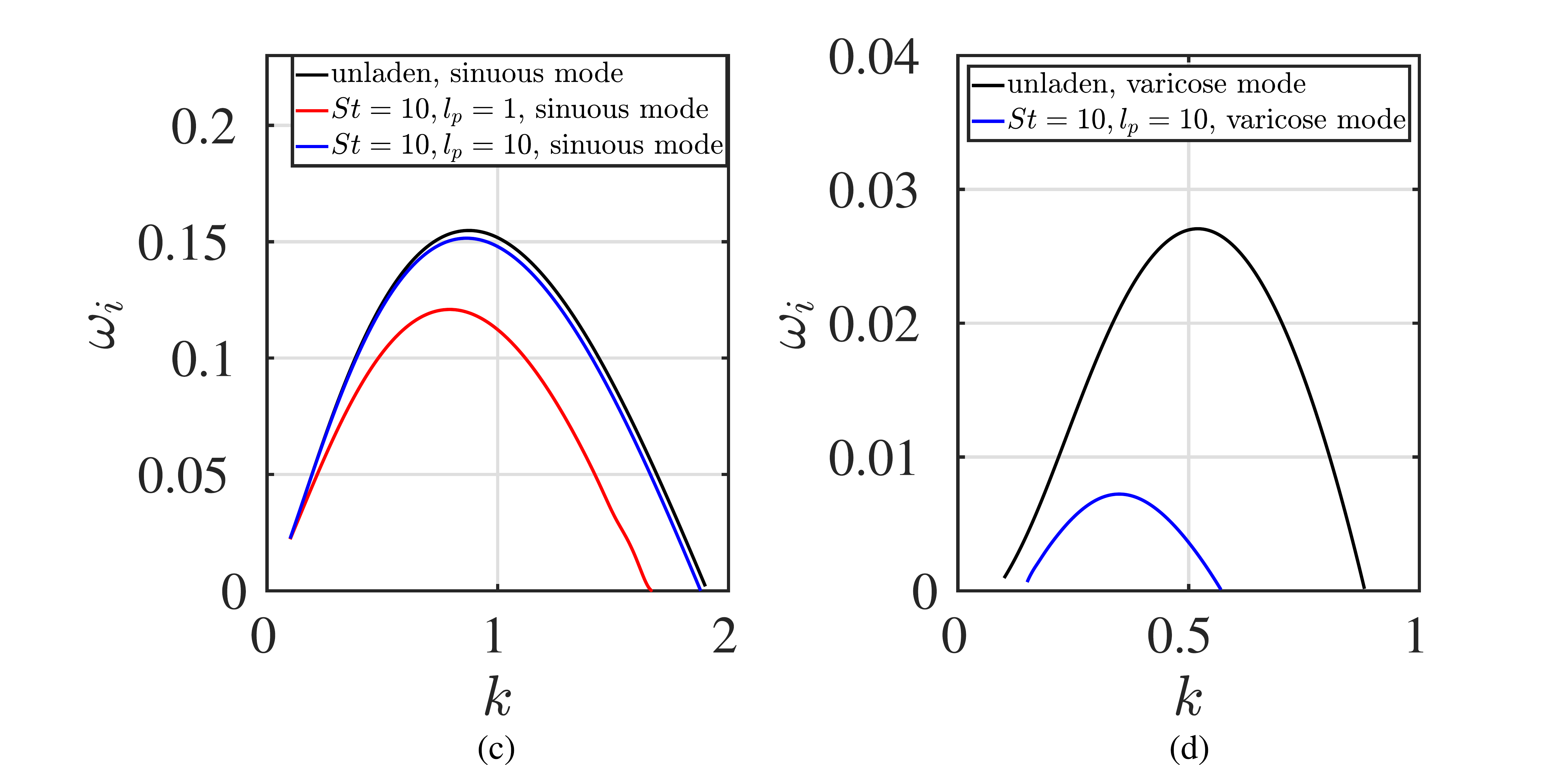}
%	\centering \includegraphics[scale=0.25]{vpl_sinuous_st_100_lp_1_10.pdf}
	\caption{Variable particle loading eigenspectra with the steepness parameter $l_{p}=1,10$. For small Stokes number $St=1\times 10^{-5}$ for $l_{p}=1$ (as in fig (a), (b)), the growth rates with particles exceed the unladen jet. For $l_{p}=10$, the growth rates are smaller than the unladen jet. In the intermediate Stokes number regime at $St=1$ and $10$ ((c) and (d)), addition of particles leads to reduction in temporal growth rates for both $l_{p}=1$ and $10$.} 
	\label{vpl_sinuous} 
\end{figure}	
\newpage
%\begin{figure}[!ht]
%	\includegraphics[width=0.7\textwidth]{varicose_low_st_lp_1_10.pdf}
%	\includegraphics[width=0.7\textwidth]{varicose_intermediate_st_lp_1_10.pdf}
%	\centering \includegraphics[scale=0.25]{varicose_vpl_st_100.pdf}
%	\caption{Variable particle loading growth rates for the varicose mode with the steepness parameter $l_{p}=1,10$. For small Stokes number $St=1\times 10^{-5}$, $St=0.1$, for $l_{p}=1$ (as in fig (a), (b)), the growth rates with particles exceed the unladen jet as seen in the constant loading case. For $l_{p}=10$, the growth rates are smaller than the unladen jet. In the intermediate Stokes number regime at $St=1$ and $10$ ((c) and (d)), addition of particles leads to reduction in temporal growth rates for both $l_{p}=1$ and $10$. At $St=10$ and $l_p=1$, the varicose mode is stabilized and is not tracked here. At large Stokes numbers ($\sim100$) (as in fig (e)) via the increasing density ratio path ($\gamma=10^{4}$) results in damping of the modes compared to the unladen flow for both values of $l_p$. } 
%	\label{vpl_varicose} 
%\end{figure}

\bibliographystyle{unsrt}  
\bibliography{references} 
%\bibliography{cas-refs} 
\end{document}